\documentclass[hidelinks]{IEEEoj}

\usepackage{amsmath,amssymb,amsfonts}
\usepackage{array}
\usepackage{textcomp}
\usepackage{pifont}
\usepackage{url}
\usepackage{verbatim}
\usepackage{graphicx}
\usepackage{cite}
\usepackage[table,dvipsnames]{xcolor} 
\usepackage{orcidlink}    
\usepackage[acronym,shortcuts]{glossaries}
\usepackage{bm}
\usepackage{rotating}
\usepackage[font=small]{caption}
\usepackage{subcaption}
\usepackage{relsize}
\usepackage{soul}
\usepackage{multirow} 
\usepackage{algorithm}
\usepackage{algpseudocode}
\usepackage{algorithmicx}
\usepackage{lipsum}
\usepackage{mathtools}
\usepackage[bottom]{footmisc}
\usepackage{tabularx}
\usepackage{bbm}
\usepackage{booktabs}
\usepackage{stfloats}
\usepackage{amsthm}
\usepackage{array}
\usepackage[none]{hyphenat}

\renewcommand{\bibliofont}{\fontfamily{\rmdefault}\fontsize{7.28}{8.21}\selectfont}

\def\BibTeX{{\rm B\kern-.05em{\sc i\kern-.025em b}\kern-.08em
T\kern-.1667em\lower.7ex\hbox{E}\kern-.125emX}}

\def\thesubsubsectiondis{\unskip\arabic{subsubsection})}

\newtheorem{remark}{Remark}

\DeclareMathOperator*{\argmax}{arg\,max}

\DeclareMathOperator{\diag}{diag}

\newacronym{6G}{6G}{sixth generation}
\newacronym{6G+}{6G+}{later versions of 6G and beyond}
\newacronym{5G}{5G}{fifth generation}
\newacronym{4G}{4G}{fourth generation}
\newacronym{OFDM}{OFDM}{orthogonal frequency division multiplexing}
\newacronym{CP-OFDM}{CP-OFDM}{cyclic prefix orthogonal frequency division multiplexing}
\newacronym{CP}{CP}{cyclic prefix}
\newacronym{OCDM}{OCDM}{orthogonal chirp division multiplexing}
\newacronym{DFT-s-OFDM}{DFT-s-OFDM}{discrete Fourier transform spread orthogonal frequency division multiplexing}
\newacronym{PAPR}{PAPR}{peak-to-average power ratio}
\newacronym{CFO}{CFO}{carrier frequency offset}
\newacronym{NTN}{NTN}{non-terrestrial networks}
\newacronym{D2D}{D2D}{device-to-device}
\newacronym{ISAC}{ISAC}{integrated sensing and communication}
\newacronym{AFDM}{AFDM}{affine frequency division multiplexing}
\newacronym{OTFS}{OTFS}{orthogonal time-frequency space}
\newacronym{ODDM}{ODDM}{orthogonal delay-Doppler multiplexing}
\newacronym{OTSM}{OTSM}{orthogonal time-sequency multiplexing}
\newacronym{FFT}{FFT}{fast Fourier transform}
\newacronym{IFFT}{IFFT}{inverse fast Fourier transform}
\newacronym{ICI}{ICI}{inter-carrier interference}
\newacronym{CPP}{CPP}{chirp-periodic prefix}
\newacronym{OOBE}{OOBE}{out-of-band emission}
\newacronym{AI}{AI}{artificial intelligence}
\newacronym{HAPS}{HAPS}{high-altitude platform station}
\newacronym{MIMO}{MIMO}{multiple-input multiple-output}
\newacronym{V2X}{V2X}{vehicle-to-everything}
\newacronym{V2V}{V2V}{vehicle-to-vehicle}
\newacronym{FT}{FT}{Fourier transform}
\newacronym{AFT}{AFT}{affine Fourier transform}
\newacronym{DFT}{DFT}{discrete Fourier transform}
\newacronym{DZT}{DZT}{discrete Zak transform}
\newacronym{IDZT}{IDZT}{inverse discrete Zak transform}
\newacronym{ISFFT}{ISFFT}{inverse symplectic finite Fourier transform}
\newacronym{HT}{HT}{Heisenberg transform}
\newacronym{IDFT}{IDFT}{inverse discrete Fourier transform}
\newacronym{DAFT}{DAFT}{discrete affine Fourier transform}
\newacronym{IDAFT}{IDAFT}{inverse discrete affine Fourier transform}
\newacronym{SISO}{SISO}{single-input single-output}
\newacronym{AWGN}{AWGN}{additive white Gaussian noise
}
\newacronym{TVIRF}{TVIRF}{time-varying impulse response function}
\newacronym{IDID}{IDID}{integer-delay-integer-Doppler}
\newacronym{IDFD}{IDFD}{integer-delay-fractional-Doppler}
\newacronym{FDFD}{FDFD}{fractional-delay-fractional-Doppler}
\newacronym{ISI}{ISI}{inter-symbol interference}
\newacronym{FIR}{FIR}{finite impulse response}
\newacronym{AoA}{AoA}{angle-of-arrival}
\newacronym{AoD}{AoD}{angle-of-departure}
\newacronym{FLOP}{FLOP}{floating point operation}
\newacronym{DAC}{DAC}{digital-to-analog conversion}
\newacronym{ADC}{ADC}{analog-to-digital conversion}
\newacronym{SotA}{SotA}{state-of-the-art}
\newacronym{ULA}{ULA}{uniform linear array}
\newacronym{UPA}{UPA}{uniform planar array}
\newacronym{PHN}{PHN}{phase noise}
\newacronym{LO}{LO}{local oscillator}\newacronym{CLO}{CLO}{common local oscillator}
\newacronym{SLO}{SLO}{separate local oscillator}
\newacronym{SE}{SE}{spectral efficiency}
\newacronym{CSI}{CSI}{channel state information}
\newacronym{LLR}{LLR}{log-likelihood ratio}
\newacronym{ZF}{ZF}{zero-forcing}
\newacronym{LMMSE}{LMMSE}{linear minimum mean-square error}
\newacronym{MMSE}{MMSE}{minimum mean-square error}
\newacronym{MRC}{MRC}{maximal ratio combining}
\newacronym{MP}{MP}{message passing}
\newacronym{BP}{BP}{belief propagation}
\newacronym{EP}{EP}{expectation propagation}
\newacronym{VN}{VN}{variable node}
\newacronym{FN}{FN}{factor node}
\newacronym{MAP}{MAP}{maximum a posteriori}
\newacronym{PMF}{PMF}{probability mass function}
\newacronym{SBC}{SBC}{symbol-to-bit converter}
\newacronym{AFDMA}{AFDMA}{affine frequency division multiple access}
\newacronym{OFDMA}{OFDMA}{orthogonal frequency division multiple access}
\newacronym{UE}{UE}{user equipment}
\newacronym{UL}{UL}{uplink}
\newacronym{DL}{DL}{downlink}
\newacronym{BER}{BER}{bit error rate}
\newacronym{SNR}{SNR}{signal-to-noise ratio}
\newacronym{RF}{RF}{radio frequency}
\newacronym{CPE}{CPE}{common phase error}
\newacronym{QPSK}{QPSK}{quadrature phase shift keying}
\newacronym{IM}{IM}{index modulation}
\newacronym{PLS}{PLS}{physical layer security}
\newacronym{SM}{SM}{spatial modulation}
\newacronym{PHY}{PHY}{physical layer}
\newacronym{STSK}{STSK}{space-time shift keying}
\newacronym{GSM}{GSM}{generalized spatial modulation}

\def\BibTeX{{\rm B\kern-.05em{\sc i\kern-.025em b}\kern-.08em
T\kern-.1667em\lower.7ex\hbox{E}\kern-.125emX}}
\AtBeginDocument{\definecolor{ojcolor}{cmyk}{0.93,0.59,0.15,0.02}}
\def\OJlogo{\vspace{-4pt}\hskip-4pt\includegraphics[height=18pt]{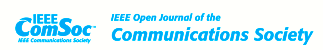}}

\begin{document}
\receiveddate{XX Month, XXXX}
\reviseddate{XX Month, XXXX}
\accepteddate{XX Month, XXXX}
\publisheddate{XX Month, XXXX}
\currentdate{January, 2026}
\doiinfo{OJCOMS.xx.xx}

\title{AFDM: Evolving OFDM Towards 6G+}

\author{
\vspace{1.5ex}
Hyeon Seok Rou\IEEEauthorrefmark{1} \IEEEmembership{(Member, IEEE)}, \\
Vincent Savaux\IEEEauthorrefmark{2} \IEEEmembership{(Senior Member, IEEE)}, ~
Zeping Sui\IEEEauthorrefmark{3} \IEEEmembership{(Member, IEEE)}, \\
Giuseppe Thadeu Freitas de Abreu\IEEEauthorrefmark{1} \IEEEmembership{(Senior Member, IEEE)}, \\
and Zilong Liu\IEEEauthorrefmark{3} \IEEEmembership{(Senior Member, IEEE)}
}
\affil{School of Computer Science and Engineering, Constructor University Bremen, Campus Ring 1, 28759 Bremen, Germany}
\affil{Institute of Research and Technology b-com, 35510 Cesson S{\'e}vign{\'e}, France}
\affil{School of Computer Science and Electronics Engineering, University of Essex, Colchester CO4 3SQ, U.K.}
\corresp{CORRESPONDING AUTHOR: Hyeon Seok Rou (e-mail: hsrou@ieee.org). \vspace{2em}}
\authornote{The work of Z. Liu was supported in part by the U.K. Engineering and Physical Sciences Research Council under Grants EP/Y037243/1 (`TITAN/REVOL6G' and `TITAN/LEGEND6G'), EP/X040569/1 (`HASC/RETHIN6G'), EP/X035352/1 (`DRIVE'), EP/Y000986/1 (`SORT'), and by the Royal Society under Grants IEC\textbackslash{}NSFC\textbackslash{}233292 and IES\textbackslash{}R1\textbackslash{}241212.}
\markboth{AFDM: Evolving OFDM Towards 6G+}{H. S. Rou \textit{et al.~}}

\vspace{-1ex}

\begin{abstract}
As \ac{6G} standardization accelerates, there is growing consensus in favor of \textit{evolutionary} waveforms that add new capabilities while preserving compatibility with the \ac{OFDM} core of 4G and 5G.
This article positions \ac{AFDM} as such a candidate, providing structural robustness for high-mobility communications and \ac{ISAC} over doubly dispersive channels while remaining backward-compatible with the legacy \ac{OFDM} air interface.
We first develop a generalized \ac{FDFD} channel model that accounts for practical pulse-shaping filters and the resulting inter-sample coupling.
Building on this model, we show that the \ac{AFDM} transceiver reuses nearly the entire \ac{OFDM} chain, adding only lightweight digital pre- and post-processing.
We then analyze the impact of hardware impairments such as phase noise and carrier frequency offset, and examine the advanced functionalities enabled by the chirp-parameter domain, including index modulation and physical-layer security.
Assessing reusability across the radio-frequency, physical, and higher layers, we conclude that \ac{AFDM} offers an efficient path toward high-fidelity \ac{6G+} communications.
\end{abstract}


\maketitle
\glsresetall

\vspace{-2ex}
\IEEEpeerreviewmaketitle

\section{Introduction}
\label{sec:introduction}

\subsection{Background}
It is widely anticipated that \ac{6G} systems will bring substantial advances across both technological and application domains, beyond the incremental gains realized in the transition from \ac{4G} to \ac{5G} systems \cite{9349624,10054381}. Besides the development of new 6G use cases, such as the integration of \ac{AI} technology and \ac{ISAC} functionalities, this evolution will also be shaped by recent decisions \cite{R1-2508043,22.870,RP-251881} reached by 3GPP, reaffirming the use of \ac{OFDM}, specifically \ac{CP-OFDM} and \ac{DFT-s-OFDM}, as the baseline waveform for \ac{6G} \cite{dahlman20205g,3gpp.36.211,3gpp.38.211}.
This choice was motivated by the maturity of \ac{OFDM} technology, which has been extensively deployed and refined across multiple wireless systems. A wide range of techniques such as \ac{PAPR} reduction \cite{4287203}, \ac{CFO} calibration \cite{han2005overview}, and advanced channel estimation and tracking methods \cite{ozdemir2007channel,6814271}, have been developed and proven effective in mitigating inherent limitations of \ac{OFDM}, further strengthening it as a solid foundation for the \acf{6G} air interface design \cite{Saad_6G}.

However, this decision should not be taken to mean that the 6G waveform is finalized. In fact, 3GPP leaves the door open for investigating new waveforms designed to address the limitations of \ac{OFDM} in future 6G use cases and to facilitate the integration of multiple functionalities, such as sensing and computing. 

Candidate waveform solutions for \ac{6G+} must remain closely aligned with the \ac{OFDM} legacy, emphasizing backward compatibility, coexistence, and hardware reusability \cite{R1-2508043}. 
This creates a stringent requirement: any \ac{6G+} waveform must reuse the existing \ac{OFDM} infrastructure as much as possible while addressing challenges that conventional \ac{OFDM} cannot overcome. 
Firstly, next-generation waveforms must cope with the deteriorating effects of high-mobility channels \cite{Bliss_DDchannel} and/or high-frequency bands (e.g., millimeter wave or Terahertz), which generally suffer from large path loss and strong phase noise and/or carrier frequency offsets. 
Secondly, next-generation waveforms are expected to offer improved sensing capabilities compared to OFDM counterparts \cite{Koivunen_waveform}.

These requirements are central to the evolution envisioned for \ac{6G+}.
For example, \ac{NTN} scenarios, consisting of satellite systems, \ac{HAPS}, and flying drones \cite{araniti2021toward}, are characterized by high moving speeds that induce Doppler shifts/spread far beyond the tolerance of \ac{OFDM}, even under advanced mitigation strategies.
Similarly, in direct \ac{D2D} and sidelink communications, which are expected to play a central role in \ac{6G} for applications such as \ac{V2V} and \ac{V2X}, or drone-to-drone connectivity \cite{noor20226g}, the non-stationary mobile channels and the lack of centralized synchronization further expose the vulnerability of \ac{OFDM} to Doppler impairments \cite{wang2006performance}.

An additional dimension arises with \ac{ISAC}, whose demand is widely present in the aforementioned scenarios \cite{Liu_ISAC}.
In \ac{ISAC} systems, the waveform must not only support reliable data transmission but also enable accurate estimation of the range and velocity for effective environment sensing \cite{Rou_DDWaveforms}.
In a highly dynamic 6G system, such a requirement may not be fulfilled by \ac{OFDM} as it lacks the intrinsic robustness to joint delay-Doppler variations.

\vspace{-2ex}
\subsection{Why AFDM?}

In the context of the above, it is clear that a new waveform paradigm is needed to truly enable \ac{6G} to its full extent, one that simultaneously preserves \ac{OFDM}'s efficiency advantages while embracing, rather than counteracting, the characteristics of doubly dispersive channels. 

To this end, \ac{AFDM} \cite{Bemani_AFDM} has emerged recently as one of the waveform paradigms proposed for this purpose.
Unlike conventional \ac{OFDM}, which modulates each information symbol onto a separate orthogonal subcarrier, \ac{AFDM} employs chirp-based subcarriers whose instantaneous frequencies increase linearly over time. 
With the \ac{AFT}, AFDM spreads each symbol across the time and frequency domains, thereby enjoying the full channel diversity as well as strong resilience to Doppler (an intrinsic property of chirps).

In addition, the chirp parameters of \ac{AFDM} are tunable, enabling flexible adaptation to the characteristics of the channel to ensure maximum diversity for optimal communications performance over a variety of channel settings \cite{AFDM_6G_Rou,11003079}, in addition to the ability to support advanced functionalities such as index modulation \cite{tao25,liu25ieeetwc,10943004,zhang26dualIM,qian25cim} and physical-layer security \cite{savaux26ieeewcl,Wang_ICC25,Rou_WCL25}.
In light of these advantages of the affine frequency domain, \ac{AFDM} has also motivated in various derivative and related schemes exploiting the chirp-based waveform structure \cite{11322797,sui2025multi,ranasinghe2025affine,yi2025non,senger2025affine}.

Furthermore, from hardware implementation perspective, \ac{AFDM} can be regarded as a direct extension and generalization of \ac{OFDM}, where the latter is reduced to \ac{AFDM} under specific parameter settings \cite{AFDM_6G_Rou}.
Namely, the transceiver processing chain of \ac{AFDM} remains largely intact, with only an additional linear-complexity pre- and post-processing step applied around the conventional \ac{IFFT}/\ac{FFT} operations of \ac{OFDM}.
Consequently, \ac{AFDM} is inherently compatible to \ac{OFDM}, allowing the reuse of key signal processing blocks such as \ac{IFFT}/\ac{FFT} modules, parallel-to-serial conversion, time-frequency resource management, and prefix handling. 
Moreover, the extensive body of \ac{OFDM} enhancements, including \ac{PAPR} reduction, pulse shaping, and spectral containment methods, can be applied to \ac{AFDM} with only minor adaptation, given the fundamental similarity in signal structure across time-frequency subcarriers.

On the other hand, competing waveform proposals such as \ac{OTFS} \cite{Wei_OTFS}, \ac{OTSM} \cite{10183832}, \ac{ODDM} \cite{9829188},
and Zak-\ac{OTFS} \cite{lampel2022otfs} adopt a distinct signaling framework.
Instead of operating in the time-frequency domain, as that in \ac{OFDM} and \ac{AFDM}, these schemes modulate the information symbols onto the two-dimensional delay-Doppler domain \cite{hong2022delay,deng25unifying}, with the aim to exploit the inherent delay-Doppler sparsity of doubly dispersive channels \cite{Bliss_DDchannel}.
As a result, they have been shown to also achieve strong theoretical performance in high-mobility scenarios \cite{li2021performance,Surabhi_OTFSdiversity}.

However, such performance comes at the cost of additional system-level modifications relative to \ac{OFDM}, including different resource allocation approaches, domain transforms, and dimensions, as well as modified multiple access schemes, particularly under practical pulse-shaping conditions. \mbox{FFT-based} simplified implementations of \ac{OTFS} (e.g., under rectangular pulse shaping) can reuse parts of the \ac{OFDM} processing chain, but full pulse-shaped implementations require 2D processing.
Consequently, their integration into existing OFDM-dominated architectures would require additional modifications at both the physical and higher layers, thereby reducing compatibility \cite{AFDM_6G_Rou}.

This requirement specifically runs counter to recent standardization decisions favoring \ac{OFDM}-based evolution and therefore complicates the adoption of such schemes within the current \ac{OFDM}-based evolutionary roadmap.

\vspace{-1ex}
\subsection{Contributions of this article}

A major objective of this article is to analyze \ac{AFDM} as a candidate for next-generation air interfaces, capable of addressing the requirements of \ac{6G}, including extreme mobility, \ac{NTN} operation, \ac{D2D} communications, \ac{ISAC}, and PHY-security, while preserving architectural compatibility with the \ac{OFDM} legacy.

Going beyond the current state-of-the-art of AFDM, reviewing its reported performance advantages over \ac{OFDM} in challenging \ac{6G} channel conditions and various use cases \cite{bemani21_iswcs,Bemani_AFDM,Rou_DDWaveforms,bemani2024integrated,luo2025target,ni2025integrated,luo2025novel,rou2025normalized,ni2025ambiguity,zhu2024afdm,zhang2025afdm,ramadan2026performance},
this article provides a comprehensive analysis of \ac{AFDM} from a practical implementation perspective. 
Specifically, we examine its feasibility in direct comparison with \ac{OFDM} across several critical dimensions:
\begin{itemize}
\item Compatibility of \ac{AFDM} with \ac{OFDM} hardware and signal processing blocks, reusing existing \ac{OFDM} modules such as \ac{FFT}, pulse shaping, prefixing, and resource management. \vspace{0.5ex}
\item Implementation complexity and overhead of \ac{AFDM} compared to \ac{OFDM}, in aspects such as transceiver processing, channel estimation, and detection. \vspace{0.5ex}
\item Robustness of \ac{AFDM} in the presence of the challenging \ac{FDFD} conditions, channel estimation errors, and hardware impairments. \vspace{0.5ex}
\item Advanced functionalities enabled by \ac{AFDM}, such as multiple access and coexistence with \ac{OFDM}, chirp-domain modulation, and physical-layer security.
\end{itemize}

Finally, we qualitatively evaluate the reusability of \ac{AFDM} across the \ac{RF}, \ac{PHY}, and higher layers, against the legacy \ac{OFDM}, indicating its potential for smooth and evolutionary integration into \ac{6G+}.

\vspace{-1ex}
\subsection{Organization of this article}
The remainder of this article is structured as follows:
Section~\ref{sec:system_model} introduces the system and channel models adopted throughout the article, including a novel general \ac{FDFD} channel input-output model based on physical pulse footprints that accurately captures inter-sample coupling in practical transceiver implementations;
Section~\ref{sec:compatibility_AFDM_OFDM} details the transceiver architecture of \ac{AFDM}, highlighting its compatibility and reusability with respect to \ac{OFDM}, considering various aspects, such as prefixing, pulse shaping, and resource management;
Section~\ref{sec:analysis_AFDMOFDM} presents a variety of thorough analyses of \ac{AFDM} in comparison with \ac{OFDM}, including implementation complexity and scalability, channel estimation and detection architecture, robustness to impairments like \ac{PHN} and \ac{CFO}, and advanced functionalities including \ac{MIMO}, chirp-domain modulation and physical-layer security;
Section~\ref{sec:analysis_AFDMOFDM} additionally concludes with a qualitative, layer-wise summary of the reusability of \ac{AFDM} compared with legacy \ac{OFDM}; Section~\ref{sec:open_challenges} then outlines the open research challenges and future directions, including synchronization and standardization pathways, as well as the concluding remarks.

\noindent \emph{Notation:} Scalars, vectors, and matrices are denoted by lowercase, bold lowercase, and bold uppercase letters, respectively, e.g., $x$, $\mathbf{x}$, and $\mathbf{X}$. 
The $i$-th element of a vector $\mathbf{x}$ is denoted by $x_i$, and the $(i,j)$-th element of a matrix $\mathbf{X}$ by $x_{i,j}$.
When subscripts become overly dense, the alternative notations $[\mathbf{x}]_i$ and $[\mathbf{X}]_{i,j}$ may be used for clarity. 
The transpose, Hermitian transpose, and inverse of a matrix $\mathbf{X}$ are denoted by $\mathbf{X}^{\mathsf{T}}$, $\mathbf{X}^{\mathsf{H}}$, and $\mathbf{X}^{-1}$, respectively. 
The operators $\otimes$ and $\odot$ respectively denote the Kronecker and Hadamard product, and $\ast$ the linear convolution.
The sets of natural, integer, rational, real, and complex numbers are denoted by $\mathbb{N}$, $\mathbb{Z}$, $\mathbb{Q}$, $\mathbb{R}$, and $\mathbb{C}$, respectively.
$\delta(\cdot)$ represents the Dirac delta function, while $\delta[\cdot]$ denotes the Kronecker delta.
The modulo-$N$ operation is denoted by $(\cdot)_N$.

\vspace{-1ex}
\section{System Model}
\label{sec:system_model}

We first present the continuous-time \ac{SISO} doubly dispersive channel model and its corresponding discrete-time equivalent representation, leading to the general \acf{FDFD} input-output relation in matrix form, which will be used throughout this article.
We note that the continuous- and discrete-time channel models and the \ac{IDID}/\ac{IDFD} special cases are standard background. The generalized \ac{FDFD} matrix decomposition of Section~\ref{sec:system_model}-\ref{subsec:matrix_form_IO} explicitly incorporates the transmit pulse-shaping kernel. To the best of our knowledge, our generalized \ac{FDFD} case is a novel contribution (see Remark~1).

\vspace{-1ex}
\subsection{Continuous-Time SISO Doubly Dispersive Channel}
\label{subsec:cont_time_SISO_model}

Consider a wireless channel with $P$ resolvable scattering propagation paths, characterized by its continuous-time \ac{TVIRF} given by \cite{Bliss_DDchannel} \vspace{-1ex}
\begin{equation}
h(t,\tau) = \sum_{p=1}^{P} h_p \, e^{j 2 \pi \nu_p t} \, \delta(\tau - \tau_p),
\label{eq:cont_time_SISO_tvirf}
\vspace{-0.25ex}
\end{equation}
where $h_p \in \mathbb{C}$ denotes the complex gain, $\tau_p \in \mathbb{R}$ denotes the propagation delay, and $\nu_p \in \mathbb{R}$ denotes the associated Doppler frequency shift of the $p$-th path, and the combined influence of the path delays $\tau_p$ (frequency selectivity) and Doppler shifts $\nu_p$ (time selectivity) gives rise to the so-called doubly dispersive fading environment.

The maximum delay spread and maximum Doppler spread of the channel are defined as $\tau_{\max} \in \mathbb{R}^+$ and $\nu_{\max} \in \mathbb{R}^+$, respectively, such that $\tau_p \in [0, \tau_{\max}]$ and $\nu_p \in [-\nu_{\max}, \nu_{\max}]$ for $p \in \{1,\ldots,P\}$, and the channel is considered underspread if $\tau_{\max} \nu_{\max} <\!\!< 1$, and overspread otherwise.

Then, given an arbitrary time-domain transmit signal $s(t)$, the corresponding input-output relation in the continuous-time domain is given by a linear convolution with the \ac{TVIRF} in \eqref{eq:cont_time_SISO_tvirf}, which is expressed as

\begin{gather} 
\nonumber \\[-6ex]
\!\!\!\!r(t) \!=\! s(t) \ast h(t,\tau) \!+\! w(t) =\!\! \int\limits_{-\infty}^{+\infty} \!\! h(t,\tau) s(t\! -\! \tau) \,d\tau \!+\! w(t),\!\!
\label{eq:cont_time_SISO_io}
\end{gather}
where $r(t)$ and $w(t)$ represent the time-domain received signal over the doubly dispersive channel and the \ac{AWGN} signal, respectively.

\vspace{-1ex}
\subsection{Discrete-Time Equivalent Model}
\label{subsec:disc_time_SISO_model}

To represent the continuous system model in a manner compatible with discrete signal processing, the continuous-time signal is sampled at a sampling interval of $T_\mathrm{s} \triangleq \frac{1}{f_\mathrm{s}}$, where $f_\mathrm{s}$ is the sampling frequency in Hz.

Then, denoting $n \in \{0,\ldots,N-1\}$ as the sample index of the discrete sequence, the discrete-time channel input-output relation is expressed as
\begin{equation}
r[n] = \sum_{p=1}^{P} h_p \, e^{j 2 \pi \nu_p n T_\mathrm{s}} \,s\!\big[n - \tfrac{\tau_p}{T_\mathrm{s}}\big] + w[n], 
\label{eq:disc_time_SISO_channel}
\end{equation}
where $r[n]$, $s[n]$, and $w[n]$ are the sampled received, transmitted, and noise signal sequences, respectively.

Then, defining the normalized delay indices and the normalized digital Doppler shift indices as \vspace{0.5ex}
\begin{equation}
\ell_p \triangleq \frac{\tau_p}{T_\mathrm{s}},\ f_p \triangleq N \nu_p T_\mathrm{s} = N\tfrac{\nu_p}{f_\mathrm{s}},
\label{eq:normalized_delay}  \vspace{0.5ex}
\end{equation}
where $N$ is the total number of samples in the sequence, the discrete-time input-output relation in \eqref{eq:disc_time_SISO_channel} can be rewritten as
\begin{equation}
r[n] = \sum_{p=1}^{P} h_p \, e^{j 2 \pi f_p \frac{n}{N}} \, s[n - \ell_p] + w[n].
\label{eq:disc_time_SISO_channel_normalized}
\end{equation}

Notice that the delay index $\ell_p$ represents the physical delay (in seconds) normalized by the sampling interval $T_\mathrm{s}$, while the digital Doppler shift index $f_p$ represents the physical Doppler shift (in Hz) normalized by the subcarrier spacing $\frac{f_\mathrm{s}}{N}$.
Therefore, depending on the channel scenario (i.e., the delay-Doppler profile) and the sampling frequency, the normalized indices $\ell_p$ and $f_p$ may take either integer or fractional values, categorized into the following three cases: 

\vspace{-2ex}
\subsubsection*{\textbf{Case 1: Integer Delay and Integer Doppler (IDID)}} 
\label{subsubsec:int_delay_int_doppler}

\noindent When both the normalized path delays and digital Doppler shifts are integer-valued, $i.e.$, $\ell_p \in \mathbb{Z}_{\ge 0}$ and $f_p \in \mathbb{Z}$, the multipath components are perfectly aligned with the discrete sampling and frequency grids.
For this case, here onwards referred to as \ac{IDID} channel, the discrete-time baseband received signal can be described by \eqref{eq:disc_time_SISO_channel_normalized} directly without concern as $\ell_p$ is an integer, and therefore the shifted samples $s[n - \ell_p]$ are well-defined.
Furthermore, the phase rotation term $e^{j 2 \pi f_p \frac{n}{N}}$ is an integer multiple of $e^{\frac{j 2 \pi}{N}}$, which corresponds to the fundamental frequency resolution of the discrete Fourier basis.

This is a common assumption adopted in most existing works, as it greatly simplifies the analysis and design of communication systems, reveals the core characteristics and behavior of each delay-Doppler tap without interference or added complexity.
However, such an assumption is often unrealistic in practical scenarios, where the delay and Doppler parameters are typically not perfectly aligned with the discrete grids, especially the Doppler shifts, necessitating more general models.

\vspace{-3ex}
\subsubsection*{\textbf{Case 2: Integer Delay and Fractional Doppler (IDFD)}}
\label{subsubsec:int_delay_frac_doppler}

\noindent In the \ac{IDFD} case, the assumption of integer Doppler shifts is relaxed to reflect practical high-mobility environments better.
Here, the normalized path delays remain integer-valued while the digital Doppler shifts are allowed to take non-integer (fractional) values, $i.e.$, $\ell_p \in \mathbb{Z}_{\ge 0}$ and $f_p \in \mathbb{R}$. 
Note that the assumption of integer delays remains relatively robust, as the high sampling rate $f_\mathrm{s}$ provides a sufficiently fine resolution $T_\mathrm{s}$ to align multipath arrivals with the discrete, integer-indexed grid.

In this case, the discrete-time model retains the structural form of \eqref{eq:disc_time_SISO_channel_normalized}, as the integer-shifted samples $s[n - \ell_p]$ are still aligned with the sampling instances. 
However, a critical distinction emerges in the phase rotation term $e^{j 2 \pi f_p \frac{n}{N}}$, where the Doppler shifts no longer correspond to integer multiples of the fundamental frequency resolution $e^{\frac{j 2 \pi}{N}}$. 
This misalignment breaks the periodic orthogonality of the discrete Fourier basis, the cornerstone of legacy waveforms such as \ac{OFDM}, thereby introducing significant \ac{ICI} and spectral leakage.

\vspace{-3ex}
\subsubsection*{\textbf{Case 3: Fractional Delay and Fractional Doppler (FDFD)}}
\label{subsubsec:frac_delay_frac_doppler}

\noindent Finally, the \ac{FDFD} model represents the most generalized and physically accurate case, where both the normalized delays and digital Doppler frequencies take arbitrary real values, $i.e.$, $\ell_p \in \mathbb{R}_{\ge 0}$ and $f_p \in \mathbb{R}$. 
While integer delays are often acceptable in systems with moderate delays, fractional delays may become significant in ultra-high-precision synchronization and sensing applications.

In this case, the time-shifted argument $(n \!-\! \ell_p)$ in the transmit sequence is no longer strictly integer-valued.
Consequently, the term $s[n - \ell_p]$ cannot be obtained by a simple index shift as in \eqref{eq:disc_time_SISO_channel_normalized}.
This \textit{fractionally delayed} version of the discrete sequence physically corresponds to the continuous-time signal being reconstructed from the discrete samples via a convolutional pulse-shaping filter $p(t)$, shifted by the continuous delay $\tau_p$, and subsequently resampled at the receiver's sampling grid.

This process is mathematically expressed through a discrete-time interpolation, given by
\begin{equation}
s[n - \ell_p] = \sum_{m=0}^{N-1} s[m] \, g\big((n - m) - \ell_p\big),
\label{eq:fractional_delay_interp}
\end{equation}
where $g(\cdot)$ denotes the effective discrete-time pulse kernel, derived from the underlying pulse-shaping filter $p(t)$ such that $g(m) = p(mT_\mathrm{s})$, with sampling interval $T_\mathrm{s}$.

Note that unlike idealized digital interpolation (e.g., sinc-based upsampling), which is independent of the waveform, the kernel $g(\cdot)$ in this model reflects the actual hardware pulse footprint (e.g., raised cosine or rectangular).

Then, by substituting~\eqref{eq:fractional_delay_interp} into the discrete-time input-output relation in \eqref{eq:disc_time_SISO_channel}, we obtain the expanded discrete-time representation for a \ac{FDFD} channel as
\begin{equation}
r[n] = \sum_{p=1}^{P} h_p \, e^{j 2 \pi f_p \frac{n}{N}} 
\sum_{m=0}^{N-1} s[m] \, g\big((n - m) - \ell_p\big) + w[n].
\label{eq:general_DDsystem_FDFD_expanded}
\end{equation}

This formulation reveals that the choice of the physical pulse shape $p(t)$ directly governs the inter-sample coupling in the discrete domain. 
Consequently, this model provides a high-fidelity foundation that simultaneously accounts for spectral leakage due to fractional Doppler and the inter-sample interference caused by fractional delays, which is essential for evaluating next-generation waveforms in \ac{6G} propagation environments.

\subsection{Generalized Matrix-form Input-Output Relationship}
\label{subsec:matrix_form_IO}

To facilitate efficient and insightful analysis, the discrete-time input-output relation for the \ac{FDFD} channel in \eqref{eq:general_DDsystem_FDFD_expanded} can be equivalently expressed using a matrix-vector formulation.
Namely, given a transmitted symbol block $\mathbf{s} \in \mathbb{C}^{N \times 1}$ of $N$ samples, the received vector $\mathbf{r} \in \mathbb{C}^{N \times 1}$ is obtained under a circular convolution framework with a general prefix. 

The signal sequence $s[n]$ is prepended with a prefix of length $N_\mathrm{cp}$, described by
\begin{equation}
s[n'] = s[N + n'] \cdot e^{j2\pi\phi_\mathrm{cp}(n')}, ~\forall n' \in \{-N_\mathrm{cp}, \ldots, -1\},
\label{eq:general_prefix}
\end{equation}
where $\phi_\mathrm{cp}(\cdot)$ is the waveform-dependent phase-offset function of the prefix samples, elaborated in Section \ref{sec:compatibility_AFDM_OFDM}.

Given the above, the overall convolution matrix $\mathbf{H} \in \mathbb{C}^{N \times N}$ represents the doubly dispersive channel as
\begin{equation}
\mathbf{r} = \mathbf{H} \mathbf{s} + \mathbf{w} = \sum_{p=1}^{P} h_p \mathbf{H}_p \mathbf{s} + \mathbf{w},
\label{eq:matrix_IO_general}
\end{equation}
where $h_p \in \mathbb{C}$ denotes the complex path gain of the $p$-th multipath component, identical to that of \eqref{eq:cont_time_SISO_tvirf}, \eqref{eq:disc_time_SISO_channel}, and \eqref{eq:general_DDsystem_FDFD_expanded}, $\mathbf{H}_p$ is the circular convolutional matrix excluding the path gain coefficient $h_p$ for the $p$-th path, $\mathbf{r} \in \mathbb{C}^{N \times 1}$ is the received signal vector collecting the $N$ useful samples of $r[n]$ after discarding the prefix, and $\mathbf{w} \in \mathbb{C}^{N \times 1}$ represents the noise received per sample.

For the \ac{IDID} and \ac{IDFD} cases, where the normalized delays $\ell_p$ are integer-valued, the per-path channel matrix $\mathbf{H}_p$ can be decomposed as the product of a diagonal phase-rotation matrix and a cyclic permutation matrix, yielding \cite{Ouyang_OCDM,Rou_DDWaveforms,Bemani_AFDM,hong2022delay,Bliss_DDchannel}
\begin{equation}
\mathbf{H}_p = \mathbf{\Phi}_p \cdot \mathbf{V}^{f_p} \cdot \mathbf{\Pi}^{\ell_p} \in \mathbb{C}^{N \times N}.
\label{eq:Hp_integerdelay}
\end{equation}

The three matrix components in \eqref{eq:Hp_integerdelay} are, respectively:
\begin{itemize}
\item[1)] diagonal phase-offset matrix $\mathbf{\Phi}_p$, which models the phase-offset of the utilized prefix in \eqref{eq:general_prefix} as
\begin{align}
\mathbf{\Phi}_p & \triangleq \mathrm{diag}(\boldsymbol{\phi}_p) \in \mathbb{C}^{N \times N} \label{eq:CPP_phase_matrix} \\[0.5ex] 
& \!\!\!\!\!\!= \mathrm{diag}\Big(\big[\underbrace{e^{-j2\pi\cdot \phi_\mathrm{cp}(\ell_p)}, \cdots, e^{-j2\pi\cdot \phi_\mathrm{cp}(1)}}_{\ell_p \;\text{terms}}, \!\underbrace{1, \cdots\!\vphantom{e^{(x)}}, 1}_{N - \ell_p\;\! \text{ones}}\!\big]\Big); \nonumber
\end{align}
\item[2)] diagonal Doppler shift matrix $\mathbf{V}^{f_p}$, which models the effect of the normalized digital Doppler shifts $f_p$ based the $N$-th roots-of-unity matrix
\begin{align}
\mathbf{V} &\triangleq \mathrm{diag}(\mathbf{v}) \in \mathbb{C}^{N \times N} \label{eq:rootofunity_matrix} \\[0.5ex]
& =\mathrm{diag}\Big(\Big[1, e^{-j2\pi\frac{1}{N}}, \cdots, e^{-j2\pi\frac{N-1}{N}}\Big]\Big), \nonumber
\end{align}
which when raised to the $f_p$-th power as $\mathbf{V}^{f_p}$, is the diagonal matrix representing the phase rotation due to the Doppler shift; and
\item[3)] forward cyclic shift matrix $\mathbf{\Pi}^{\ell_p}$, which models the integer delay shift of $\ell_p$ samples, defined based on the single-step forward cyclic shift matrix $\mathbf{\Pi}$
\begin{equation}
\label{eq:forwardcyclicshift_matrix}
\mathbf{\Pi} = \text{\scalebox{0.85}{$
\begin{bmatrix}
0      & 0      & \cdots & 0      & 1      \\
1      & 0      & \cdots & 0      & 0      \\
0      & 1      & \ddots & \vdots & \vdots \\
\vdots & \ddots & \ddots & 0      & 0      \\
0      & \cdots & 0      & 1      & 0
\end{bmatrix}$}} \in \mathbb{C}^{N \times N},
\end{equation}
such that left-multiplying a vector by $\mathbf{\Pi}$, $i.e.$, $\mathbf{\Pi} \mathbf{x}$, corresponds to a forward shift of $\mathbf{x}$, and therefore $\mathbf{\Pi}^{\ell_p}$ corresponds to a cyclic forward shift of $\ell_p$ indices.
\end{itemize}

As can be seen with the above formulation, the \ac{IDID} and \ac{IDFD} cases benefit from a highly sparse channel matrix $\mathbf{H}_p$, which consists of only a single non-zero element per row and column, due to the perfect alignment of the delays $\ell_p$ with the sampling grid (diagonal phase matrices and permutation matrix structure).

However, in the \ac{FDFD} case with \textbf{non-integer} delays $\ell_p$, it can be observed that such efficient formulations in \eqref{eq:CPP_phase_matrix} and \eqref{eq:forwardcyclicshift_matrix} based on integer delays $\ell_p$ are no longer valid, in a similar construction to \eqref{eq:general_DDsystem_FDFD_expanded}.

Namely, the separation of the phase-offset matrix and the shift matrix into independent diagonal and permutation operators $\mathbf{\Phi}_p$ and $\mathbf{\Pi}^{\ell_p}$ is no longer possible due to the inter-sample interference caused by fractional delays and the effective discrete-time pulse kernel $g(\cdot)$, and the generalized per-path \ac{FDFD} channel matrix can be expressed as
\begin{equation}
\mathbf{H}_p \triangleq \mathbf{V}^{f_p} \cdot \mathbf{\Psi}(\ell_p) \in \mathbb{C}^{N \times N},
\label{eq:Hp_FDFD_generalized}
\end{equation}
where $\mathbf{V}^{f_p}$ models the sample-wise phase rotations due to the digital Doppler shift $f_p$, identical to the \ac{IDID} and \ac{IDFD} cases, while the new \textit{chirp-periodically circulant} fractional delay matrix $\mathbf{\Psi}(\ell_p)$ captures the combined effect of the fractional delay and the corresponding prefix phase-offset, defined as
\begin{equation}
\mathbf{\Psi}(\ell_p) \triangleq \big( \mathbf{G}(\ell_p) + \mathbf{\Phi}(\ell_p) \big) \in \mathbb{C}^{N \times N},
\label{eq:Hp_FDFD_modular}
\end{equation}
with the constituting effective interpolated delay matrix $\mathbf{G}(\ell_p)$ and the effective interpolated delay \ac{CPP} matrix $\mathbf{\Phi}_p(\ell_p)$ defined element-wise as
\begin{align}
\hspace{-1.5ex}[\mathbf{G}(\ell_p)]_{n,m} &= g(n - m - \ell_p),\label{eq:fracdel_interp_matrix}
\\[1ex]
\hspace{-1.5ex}[\mathbf{\Phi}(\ell_p)]_{n,m} &= g(n - m - \ell_p + N) \cdot e^{j2\pi\phi_\mathrm{cp}(n - m - \ell_p)},\!\!\!
\label{eq:fracCPP_interp_matrix}
\end{align}
where $\mathbf{G}(\ell_p)$ is generally dense and Toeplitz.

\begin{figure*}[b]
\vspace{0.5em}
\centering
\begin{subfigure}[b]{0.3\textwidth}
\centering
\includegraphics[width=\textwidth]{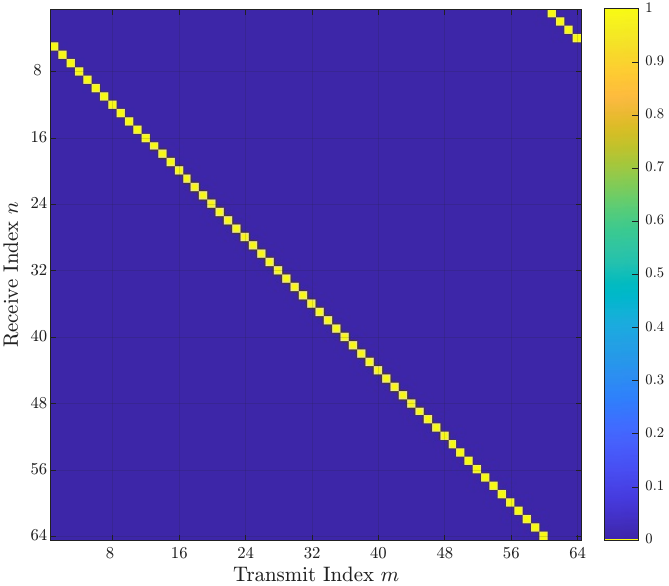}
\caption{Conventional IDID model following \eqref{eq:Hp_integerdelay}.}
\label{fig:idid_classic}
\end{subfigure}
\hfill
\begin{subfigure}[b]{0.3\textwidth}
\centering
\includegraphics[width=\textwidth]{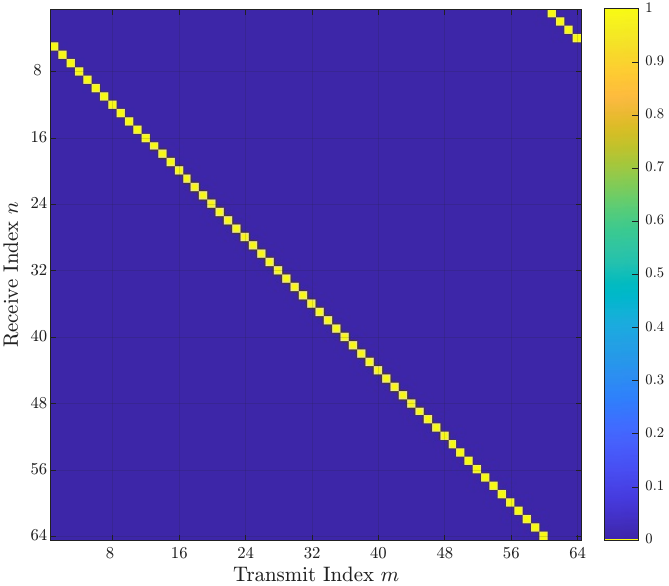}
\caption{Generalized IDID model following \eqref{eq:Hp_FDFD_generalized}.}
\label{fig:idid_modular}
\end{subfigure}
\hfill
\begin{subfigure}[b]{0.3\textwidth}
\centering
\includegraphics[width=\textwidth]{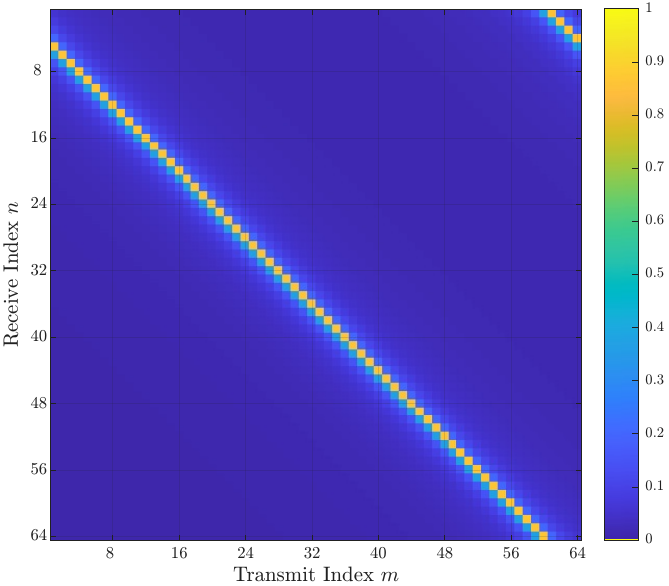}
\caption{Generalized FDFD model following \eqref{eq:Hp_FDFD_generalized}.}
\end{subfigure}

\vspace{-1.5ex}
\caption{Illustration of a single path within the channel matrix under IDID and FDFD scenarios, with $N=64$.
(a) shows the conventional IDID sparse permutation model of eq. \eqref{eq:Hp_integerdelay} with $\ell_p = 4$ and $f_p = 2$; 
(b) shows the proposed generalized FDFD formulation of eq. \eqref{eq:Hp_FDFD_generalized} but under the same integer conditions; 
and (c) shows the proposed generalized FDFD formulation with fractional delay ($\ell_p = 4.3$) and fractional Doppler ($f_p = 2.1$), revealing the inter-sample coupling effects. The FDFD models in (b) and (c) are constructed based on the ideal band-limited pulse (Sinc interpolation).}
\label{fig:FDFD_chan}
\vspace{-1ex}
\end{figure*}

As illustrated in Fig. \ref{fig:FDFD_chan}, for integer delays, the sum $\mathbf{G}(\ell_{p}) + \mathbf{\Phi}_{p}(\ell_{p})$ perfectly reconstructs the sparse, phase-rotated cyclic permutation matrix used in legacy \ac{OFDM} and \ac{AFDM} literature, as evidenced by the identity between Fig. \ref{fig:FDFD_chan}(a), based on \eqref{eq:Hp_integerdelay}, and Fig. \ref{fig:FDFD_chan}(b), based on \eqref{eq:Hp_FDFD_generalized} and \eqref{eq:Hp_FDFD_modular}.
Furthermore, the introduction of fractional delays in Fig.~\ref{fig:FDFD_chan}(c) reveals the dense inter-sample coupling inherent to true \ac{6G} doubly dispersive channels\footnotemark, where each received sample becomes a linear combination of all $N$ transmitted samples. 

\begin{figure*}[b]
\centering
\begin{subfigure}[b]{0.3\textwidth}
\centering
\includegraphics[width=\textwidth]{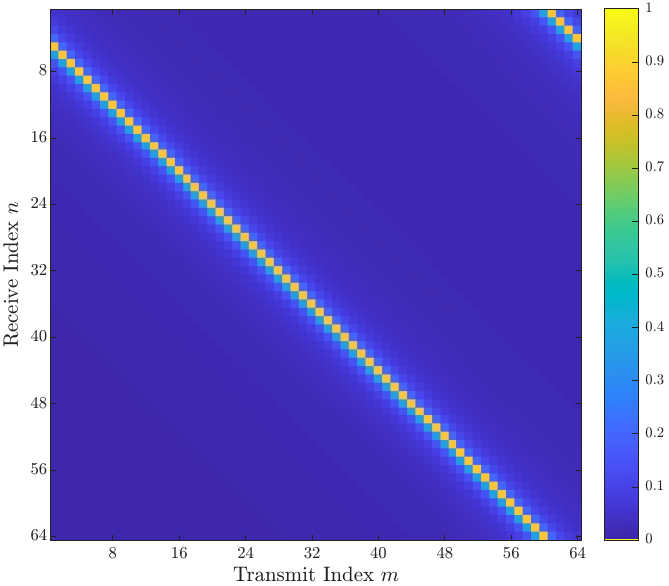}
\caption{Sinc (Ideal Band-limited)}
\label{fig:pulse_sinc}
\end{subfigure}
\hfill
\begin{subfigure}[b]{0.3\textwidth}
\centering
\includegraphics[width=\textwidth]{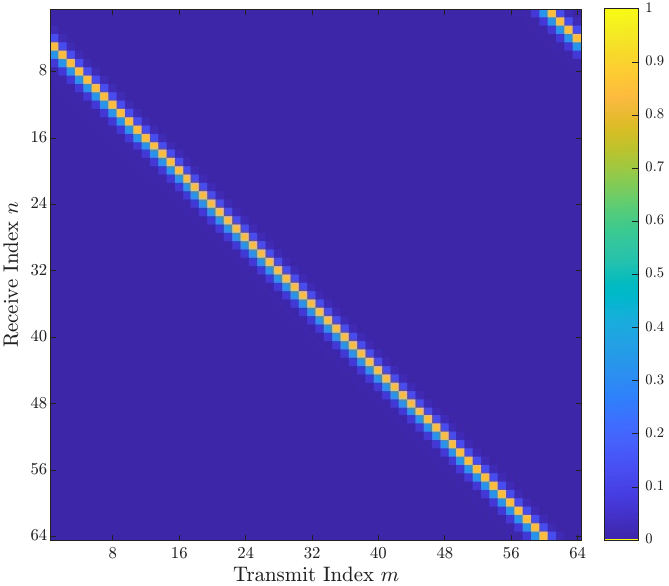}
\caption{Raised Cosine ($\alpha = 0.5$)}
\label{fig:pulse_rc}
\end{subfigure}
\hfill
\begin{subfigure}[b]{0.3\textwidth}
\centering
\includegraphics[width=\textwidth]{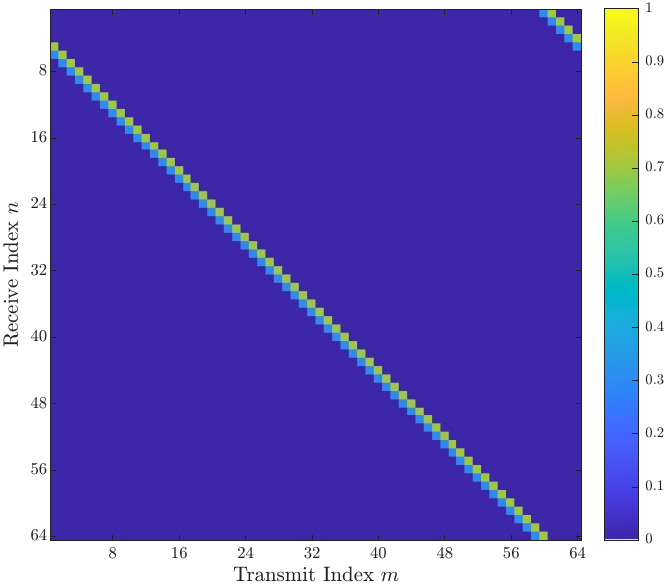}
\caption{Rectangular (No Pulse Shaping)}
\label{fig:pulse_rect}
\end{subfigure}

\vspace{-1.5ex}
\caption{Impact of different transmit pulse shapes $g(\cdot)$ on the single-path \ac{FDFD} channel matrix $\mathbf{H}_p$ ($N=64, \ell_p=4.3, f_p=2.1$), where the choice of the transmit filter governs the sparsity of the inter-sample coupling and the thickness of the diagonal band, while the main position of the shifted components remains unchanged and deterministic to the integer delay part $\lfloor \ell_p \rceil$.}
\label{fig:pulse_comparison}
\vspace{-1em}
\end{figure*}

This framework inherently therefore accommodates the transmit pulse-shape's influence on channel coupling, where the pulse-shape kernel $g(\cdot)$ affects inter-sample coupling. 
As illustrated in Fig.~\ref{fig:pulse_comparison}, under the ideal band-limited assumption (sinc interpolation kernel), the channel matrix exhibits the true inter-sample smearing and a consequently denser structure, distributing symbol energy across a wide band diagonal. 
In contrast, a rectangular pulse (non-pulse-shaped) yields a significantly sparser matrix with minimal inter-sample coupling, albeit with non-ideal spectral leakage and \ac{ICI}.
In between, a practical raised cosine pulse achieves better localization than the sinc kernel while capturing sufficient inter-sample coupling, and still obtains a sufficiently sparse structure exploitable for efficient estimation and detection algorithms.

\footnotetext{The proposed core \ac{FDFD} channel model is readily extensible to broader doubly-dispersive channel scenarios \cite{ranasinghe2025doubly,11157883,ranasinghe2025flexible}.}

\vspace{-0.5ex}
\begin{remark}
To the best of our knowledge, the decomposition $\mathbf{\Psi}(\ell_p) = \mathbf{G}(\ell_p) + \boldsymbol{\Phi}(\ell_p)$ in \eqref{eq:Hp_FDFD_modular}--\eqref{eq:fracCPP_interp_matrix} capturing the fractional delay effects with respect to the transmit pulse-shaping kernel in its explicit matrix form decoupled from the phase correction is a new contribution to the literature.
\end{remark}

Although existing \ac{FDFD} channel models are completely correct and identical numerically, the effects of fractional delays and the pulse-shaping filter components are absorbed into non-explicit element-wise descriptions of the channel elements. Consequently, these effects are not explicitly separated from the prefix-induced phase correction, and are often omitted in the analysis of \ac{AFDM} and other waveforms proposed for doubly dispersive channels.

On the other hand, the proposed formulation explicitly incorporates the transmit pulse-shaping kernel $g(\cdot)$ as the Toeplitz matrix $\mathbf{G}(\ell_p)$, cleanly separating its contribution from the prefix-induced phase correction $\boldsymbol{\Phi}(\ell_p)$.
This modular structure therefore unifies the \ac{IDID}, \ac{IDFD}, and \ac{FDFD} cases under a single parametric framework (verified in Fig.~\ref{fig:FDFD_chan}(a)-(b)). It directly exposes the pulse-dependent sparsity of the effective channel matrix, which governs the complexity and performance of the sparse detectors as will be further discussed in Section~\ref{sec:analysis_AFDMOFDM}.

\subsection{Extension to MIMO Channels}
\label{subsec:mimo_model}

The \ac{SISO} \ac{FDFD} model above extends naturally to \ac{MIMO} channels with $N_\mathrm{t}$ transmit and $N_\mathrm{r}$ receive antennas via a standard Kronecker-product construction.
For a \ac{ULA} geometry, the per-path spatial signature is captured by the outer product $\mathbf{\Theta}_p = \mathbf{a}_\mathrm{r}(\theta_p^{\mathrm{r}}) \mathbf{a}_\mathrm{t}^{\mathsf{H}}(\theta_p^{\mathrm{t}}) \in \mathbb{C}^{N_\mathrm{r} \times N_\mathrm{t}}$, where $\mathbf{a}_\mathrm{t}(\theta_p^{\mathrm{t}})$ and $\mathbf{a}_\mathrm{r}(\theta_p^{\mathrm{r}})$ denote the transmit and receive array response vectors, respectively.
Stacking the $N$ transmit sample vectors into $\bar{\mathbf{s}} \in \mathbb{C}^{N_\mathrm{t} N \times 1}$ and the received sample vectors into $\bar{\mathbf{r}} \in \mathbb{C}^{N_\mathrm{r} N \times 1}$, the blockwise \ac{MIMO} \ac{FDFD} model can be formulated as \vspace{1ex}
\begin{align}
\bar{\mathbf{r}} & = \bigg(\sum_{p=1}^{P} h_p \cdot \mathbf{\Theta}_p
\otimes\Big( \mathbf{V}^{f_p} \cdot \mathbf{\Psi}(\ell_p) \Big) \bigg) \cdot \bar{\mathbf{s}} + \bar{\mathbf{w}}, \label{eq:MIMO_vec_FDFD}
\end{align}
where $\mathbf{\Psi}(\ell_p)$, $\mathbf{V}^{f_p}$ are as defined in \eqref{eq:Hp_FDFD_generalized}.
The channel structure exhibits the same per-block interference characteristics as in the \ac{SISO} case, scaled by the per-path spatial signatures.
Extensions to other array geometries (\acp{UPA}, non-uniform arrays) and practical antenna correlation models \cite{chi26mamp} follow standard procedures and are omitted from the scope of this article.

\section{On the Compatibility of AFDM over OFDM}
\label{sec:compatibility_AFDM_OFDM}

This section examines the \ac{AFDM} modulator and demodulator structures in detail and shows that they require virtually no modifications to the underlying hardware architecture relative to legacy \ac{OFDM}, particularly for the core transform processing, pulse shaping, resource allocation, and prefixing.
All conventional \ac{OFDM} modulator blocks can be reused directly, with only minimal additions in baseband processing.
The \ac{OFDM} and \ac{AFDM} signal models are well-known background. The novel contribution of this section is the compatibility analysis in Section~\ref{sec:compatibility_AFDM_OFDM}-D built on top of them, establishing that OFDM and AFDM share the same underlying hardware and processing chain.

To that end, let us first introduce the input-output system model of the conventional \ac{OFDM} and the \ac{AFDM} modulator/demodulator, given the model of the discrete-time doubly dispersive channel.

\subsection{OFDM Signal Model}
\label{subsec:signal_model_OFDM}

Let the $k$-th data symbol be denoted by $x_k \in \mathcal{X}$ for $k \in \{0,\ldots,N-1\}$,  where $\mathcal{X} \subset \mathbb{C}$ is the complex constellation set, with cardinality $|\mathcal{X}| = M$, $i.e.$, $M$-QAM or $M$-PSK.

Then, the \ac{OFDM} transmit signal block with $N$ samples in the time domain is generated via the \ac{IDFT} as $\mathbf{s} = \mathbf{F}^\mathsf{H} \cdot \mathbf{x} \in \mathbb{C}^{N \times 1}$, where $\mathbf{s} \in \mathbb{C}^{N \times 1}$ is the transmit signal vector collecting one \ac{OFDM} block in the time domain, $\mathbf{x} = [x_0, \ldots, x_{N-1}]^{\mathsf T} \in \mathbb{C}^{N \times 1}$ is the vector collecting the $N$ data symbols, and $\mathbf{F} \in \mathbb{C}^{N \times N}$ is the $N$-point \ac{DFT} matrix.

For \ac{OFDM}, the prefix required is a \acf{CP}, and does not need a phase-offset per sample, and therefore, $\phi_\mathrm{cp}( \cdot) = 0$ and $\mathbf{\Psi}(\ell_p) =  \mathbf{G}(\ell_p)$ in \eqref{eq:Hp_FDFD_modular}, yielding
\begin{equation}
\mathbf{r} = \Big( \sum_{p=1}^{P} h_p \!\cdot\! \mathbf{V}^{f_p} \!\cdot\! \mathbf{G}(\ell_p) \Big) \mathbf{F}^\mathsf{H} \!\cdot\! \mathbf{x} + \mathbf{w} \in \mathbb{C}^{N \times 1}.
\label{eq:OFDM_rx_channel}
\end{equation}

On the receiver side, the \ac{OFDM} demodulator applies the forward $N$-point normalized \ac{DFT} to the received signal, yielding the frequency-domain received vector as
\begin{align}
\mathbf{y}
&= \mathbf{F} \cdot \mathbf{r}
= \underbrace{\mathbf{F} \Big( \sum_{p=1}^{P} h_p \!\cdot\! \mathbf{V}^{f_p} \!\cdot\! \mathbf{G}(\ell_p) \Big) \mathbf{F}^\mathsf{H}}_{\triangleq \,\mathbf{\Xi}^\mathrm{OFDM}\, \in\, \mathbb{C}^{N \times N}} \cdot \mathbf{x} + \mathbf{F}\mathbf{w} \in \mathbb{C}^{N \times 1}, \nonumber \\[-3ex] 
&     \label{eq:OFDM_rx_freq} 
\end{align}

\noindent where, since the modulator and demodulator are independent of the path index $p$, the effective channel matrix describing the input-output relation of the data symbols can be expressed as the sum of the per-path effective channels, as
\begin{equation}
\mathbf{\Xi}^\mathrm{OFDM} \triangleq \sum_{p=1}^{P} h_p \cdot \mathbf{\Xi}^\mathrm{OFDM}_p \in \mathbb{C}^{N \times N},
\end{equation}
with $\mathbf{\Xi}^\mathrm{OFDM}_p \triangleq \mathbf{F} \!\cdot\!  \mathbf{V}^{f_p} \!\cdot\!  \mathbf{G}(\ell_p) \!\cdot\!  \mathbf{F}^\mathsf{H} \in \mathbb{C}^{N \times N}$.

With integer delays and \underline{no} Doppler shift, $i.e.$, $\mathbf{G}(\ell_p) = \mathbf{\Pi}^{\ell_p}$ and $\mathbf{V}^{f_p} = \mathbf{I}_N$, the effective channel matrix reduces to a completely diagonal matrix, which preserves the orthogonality among subcarriers and avoids inter-carrier interference: the defining characteristic of \ac{OFDM} in static multipath channels.

However, when Doppler shifts are present, it is well known that the orthogonality among \ac{OFDM} subcarriers is disrupted, leading to \ac{ICI}.
This effect is observed in the above model with the presence of $\mathbf{V}^{f_p} \neq \mathbf{I}_N$ introducing phase-rotations in the subcarriers, which no longer align with the Fourier basis.
Therefore, off-diagonal elements appear in the effective channel matrix even when $f_p$ is integer-valued, which are indistinguishable between paths.

For fractional Doppler or delay values, the effective channel becomes extremely dense, further increasing interference.

\begin{figure*}[b]
\vspace{-1ex}
\centering
\begin{subfigure}[b]{0.3\textwidth}
\centering
\includegraphics[width=\textwidth]{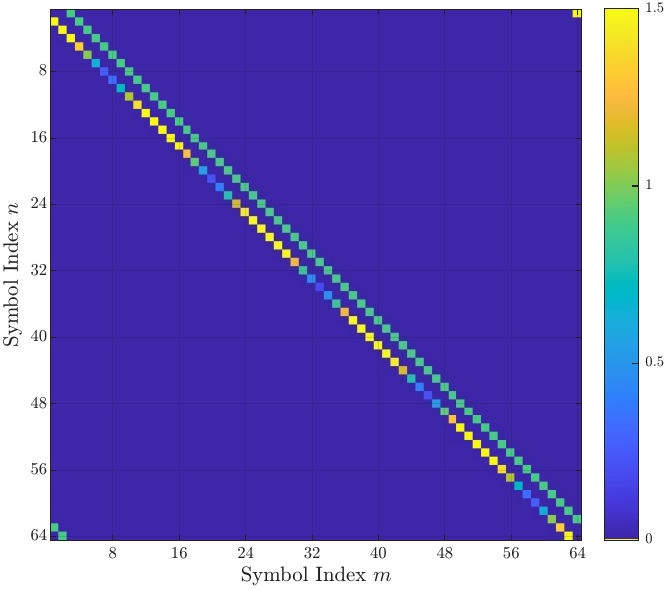}
\caption{OFDM effective channel (IDID).}
\label{fig:ofdm_3p_idid}
\end{subfigure}
\hfill
\begin{subfigure}[b]{0.3\textwidth}
\centering
\includegraphics[width=\textwidth]{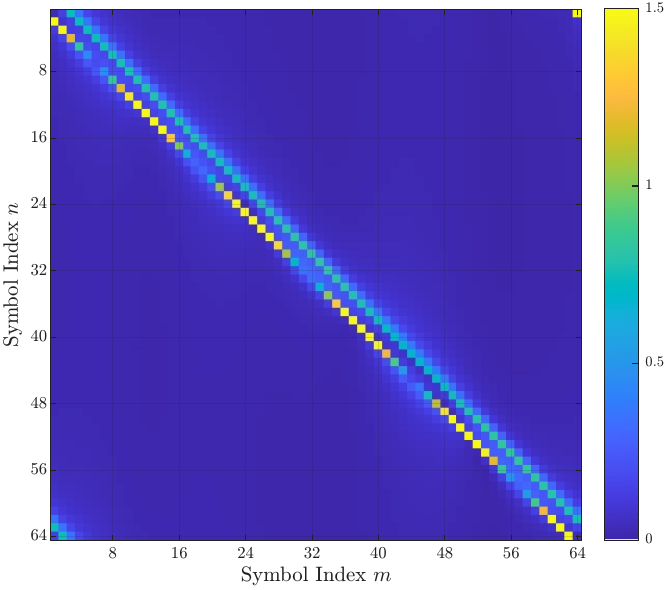}
\caption{OFDM effective channel (IDFD).}
\label{fig:ofdm_3p_idfd}
\end{subfigure}
\hfill
\begin{subfigure}[b]{0.3\textwidth}
\centering
\includegraphics[width=\textwidth]{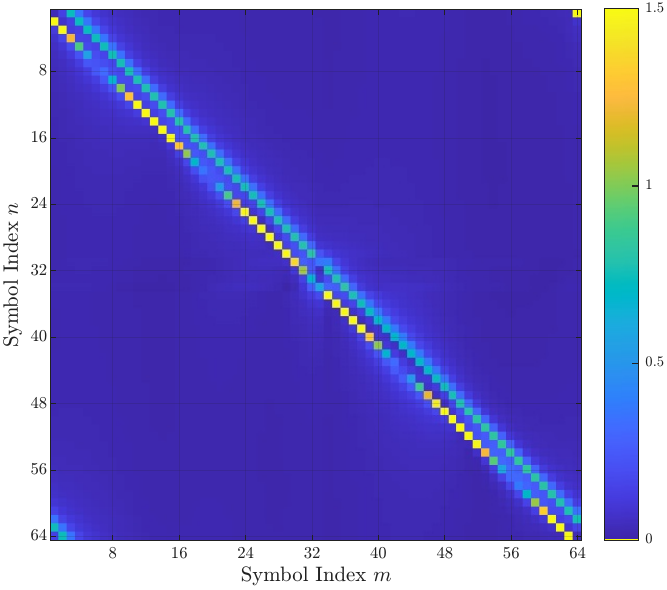}
\caption{OFDM effective channel (FDFD).}
\label{fig:ofdm_3p_fdfd}
\end{subfigure}

\vspace{0.05em}

\begin{subfigure}[b]{0.3\textwidth}
\centering
\includegraphics[width=\textwidth]{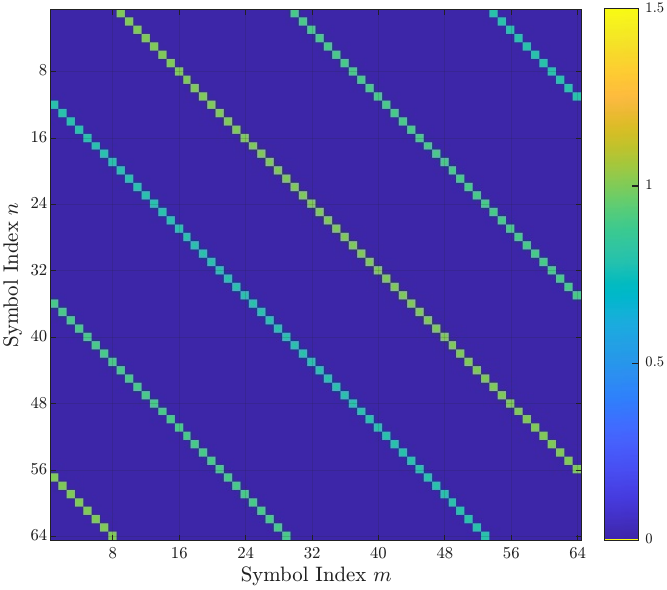}
\caption{AFDM effective channel (IDID).}
\label{fig:afdm_3p_idid}
\end{subfigure}
\hfill
\begin{subfigure}[b]{0.3\textwidth}
\centering
\includegraphics[width=\textwidth]{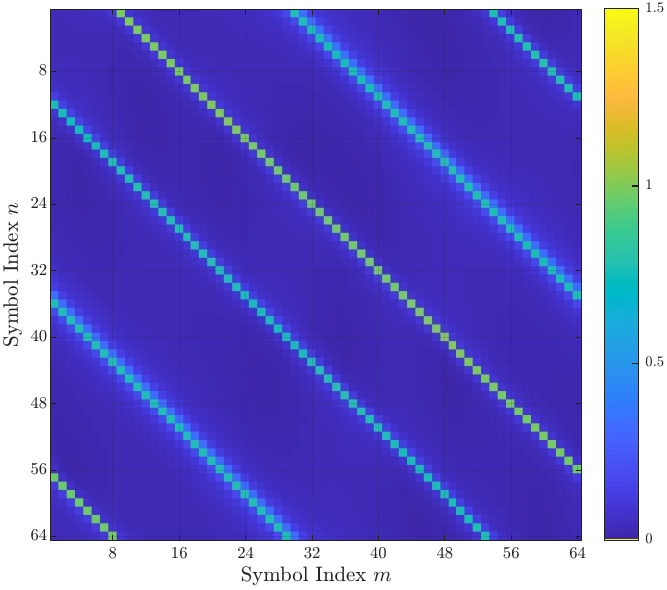}
\caption{AFDM effective channel (IDFD).}
\label{fig:afdm_3p_idfd}
\end{subfigure}
\hfill
\begin{subfigure}[b]{0.3\textwidth}
\centering
\includegraphics[width=\textwidth]{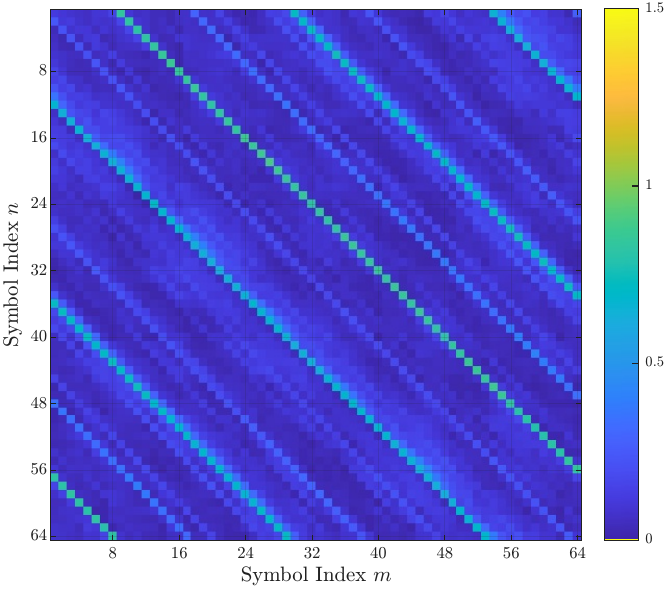}
\caption{AFDM effective channel (FDFD).}
\label{fig:afdm_3p_fdfd}
\end{subfigure}

\vspace{-1.5ex}
\caption{Effective channel matrices $\mathbf{\Xi}$ for \ac{OFDM} and \ac{AFDM}, respectively on the first and second rows, in a 3-path doubly dispersive scenario with $N=64$.
The physical channel parameters are defined by path coefficients $\{1, 0.9, 0.8\}$, normalized delays $\{1.3, 3.25, 5.96\}$, and normalized digital Dopplers $\{1.1, -2.3, 0.85\}$, where for the \ac{IDID} and \ac{IDFD} cases, the delays and Dopplers are rounded to the nearest integers as relevant.
For the \ac{AFDM} implementation, chirp parameters are set to $c_1 = (2(f_{\max} + 1) + 1)/2N$, with $f_{\max} = 3$, and $c_2 = 1/2N$.
It can be seen that on top of the discussed effects of the fraction taps in Section \ref{sec:system_model}-\ref{subsec:matrix_form_IO}, the different paths of the \ac{OFDM} effective channel matrices are indistinguishable under Doppler shifts, leading to significant inter-carrier interference, while \ac{AFDM} maintains a more distinguishable effective channel matrix, preserving better orthogonality.}
\label{fig:channel_matrices_comparison}
\vspace{-1.5em}
\end{figure*}

\subsection{AFDM Signal Model}
\label{subsec:signal_model_AFDM}

To address the challenges posed by Doppler shifts identified above, the \ac{AFDM} was proposed to convert the static subcarriers of \ac{OFDM} into chirp-based subcarriers in the \ac{DAFT} basis.

The \ac{AFDM} transmit block with $N$ samples in the time domain is generated via the \ac{IDAFT} as
\begin{equation}
\mathbf{s} = \mathbf{A}^\mathsf{H} \cdot \mathbf{x} \in \mathbb{C}^{N \times 1},
\label{eq:AFDM_tx_vec}
\vspace{-0.5ex}
\end{equation}

\noindent where $\mathbf{A} \in \mathbb{C}^{N \times N}$ is the forward normalized \ac{DAFT} matrix, which is unitary, described by
\begin{equation}
\label{eq:DAFTmatrix}
\mathbf{A} \triangleq \mathbf{\Lambda}_{\lambda_2} \mathbf{F}_N \mathbf{\Lambda}_{\lambda_1} \in \mathbb{C}^{N \times N},
\end{equation}
where $ \mathbf{F}_N \in \mathbb{C}^{N \times N}$ is the $N$-point \ac{DFT} matrix, and $\mathbf{\Lambda}_{\lambda_i} \triangleq \mathrm{diag}[e^{-j2\pi \lambda_i (0)^2}, \cdots, e^{-j2\pi \lambda_i (N-1)^2}] \in \mathbb{C}^{N \times N}$ is a diagonal chirp matrix defined by a central digital frequency $\lambda_i$, with two parametrizable chirp parameters $\lambda_1$ and $\lambda_2$.

Note that setting $\lambda_1 = \lambda_2 = 0$ yields the standard \ac{DFT}, while $\lambda_1 = \lambda_2 = \frac{1}{2N}$ corresponds to the discrete Fresnel transform matrix utilized in \ac{OCDM}.
Furthermore, the first chirp frequency $\lambda_1$ is a key parameter which can be optimized based on the channel statistics to ensure full diversity in doubly dispersive channels, namely by satisfying \cite{Bemani_AFDM,Rou_DDWaveforms}
\begin{equation}
\label{eq:AFDM_lambda1_condition}
\lambda_1 \geq \frac{2(f_{\max} + \xi) + 1}{2N},
\end{equation}
where $\lfloor f_{\max} \rceil \in \mathbb{Z}^+$ is the closest integer value of the maximum normalized Doppler shift $f_{\max}$, and $\xi \in \mathbb{Z}^+$ is a design parameter typically referred to as the \textit{guard width} that provides a margin for robustness in Doppler.

On the other hand, the second chirp parameter $\lambda_2$ does not influence the orthogonality of \ac{AFDM} over doubly dispersive channels. 
Instead, it serves as a flexible degree of freedom for tailoring time-domain waveform properties, such as the ambiguity function and \ac{PAPR} characteristics. 

To prevent spectral aliasing and ensure a unique signal representation, $\lambda_2$ is typically chosen such that $\lambda_2 \ll 1$ or $\lambda_2 \in \mathbb{R} \setminus \mathbb{Q}$ (\textit{i.e.}, an irrational value), so that the phase progression of the second chirp does not periodically align with the discrete sampling grid.
The necessary prefix in \ac{AFDM} is a \ac{CPP} to ensure circular convolution under chirp-domain periodicity, determined by the phase-offset function $\phi^\mathrm{AFDM}_{\mathrm{cp}}(n) = \lambda_1(N^2 + 2Nn)$, making $\mathbf{\Phi}_p$ in \eqref{eq:CPP_phase_matrix} non-identity.

However, it is also known that for specific configurations of the \ac{AFDM} parameters, namely when $2N\lambda_1$ is an even integer, the \ac{CPP} reduces to a conventional \ac{CP} without phase offsets, $i.e.$, $\phi^\mathrm{AFDM}_{\mathrm{cp}}(n) = 0$ for all $n$, and therefore $\mathbf{\Phi}_p = \mathbf{I}_N$.

The received signal after passing through the doubly dispersive channel in the general \ac{FDFD} case is then expressed as
\begin{equation}
\mathbf{r} = \Big( \sum_{p=1}^{P} h_p \!\cdot\! \mathbf{\Phi}_p \!\cdot\! \mathbf{V}^{f_p} \!\cdot\! \mathbf{G}(\ell_p) \Big) \mathbf{A}^\mathsf{H} \!\cdot\! \mathbf{x} + \mathbf{w} \in \mathbb{C}^{N \times 1}.
\label{eq:AFDM_rx_channel}
\end{equation}

At the receiver, the demodulator applies the matched forward \ac{DAFT}, $i.e.$, with the same parametrization of $\lambda_1$ and $\lambda_2$, yielding the received vector $\mathbf{y}
= \mathbf{A} \cdot \mathbf{r}$ in the affine-frequency domain as
\begin{align}
\mathbf{y}= \mathbf{A} \Big( \sum_{p=1}^{P} h_p \!\cdot\! \mathbf{\Phi}_p \!\cdot\! \mathbf{V}^{f_p} \!\cdot\! \mathbf{G}(\ell_p) \Big) \mathbf{A}^\mathsf{H} \cdot \mathbf{x} + \mathbf{A}\mathbf{w}       \label{eq:AFDM_rx_freq}
\end{align}
with $\mathbf{\Xi}^\mathrm{AFDM} \triangleq \sum_{p=1}^{P} h_p \cdot \mathbf{\Xi}^\mathrm{AFDM}_p$, where $\mathbf{\Xi}^\mathrm{AFDM}$ represents the structural effective channel matrix describing the input-output relation of the data symbols between the \ac{AFDM} modulator and demodulator, which, as in the \ac{OFDM} case, can be expressed as the sum of the per-path components and the channel fading coefficients, with $\mathbf{\Xi}^\mathrm{AFDM}_p \triangleq \mathbf{A} \!\cdot\! \mathbf{\Phi}_p \!\cdot\! \mathbf{V}^{f_p} \!\cdot\! \mathbf{G}(\ell_p) \!\cdot\! \mathbf{A}^\mathsf{H} \in \mathbb{C}^{N \times N}. $

The \ac{OFDM} and \ac{AFDM} effective channels have been illustrated for the \ac{IDID}, \ac{IDFD}, and \ac{FDFD} cases in Fig.~\ref{fig:channel_matrices_comparison}.

\begin{figure*}[b]
\vspace{-1ex}
\centering
\subfloat[Modulator block structure of OFDM.]
{
\includegraphics[width=0.925\textwidth]{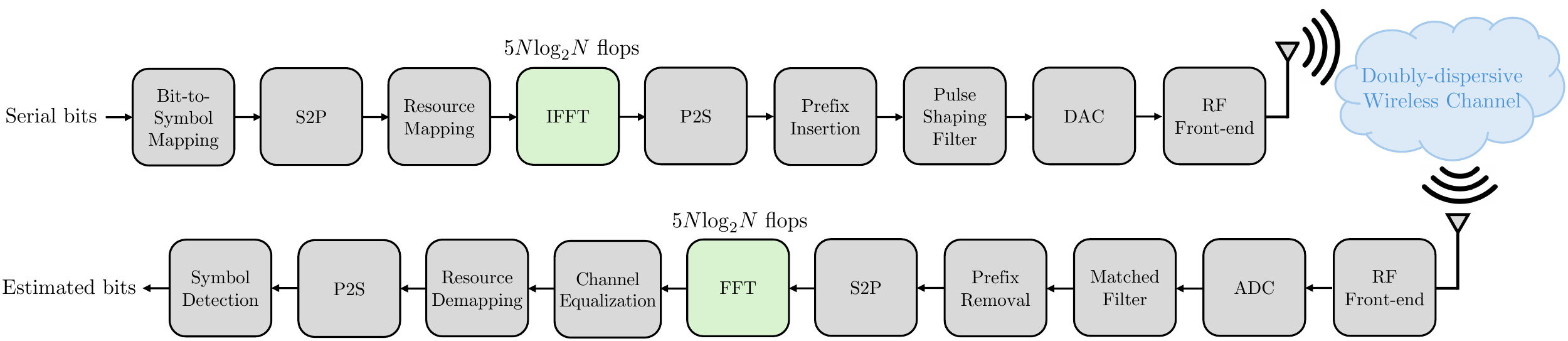}
\label{fig:OFDM_blockdiagram}
}
\\[0.2em] 
\subfloat[Modulator block structure of AFDM.]
{
\includegraphics[width=0.925\textwidth]{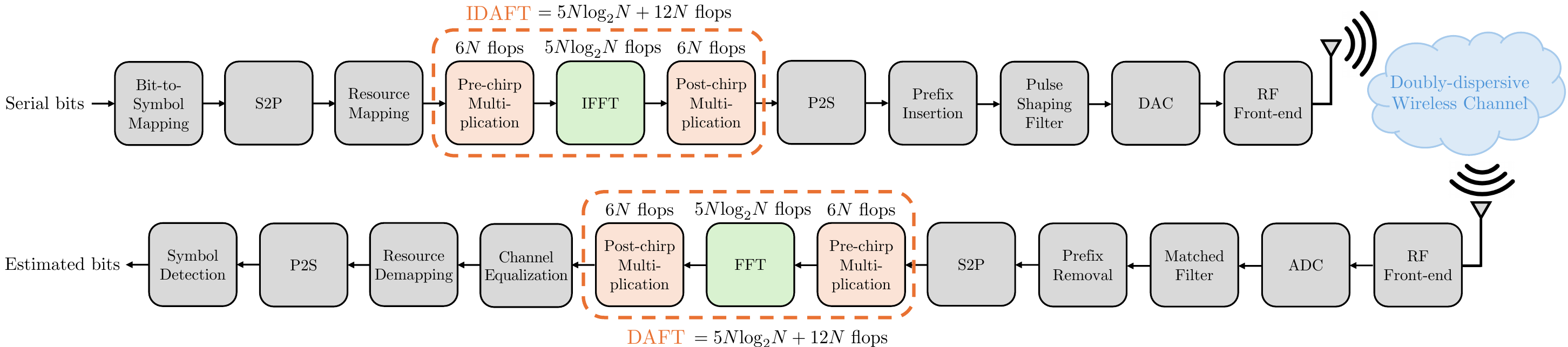}
\label{fig:AFDM_blockdiagram}
}
\vspace{-0.5em}
\caption{Comparison of \ac{OFDM} and \ac{AFDM} transceiver structures, which highlights that the only structural difference lies in the two element-wise chirp-phase rotations (pre- and post-multiplications) surrounding the \ac{IFFT}/\ac{FFT} blocks from the conventional \ac{OFDM} transceivers.}
\label{fig:OFDM_AFDM_blockdiagram}
\vspace{-1ex}
\end{figure*}

\subsection{MIMO Extension of OFDM and AFDM Signal Models}

The \ac{MIMO} signal model of \ac{OFDM} and \ac{AFDM} follows from a block-wise application of the modulator and demodulator to a stacked model, replacing the \ac{SISO} channel part in \eqref{eq:MIMO_vec_FDFD} with the effective channel including the modulator operations, i.e., \vspace{-1ex}
\begin{equation}
\bar{\mathbf{y}}^\mathrm{MIMO\text{-}OFDM} \!=\!\! \bigg(\sum_{p=1}^{P} h_p \!\cdot\! \mathbf{\Theta}_p \otimes \mathbf{\Xi}_p^\mathrm{OFDM} \bigg) \bar{\mathbf{x}} + \tilde{\mathbf{w}},
\label{eq:MIMO_OFDM_freq}
\end{equation}
\begin{equation}
\bar{\mathbf{y}}^\mathrm{MIMO\text{-}AFDM} \!=\!\! \bigg(\sum_{p=1}^{P} h_p \!\cdot\! \mathbf{\Theta}_p \otimes \mathbf{\Xi}_p^\mathrm{AFDM} \bigg) \bar{\mathbf{x}} + \tilde{\mathbf{w}}.
\label{eq:MIMO_AFDM_freq} \vspace{-1ex}
\end{equation}

Trivially, the channel structure exhibits the same per-block interference characteristics as in the \ac{SISO} case, but with the corresponding antenna response coefficients across antennas.

\vspace{-0.5ex} 
\subsection{A Note on OFDM and AFDM Compatibility}

For the modulator/demodulator, the fundamental \ac{DAFT}/\ac{IDAFT} steps require only a core \ac{FFT}/\ac{IFFT} block and two subcarrier-wise (element-wise) phase-rotation blocks, as described in \eqref{eq:DAFTmatrix}. Thus, for any parametrization of the \ac{AFDM}, these two phase-rotation blocks (highlighted in blue) are placed immediately before and after the \ac{FFT}/\ac{IFFT} block (highlighted in green), as illustrated in Fig.~\ref{fig:OFDM_AFDM_blockdiagram}. 

This phase-rotation mechanism is already widely adopted in conventional \ac{OFDM} systems and techniques, such as: 
\textit{Selected Mapping (SLM)} method for \ac{PAPR} reduction, where each OFDM symbol is multiplied by a unit-magnitude phase vector before the \ac{IFFT} \cite{Baeuml1996SLM},  
\textit{Cyclic Delay Diversity (CDD)} technique adopted in LTE, where transmit antennas apply linearly varying per-subcarrier phase rotations to achieve frequency diversity \cite{3gpp36211}, 
\textit{pilot phase randomization} in IEEE~802.11 WLANs, where predefined phase sequences modulate pilot subcarriers to aid carrier tracking \cite{IEEE80211}, and 
\textit{pilot scrambling} in DBV-T, where pilot tones are multiplied by a pseudo-random binary phase to reduce interference \cite{ETSI300744}.

Furthermore, the remaining baseband processing blocks, such as resource mapping for pilots and subcarrier allocation, can be directly reused, since the resources in \ac{AFDM} are indexed in a one-dimensional (affine) frequency domain. 
Similarly, the prefix is prepended along this single dimension, followed by a time-domain windowing block applied identically to that of conventional \ac{OFDM}.
In summary, \ac{AFDM} does not require any changes to the physical transceiver hardware, introducing only lightweight baseband chirp rotations and \ac{CPP} phase adjustments. 

This high level of structural compatibility between \ac{AFDM} and \ac{OFDM} contrasts with \ac{OTFS}, which requires several structural modifications beyond standard \ac{OFDM} for a mathematically compliant implementation.
The \ac{OTFS} waveform is inherently defined in the delay-Doppler domain and relies on either an \ac{ISFFT} combined with the \ac{HT}, or the \ac{IDZT}, to transform the delay-Doppler symbols into a time-domain sequence \cite{deng25unifying}.
When assuming the non-ideal rectangular pulses for interpolation and upsampling, this combination of \ac{ISFFT} and \ac{HT} collapses into the \ac{IDZT}, which can be efficiently implemented using interleavers and multiple lower-dimensional \acp{IDFT}. 
However, in practical doubly dispersive channels with non-integer Dopplers and delays, the necessary bi-orthogonality property of the \ac{OTFS} basis is not fully satisfied under such non-ideal rectangular pulses, and therefore complex pulse-shaping is necessary as will be further elaborated in the following.

\vspace{-1ex}
\subsection{Pulse-Shaping, Windowing, and Prefixing Blocks}

Pulse-shaping is an important criterion in the discrete-time modeling and practical realization of multicarrier waveforms in doubly dispersive channels.
The legacy \ac{OFDM} systems (e.g., LTE, 5G NR, Wi-Fi) traditionally employ an implicit rectangular pulse (i.e., no pulse-shaping) alongside a cyclic prefix, which still ensures orthogonality in static multipath channels. In a general case, however, additional spectral containment may be desired, especially under high-mobility channels, implemented by a time-domain windowing framework (sometimes simply called OFDM pulse-shaping), where a real-valued filter is applied sample-wise prior to the \ac{DAC}, i.e., \vspace{-1ex}
\begin{equation}
\label{eq:OFDM_pulse_shaping}
\mathbf{s}_\mathrm{ps} = \mathrm{diag}(\mathbf{p})\cdot \, \mathbf{s}_{\mathrm{cp}} = \mathbf{p} \odot \mathbf{s}_{\mathrm{cp}} \in \mathbb{C}^{(N + N_\mathrm{cp}) \times 1},
\end{equation}

\noindent where $\mathbf{s}_{\mathrm{cp}} \in \mathbb{C}^{(N + N_\mathrm{cp}) \times 1}$ is the modulator output vector including a cyclic prefix of length $N_\mathrm{cp}$, and $\mathbf{p} \in \mathbb{R}^{(N + N_\mathrm{cp}) \times 1}$ is the real-valued windowing filter, and $\mathbf{s}_\mathrm{ps} \in \mathbb{C}^{(N + N_\mathrm{cp}) \times 1}$ is the windowed transmit signal vector.

For the \ac{AFDM}, the modulator and demodulator architecture admit the same generic pulse-shaping operation without altering hardware, employing the same digital filter in \eqref{eq:OFDM_pulse_shaping} as \ac{OFDM} prior to transmission, as the subcarrier orthogonality in time-frequency domain is guaranteed by the underlying unitary basis function \ac{DAFT} and the \ac{CPP} ensures circular convolutional structure under chirp-periodicity, just as the \ac{OFDM}'s \ac{DFT} basis with \ac{CP}, although some non-ideal spectral properties may arise \cite{10901415}.

However, as mentioned before, the more critical case is \ac{OTFS}, whose theoretically ideal bi-orthogonality in delay-Doppler cannot be fully satisfied in practical systems with finite signal support, especially under rectangular pulses.
In such cases, the practical \ac{OTFS} input-output relation departs from the theoretical delay-Doppler convolution model, suffering a loss of bi-orthogonality that necessitates complex equalization.
Therefore, \ac{OTFS} generally requires dedicated pulse-shaping in the interpolating upsampler to mitigate the loss of bi-orthogonality (which is approached but not exactly achieved under finite support), a limitation not present in \ac{OFDM} and \ac{AFDM} whose channel and modulation basis remain mutually orthogonal for arbitrary pulse shapes.

One approach of pulse shaping for \ac{OTFS} is to apply a time-domain pulse-shaping \ac{HT} after the \ac{ISFFT}, which can mitigate some non-idealities but remains fundamentally limited by finite support and the Balian-Low constraints.
However, the non-rectangular pulse-shaping \ac{HT} implies that the \ac{OTFS} modulator requires three consecutive \ac{FFT}/\ac{IFFT} operations with interleaving and reshaping, which cannot be reduced further as commonly assumed for the rectangular pulse case, leading to increase in complexity and latency.

Alternatively, recent work on Zak-\ac{OTFS} introduces a direct delay-Doppler domain pulse-shaping approach, which eliminates the need for the \ac{ISFFT} and \ac{HT}, but instead requires a two-dimensional convolution in the delay-Doppler domain for the pulse shaping before the single \ac{IDZT}.
This DD-domain filtering enables more controlled localization and yields the closest practical approximation to the ideal twisted convolution, although perfect bi-orthogonality remains unattainable under finite time-bandwidth constraints. 

Thus, careful pulse shaping is necessary for \ac{OTFS} to retain its desired structural characteristics in practical doubly-dispersive channels, whereas \ac{AFDM} and \ac{OFDM} preserve their orthogonality and convolutional forms even with simple rectangular pulses.

Furthermore, unlike \ac{AFDM}, where a one-dimensional cyclic prefix (with chirp periodicity) is sufficient to ensure robustness, \ac{OTFS} has recently been shown to require a so-called double cyclic prefix (in both time and frequency dimensions) to improve its robustness on top of the single time-domain cyclic prefix \cite{10531762}.
This unique prefixing strategy necessitates a distinct resource allocation and indexing scheme, and also leads to increased signaling overhead and complexity.

Overall, this section shows that \ac{AFDM} preserves the core \ac{OFDM} hardware pipeline and one-dimensional framing structure, requiring only lightweight digital chirp rotations and a re-parameterized cyclic prefix, without introducing new baseband modules or RF redesign. 
In contrast, \ac{OTFS} operates in a distinct modulation domain and, under practical pulse-shaping conditions for doubly dispersive channels, requires 2D processing and non-native prefixing and resource grids, leading to higher integration complexity relative to \ac{OFDM} (see Table~\ref{tab:complexity_summary}). While simplified rectangular-pulse \ac{OTFS} implementations can partially reuse \ac{OFDM} hardware, full bi-orthogonal implementations incur the multi-stage overhead quantified above.
Consequently, \ac{AFDM} is shown to be a hardware- and specification-efficient evolutionary waveform candidate for next-generation systems.

\section{Analysis of the AFDM Transceiver over OFDM}
\label{sec:analysis_AFDMOFDM}

Each subsection surveys existing \ac{AFDM} algorithms and techniques to provide a unified picture of the transceiver.
The distinguishing contribution of this section is the systematic analysis of each layer's compatibility with the \ac{OFDM} legacy, and its behavior under the generalized \ac{FDFD} channel model of Section~II. Where specific results are drawn from prior works, they are cited accordingly. The FDFD-contextualized analyses, the layer-by-layer comparison against \ac{OFDM} counterparts, and the structural observations on sparsity and reusability are new contributions of this article.

\subsection{Modulation Complexity and Scalability}

Drawing from the previous section, the additional computational burden of the \ac{AFDM} modulator/demodulator relative to conventional \ac{OFDM} stems only from the two chirp-domain element-wise phase rotations applied immediately before and after the \ac{FFT}/\ac{IFFT} stage; all other blocks are identical and can be reused directly (excluding channel equalization), as shown in Fig.~\ref{fig:OFDM_AFDM_blockdiagram}.
Thereby, each complex phase rotation over $N$ samples (equivalent to the number of subcarriers) is implemented simply via $N$ element-wise complex multiplications, which require a total of $6N$ \acp{FLOP}. 
Hence, the two rotations introduce an extra cost of $12N$ \acp{FLOP} per \ac{AFDM} symbol block of $N$ complex symbols.

With the conventional $Q$-point \ac{FFT}/\ac{IFFT} cost given by $C_{\mathrm{FFT},Q} \triangleq 5Q\log_2 Q$ \acp{FLOP}, the total per-block modulator/demodulator complexity for \ac{AFDM} with $N$ symbols is then obtained as
\begin{equation}
\label{eq:afdm_cost}
C_{\mathrm{AFDM},N} = 5N\log_2 N + 12N \quad \text{[FLOPs]}.
\end{equation}

By comparison, the conventional \ac{OFDM} modulator cost is trivially given by the complexity of the $N$-point \ac{IFFT}/\ac{FFT} blocks directly, i.e., $C_{\mathrm{OFDM},N} = C_{\mathrm{FFT},N}$.
Therefore, the relative complexity overhead of \ac{AFDM} over \ac{OFDM} is
\begin{equation}
\label{eq:relative_overhead}
\frac{C_{\mathrm{AFDM},N} - C_{\mathrm{OFDM},N}}{C_{\mathrm{OFDM},N}}
\;=\;
\frac{12N}{5N\log_2 N}
\;=\;
\frac{12}{5\log_2 N}.
\end{equation}

For practical \ac{FFT}/\ac{IFFT} sizes (number of subcarriers) considered in \ac{6G}, the overhead in \eqref{eq:relative_overhead} evaluates to modest values, with $N = 256$ amounting to a $30\%$ overhead, $N = 1024$ to $24\%$, and $N = 4096$ to $20\%$.
Hence, although the chirp rotations are neither cost-free nor negligible, their asymptotic impact diminishes logarithmically as the number of subcarriers grows, and the \ac{FFT}/\ac{IFFT} remains the dominant complexity term. 

Next, as a reference and comparison, we provide a similar analysis for the \ac{OTFS}.
For the \ac{ISFFT} + \ac{HT} approach with the \ac{OTFS} delay-Doppler frame of size $K \times L$, where we set $N = KL$ to fairly compare the different waveforms in terms of the resource usage (i.e., number of discrete symbol resources), the total complexity of the \ac{ISFFT}, which is implemented using $L$ IFFTs of size $K$, and $K$ FFTs of size $L$, is given by
\begin{align}
\!\!\!\!\!\!C_{\mathrm{ISFFT},KL} &= L (5K \log_2 K) + K (5L \log_2 L)  ~\text{[FLOPs]}\! \nonumber \\
& =  5KL (\log_2 KL) = 5N \log_2 N = C_{\mathrm{FFT},N}, 
\end{align}
i.e., the same as a single $N$-point \ac{IFFT} for $N = KL$. 

Then, a pulse-shaped \ac{HT} is performed, consisting of $K$ successive \ac{IFFT} operations of length $L$, each followed by a complex-real time-domain scaling that applies the prototype pulse, with an additional computational complexity of
\begin{align}
C_{\mathrm{HT},KL} &= K(5L \log_2 L) + 2KL  \quad \text{[FLOPs]} \\
& = N(5\log_2L + 2). \nonumber
\end{align}

In all, the total complexity of the \ac{ISFFT}-\ac{HT} based \ac{OTFS} modulator/demodulator is then given by
\begin{align} \label{eq:isfftotfs_cost}
C_{\mathrm{OTFS},KL}
&= C_{\mathrm{ISFFT},KL} + C_{\mathrm{HT},KL} \\
&= 5N(\log_2NL) + 2N. ~~\text{[FLOPs]} \nonumber 
\end{align}

Alternatively, if pulse-shaping is performed directly in the delay-Doppler domain via 2D convolution in the prior, a single \ac{IDZT} approach may be employed.
The \ac{IDZT} on a $K \times L$ delay-Doppler block is implemented using $K$ \acp{IFFT} of length $L$ with some sample interleaving operation, where the latter is approximated to have negligible computation complexity (only space complexity), such that the total complexity required for the \ac{IDZT} is
\begin{equation}
\label{eq:zak_cost}
C_{\mathrm{Zak}} = K \big(5L \log_2 L\big) = 5N \log_2L. ~~\text{[FLOPs]}
\end{equation}

The {2D} convolution for the pulse shaping in the delay-Doppler grid of size $K \times L$, is efficiently implemented using an FFT-based approach instead of the inefficient naive convolution, which involves a forward 2D \ac{FFT}, a point-wise multiplication with the pulse's 2D \ac{FFT}, and an inverse 2D \ac{FFT} back to the delay-Doppler, i.e., 
\begin{equation}
\mathbf{X}_{\mathrm{DD,ps}} = \mathrm{IDFT_{\mathrm{2D}}}\Big( \mathrm{DFT}_{\mathrm{2D}}(\mathbf{X}_\mathrm{DD}) \odot \mathrm{DFT}_{\mathrm{2D}}(\mathbf{P}) \Big),
\end{equation}
where $\mathbf{X}_\mathrm{DD} \in \mathbb{C}^{K \times L}$ is the delay-Doppler symbol block, $\mathbf{P} \in \mathbb{R}^{K \times L}$ is the real-valued pulse shaping filter in the delay-Doppler domain, and $\mathbf{X}_{\mathrm{DD,ps}} \in \mathbb{C}^{K \times L}$ is the pulse-shaped delay-Doppler symbol block.

We may assume that the complexity of obtaining the \ac{FFT} of the pulse filter, $\mathrm{DFT}_{\mathrm{2D}}(\mathbf{P})$, is negligble as it can be computed offline once, therefore the computational complexity is obtained for one FFT of size $K \times L$, one IFFT of size $K \times L$, and the element-wise multiplication of two matrices of size $K \times L$, yielding a total complexity of
\begin{align}
C_{\mathrm{DDps},KL} &= 2 \cdot (5KL \log_2 KL) + 6KL \quad \text{[FLOPs]}  \\
&= 10N \log_2 N + 6N.\nonumber
\end{align}

Therefore, in all, the total complexity of the Zak-\ac{OTFS} modulator/demodulator with delay-Doppler pulse shaping is approximated as
\begin{align}
\label{eq:zakotfs_cost}
C_{\mathrm{Zak}\text{-}\mathrm{OTFS}}&= C_\mathrm{Zak} + C_{\mathrm{DDps},KL} ~~\text{[FLOPs]}\\
& = 5N \log_2 L + 10N \log_2 N + 6N \nonumber \\
& = 5N\log_2 (L N^2) + 6N \nonumber
\end{align}

Comparing the modulator complexities including the pulse-shaping overheads of \ac{OFDM}, \ac{AFDM}, and the two \ac{OTFS} implementations in \eqref{eq:isfftotfs_cost} and \eqref{eq:zakotfs_cost}, which are summarized in Table \ref{tab:complexity_summary}, it is evident that both \ac{OTFS} approaches introduce significantly higher computational burdens compared to the \ac{OFDM} due to their multi-stage transform structures and 2D processing requirements under practical systems\footnote{Regarding \ac{FFT} dimension flexibility, \ac{OFDM} and \ac{AFDM} support arbitrary \ac{FFT}/\ac{IFFT} sizes $N$ with one-dimensional frequency-domain processing, enabling flexible numerology and resource allocation. 
In contrast, \ac{OTFS} imposes strict integer divisibility constraints on the delay-Doppler grid dimensions $K, L \in \mathbb{N}$ such that $N = KL$, limiting design flexibility and potentially resulting in awkward \ac{FFT} sizes for practical implementations and efficient computations.}, while \ac{AFDM} maintains a modest overhead over \ac{OFDM} and therefore represents a more computationally efficient alternative.

In addition to arithmetic complexity, memory movement and on-chip buffering often dominate latency and energy consumption in practical modems, especially in high-throughput systems. 
In this perspective, the \ac{AFDM} waveform requires only local element-wise multiplications for its additional processing, which do not necessitate global data reshaping beyond the existing FFT/IFFT buffering inherent in the system.
Conversely, \ac{OTFS} implementations may involve more complex data handling due to frequent reshaping operations under practical, non-ideal conditions, such as interleaving and reshaping required in the \ac{ISFFT} and pulse shaping stages. 

Beyond per-block complexity, we consider how \ac{AFDM} and \ac{OTFS} scale with the number of subcarriers and antennas, especially for practical \ac{MIMO} systems at high bandwidths and data rates.
Since the computational complexity and memory requirements of both \ac{AFDM} and \ac{OTFS} scale linearly with the antenna count in \ac{MIMO} systems, the relative overheads identified above remain consistent regardless of that count, so the previous analysis applies directly to massive \ac{MIMO} deployments for \ac{6G+}.

\begin{table}[t]
\vspace{-2.5ex}
\centering
\caption{Computational complexity of key operations and total modulator/demodulator complexities in \acp{FLOP}, with $N = KL$.}
\label{tab:complexity_summary}
\footnotesize
\begin{minipage}{0.45\textwidth}
\centering
\begin{tabular}{|c|c|}
\hline
\textbf{Operation} & \textbf{Complexity (FLOPs)} \\
\hline
$\mathrm{IDFT}_N$ & $5 N \log_2 N$ \\
$\mathrm{IDAFT}_N$ & $5 N \log_2 N$ \\
$\mathrm{ISFFT}_{K,L}$ & $5 N \log_2 N$ \\
$\mathrm{HT}_{K,L}$ & $5N \log_2 L + 2N$ \\
$\mathrm{IDZT}_{K,L}$ & $5 N \log_2 L$ \\
1D pulse shaping & $12 N$ \\
DD pulse shaping & $10 N \log_2 N + 6 N$ \\
\hline
\end{tabular}
\end{minipage}%

\vspace{3ex}

\begin{minipage}{0.45\textwidth}
\centering
\footnotesize
\begin{tabular}{|c|c|}
\hline
\textbf{Pulse-shaped Waveform} & \textbf{Complexity (FLOPs)} \\
\hline
OFDM & $5 N \log_2 N$ \\
AFDM & $5 N \log_2 N + 12 N$ \\
OTFS & $5 N \log_2 N + 5N \log_2 L + 2N$ \\
Zak-OTFS & $10 N \log_2 N + 5 N \log_2 L + 6 N$ \\
\hline
\end{tabular}
\end{minipage}
\vspace{-2ex}
\end{table}

\subsection{Pilot Scheme and Channel Estimation} 
\label{subsec:chanest}

In this section, we discuss channel estimation strategies for \ac{AFDM} in doubly dispersive channels, focusing on pilot design and estimation algorithms. 
Given the structural similarities between \ac{AFDM} and \ac{OFDM}, many established pilot schemes can be adapted to \ac{AFDM}, albeit with modifications to account for the unique characteristics of the affine-frequency domain.

\vspace{-3ex}
\subsubsection{Pilot Schemes}

Due to the structure of the effective channel matrix $\mathbf{\Xi}^\mathrm{AFDM}$ as described in Section~\ref{sec:compatibility_AFDM_OFDM}-\ref{subsec:signal_model_AFDM}, the usual pilot schemes for \ac{AFDM} involve one or multiple pilot subcarriers \cite{yin22,Bemani_AFDM}, each of which is isolated from other subcarriers by guard bands (in the DAFT domain) to capture the contribution of each path without interference properly. 

In the case of the conventional single-pilot scheme \cite{Bemani_AFDM}, the transmitted symbol vector $\mathbf{x} \in \mathbb{C}^{N \times 1}$ in the affine-frequency domain is designed as

\begin{figure}[b]
\vspace{-1ex}
\centering{\includegraphics[width=0.95\columnwidth]{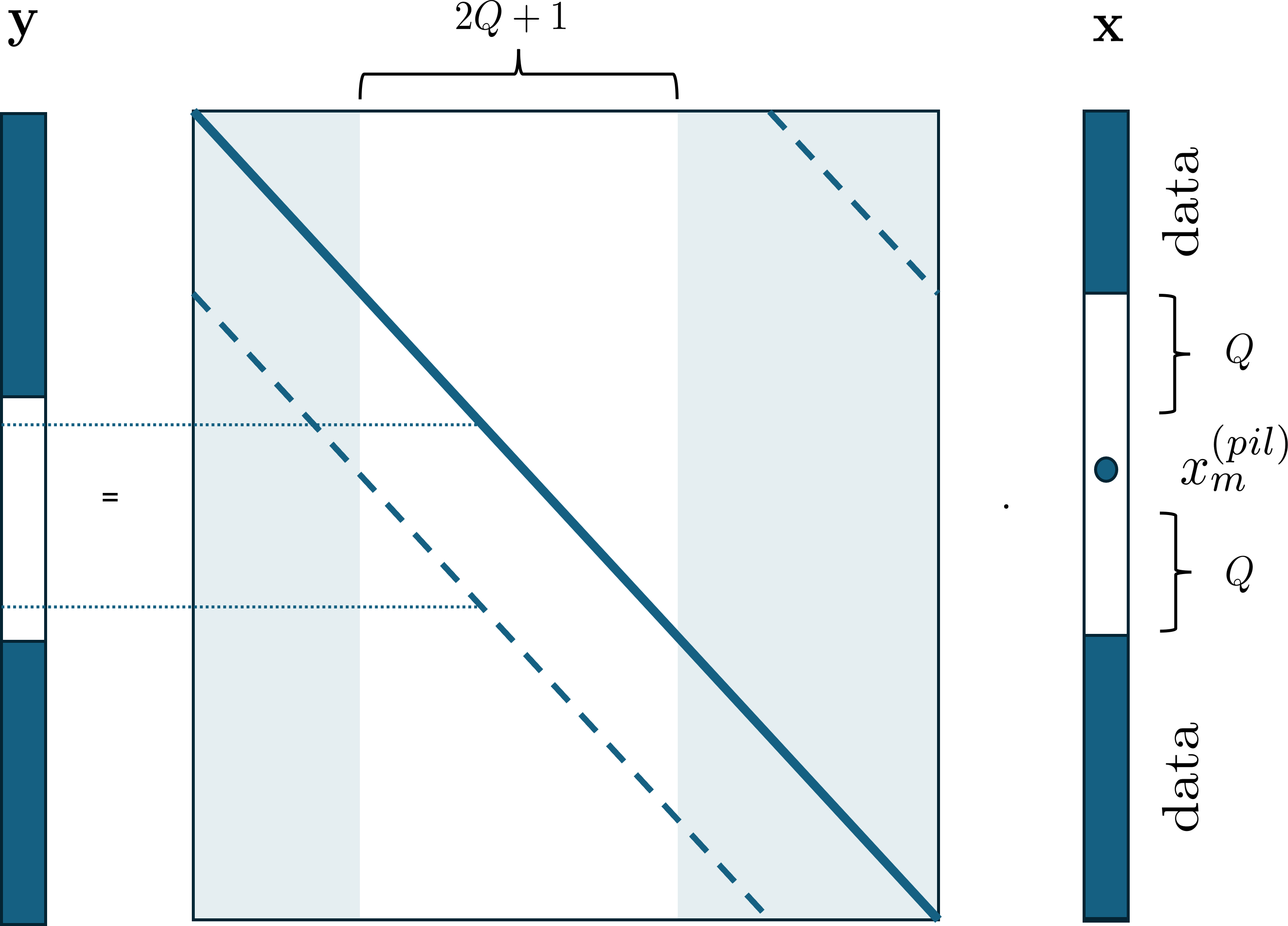}}
\vspace{1ex}
\caption{Transmitted pilot vector multiplexed with data $\mathbf{x}$ (as in \ref{eq:pilotsap}) in the affine-frequency domain and received signal $\mathbf{y}$.}
\label{fig:pilot}
\vspace{-2ex}
\end{figure}

$~$ \vspace{-3ex}
\begin{align}
~~\mathbf{x} = [x_0, \ldots, x_{m-Q}, & \label{eq:pilotsap} \\[0.5ex] 
& \hspace{-5ex}\underbrace{0,\ldots,0}_{Q}, x_m^{\mathrm{(pil)}}\!, \underbrace{0,\ldots,0}_Q, x_{m+Q}, \ldots ,x_{N-1}]^\mathsf{T},  \nonumber
\end{align}
where $x_m^{\mathrm{(pil)}}$ is the pilot at position $m$ in the vector, surrounded by $Q$ null subcarriers that serve as a guard band, and the remaining subcarriers are used to transmit data symbols, as shown in Fig. \ref{fig:pilot}. 

Note that to avoid interference from the data subcarriers, the guard band width $Q$ must satisfy \cite{yin22,Bemani_AFDM} 
\begin{equation}
Q \geq (\ell_{max}+1)( 2f_{\max}+1)-1,  
\label{eq:defQ}
\end{equation}
where $\ell_{max}$ and $f_{\max}$ are the maximum normalized delay and maximum normalized digital Doppler shift of the doubly dispersive channel, respectively.

In the case of a multiple-pilot aided scheme, the pilot subcarriers must be separated by a distance of at least $Q$, which can improve the channel estimation performance (through the diversity of pilots), at the expense of a loss of \ac{SE}, since each additional pilot involves at least $Q+1$ subcarriers not used for data transmission. 

To circumvent the \ac{SE} loss, a guard interval-free pilot scheme has been proposed in \cite{zhou24}, where the $2Q$ null subcarriers in \eqref{eq:pilotsap} are substituted by data. 
Alternatively, a superimposed pilot scheme is also described in \cite{zheng25}, theoretically optimizing the SE. 
In this case, pilot subcarriers $x_m^{\mathrm{(pil)}}$ are superimposed to the data $x_m$ (\emph{i.e.} $x_m + x_m^{\mathrm{(pil)}}$), with a distance $Q$ between the added pilots. 
However, in both cases, channel estimation and data detection require specific processing at the receiver side, as discussed in the following. 

For a negligible Doppler shift, or when all paths share a common Doppler shift and integer delays (a usual assumption in OFDM systems), the channel frequency response and the Doppler shift can be estimated separately in the frequency and time domains, respectively \cite{savaux25_PHYCOM}, while estimation in the DAFT domain remains practicable. 
To this end, a single pilot aided scheme as in \eqref{eq:pilotsap} can be considered without data multiplexing, {i.e.} $\mathbf{x}$ consists of $N-1$ null subcarriers with only one pilot $x_m^{\mathrm{(pil)}}$ at the $m$-th position.

In the frequency domain, the pilot vector is given by $\mathbf{x}_f = \mathbf{F} \cdot \mathbf{A}^\mathsf{H} \cdot \mathbf{x}$, which is not tractable in general. 
However, according to \cite{savaux25_PHYCOM,Savaux_DFTAFDM}, $\mathbf{x}_f$ has a simple and tractable expression in the special case of $q=2N\lambda_1 \in \mathbb{Z}$, where the non-zero elements of $\mathbf{x}_f$ are then constant modulus and equi-spaced with a distance $|q|$ between two consecutive pilot subcarriers in the frequency domain, starting with subcarrier $m_q = (m)_{|q|}$. 
Interestingly, setting $|q|\in\{2,4\}$, the pilot scheme is very similar to the pilot distribution described for the \ac{5G} signals \cite{3GPP_38211r16}, so that the estimation processing used for \ac{5G} signals can be used in AFDM. 
Furthermore, to limit the \ac{SE} loss, it is also possible to multiplex pilot and data in the frequency domain or in the \ac{DAFT} domain as in \eqref{eq:pilotsap}.

\vspace{-2.5ex}
\subsubsection{Channel Estimation in the DAFT Domain} 

In this section, we present the basics of channel estimation in the DAFT domain based on the pilot scheme described in (\ref{eq:pilotsap}). Other methods will be briefly discussed afterwards. 


\paragraph{Integer Delay and Doppler} 
The estimation of the channel necessitates the estimation of $3P$ unknown parameters corresponding to the delay $\ell_p$, the Doppler shift $f_p$, and the complex gain $h_p$ for each of the paths $p= \{1,\ldots,P\}$. As raised by the authors in \cite{Bemani_AFDM}, the common maximum likelihood (ML) estimation of the $3P$ parameters is intractable in practice. However, it is possible to estimate the parameters separately, as described below. First, it must be noticed that by substituting (\ref{eq:pilotsap}) into (\ref{eq:AFDM_rx_freq}), we find that the elements $y_k$ of $\mathbf{y}$, for $k \in \big\{  \big(m-(Q-f_{\max})\big)_N, \,  \big(m+f_{\max} \big)_N \big\}$ 
are given by \cite{bemani21_iswcs}
\begin{equation}
y_k = \begin{cases}
[\mathbf{\Xi}^\mathrm{AFDM}]_{k,m} \cdot x_m^{\mathrm{(pil)}} + [\mathbf{A}\mathbf{w}]_k& \\[-1ex]
&\hspace{-15ex}\text{ if } k = \big(m+N - \text{loc}_p\big)_N, \\[-0.5ex]
[\mathbf{A}\mathbf{w}]_k & \hspace{-15ex} ~\text{otherwise},
\end{cases}
\label{eq:ykchest}
\end{equation}
where $\text{loc}_p = \big(f_p +\ell_p(2 f_{\max} + 1)\big)_N$. 

This is illustrated in Fig. \ref{fig:pilot}. Then, for any $\ell= \{0,1,\ldots,\ell_{\max}\}$, the ML estimation of $f_\ell$ is given by

\begin{equation}
\hat{f}_\ell = \argmax_{k' \in \Omega_{k'}} \hspace{1mm} |y_k|^2, 
\label{eq:flestimation}
\end{equation}
where $\Omega_{k'} = [ -f_{\max},f_{\max}]$ and $k = \big(l \cdot 2f_{\max} + k'\big)_N $. 

If the number of paths $P$ is known \emph{a priori}, it corresponds to the $P$ highest values of $|y_k|^2$ among the $\ell_{\max}$ calculated in (\ref{eq:flestimation}). 
If $P$ is unknown, it is suggested in \cite{yin22} to set a threshold $\zeta$ depending on the SNR and to keep the $P$ paths whose value $|y_k|^2$ exceeds $\zeta$. 

Once $(f_p,\ell_p)$, for $p=\{1,\ldots,P\}$ is estimated, the estimate of the complex gains $h_p$ can be obtained through
\begin{equation}
\hat{h}_p = \frac{\big[\mathbf{\Xi}^\mathrm{AFDM'}\big]_{k_p,m}^* \cdot y_{k_p}}{x_m^{\mathrm{(pil)}}}
\label{eq:hkpestimation}
\end{equation}
where $k_p = \big(\ell_p \cdot 2f_{\max} + f_p \big)_N $, and the value $[\mathbf{\Xi}^\mathrm{AFDM'}]_{k_p,m}$ is generated using the estimates $(\hat{f}_p,\hat{\ell}_p)$.   


\paragraph{Integer Delay and Fractional Doppler}

In this part, we denote the integer and fractional parts of $f_p$ by $f_p^{(i)}=\lfloor f_p \rceil$ and $f_p^{(f)} = f_p - f_p^{(i)}$, respectively. 
As introduced in \cite{Bemani_AFDM}, the estimation of the $3P$ channel parameters in the case of fractional Doppler starts by the estimation of the integer part $f_p^{(i)}$ as in (\ref{eq:ykchest})-(\ref{eq:flestimation}). 
Then, the fractional part $f_p^{(f)}$ is estimated through 
\begin{equation}
\hat{f}_p^{(f)} = \argmax_{f_p^{(f)} \in [-\frac{1}{2},\frac{1}{2}]} \hspace{1mm} \frac{\sum_{k' \in \Omega_{k'}} \big|\big[\mathbf{\Xi}^\mathrm{AFDM'}\big]_{k,m}^*y_{k}\big|^2}{\sum_{k' \in \Omega_{k'}} \big|\big[\mathbf{\Xi}^\mathrm{AFDM'}\big]_{k,m}\big|^2}, 
\label{eq:ffracestimation}
\end{equation}
where $k = \big(\ell_p \cdot 2f_{\max} + k'\big)_N $, and the value $[\mathbf{\Xi}^\mathrm{AFDM'}]_{k_p,m}$ is generated using the estimates $(\hat{f}_p = \hat{f}_p^{(i)} + f_p^{(f)},\hat{\ell}_p)$. 

Moreover, in practice, (\ref{eq:ffracestimation}) is performed using a search on a fine discretization of $[-\frac{1}{2},\frac{1}{2}]$, such that ultimately, the estimate of the complex gains $h_p$ can be obtained through
\begin{equation}
\hat{h}_p = \frac{\sum_{k' \in \Omega_{k'}} \big[\mathbf{\Xi}^\mathrm{AFDM'}\big]_{k,m}^*y_{k}}{x_m^{\mathrm{(pil)}}}, 
\label{eq:hkpestimationfrac}
\end{equation}  
where the value $[\mathbf{\Xi}^\mathrm{AFDM'}]_{k_p,m}$ is generated using the estimates $(\hat{f}_p = \hat{f}_p^{(i)} + \hat{f}_p^{(f)},\hat{\ell}_p)$. 


\paragraph{Fractional Delay and Doppler}

Finally, the problem of estimating both fractional delay and Doppler parameters can be particularly difficult or even untractable if $\mathbf{G}(\ell_p)$ is dense and Toeplitz as suggested in \eqref{eq:fracdel_interp_matrix}. 
However, according to the finite interpolation kernel, $\mathbf{G}(\ell_p)$ (and as illustrated in Figure~\ref{fig:pulse_comparison}) is a banded matrix with a relatively tight band diagonal of defined width $2B+1$. 

Furthermore, the matrix $\mathbf{G}(\ell_p)$ can be approximated to be circulant instead of Toeplitz, which becomes exact if a cyclic suffix is also used in addition to the CPP, or if a negative delay is applied previous to the demodulation process. 
In light of the above, by denoting $[g_{-B},..,g_0,..,g_B]$ as the non-zero coefficients of the band diagonal of $\mathbf{G}(\ell_p)$, where $g_q = g(q-\ell_p)$, $p=-B,..,B$, $\mathbf{G}(\ell_p)$ can be rewritten as 
\begin{equation}
\mathbf{G}(\ell_p) = \sum_{q=-B}^{B} g_q \cdot \mathbf{\Pi}^{\ell_p + q}.  
\label{eq:Glpfrac}
\end{equation}

Then, the discrete input-output relation is expressed as
\begin{equation}
\mathbf{r} = \bigg( \sum_{p=1}^{P}\sum_{q=-B}^{B} h_p \cdot \mathbf{\Phi}_p \cdot \mathbf{V}^{f_p} \cdot g_q \cdot \mathbf{\Pi}^{\ell_p + q} \bigg) \cdot \mathbf{s} + \mathbf{w}.
\label{eq:matrix_DDsystem_FDFD2}
\end{equation}

Ordering the $P(2B+1)$ tuples $(p,q)$ as $\big\{(p\!=\!1,q\!=\!-B),$ $\ldots,(p\!=\!1,q\!=\!B),(p=2,q=-B),\ldots,(p=P,q=B)\big\}$, and defining $u=\{1,\ldots,P(2B+1)\}$ as the corresponding indices, \eqref{eq:matrix_DDsystem_FDFD2} can be rewritten in a final compact form as  
\begin{equation}
\mathbf{r} = \Bigg( \sum_{u=1}^{P(2B+1)}\!\!\! h_u' \cdot \mathbf{\Phi}_u \cdot \mathbf{V}^{f_u} \cdot  \mathbf{\Pi}^{\ell_u'} \Bigg) \cdot \mathbf{s} + \mathbf{w}, 
\label{eq:matrix_DDsystem_FDFD3}
\end{equation}
where $h_u' = h_p \cdot g_q$ and $\ell_u' = \ell_p + q$ are defined to highlight the similarity of \eqref{eq:matrix_DDsystem_FDFD3} to the conventional IDID doubly dispersive channel model of \eqref{eq:Hp_integerdelay}.

Under this reformulation, it follows that the channel estimation methods that have been previously described for the IDID doubly dispersive channel, can be adapted to the fractional delay and Doppler case, under the new virtual path parameters $\{h_u',\ell_u',f_u\}$ for $u=\{1,\ldots,P(2B+1)\}$ and the deterministic structure now available through \eqref{eq:Hp_FDFD_modular}.
This shows that \ac{AFDM} allows accurate estimation of the Doppler shift and delay beyond the Nyquist rate, even though the signal is sampled at the Nyquist rate.

\vspace{-2ex}
\subsubsection{Special Case: OFDM-like Estimation Under Low Doppler}  
\label{subsec:speccaseestim}

The previous sections presented the channel estimation explicitly in the \ac{DAFT} domain native to the \ac{AFDM}, considering fully doubly dispersive channels. 
However, in condition of low Doppler shift, the latter can be omitted, leading to $\mathbf{V}^{f_p} \approx \mathbf{I}_N$, and then $\mathbf{\Xi}^\mathrm{OFDM}$ is diagonal with elements $(\mathbf{\Xi}^\mathrm{OFDM})_{n,n} = \sum_{p=1}^{P} h_p e^{-2j\pi \frac{\ell_p n }{N}}$. 

Note that these conditions correspond to the channels usually assumed in \ac{OFDM} systems, so prior \ac{OFDM} estimation methods remain valid and are especially relevant when only delay estimation is required.
It has been shown in \cite{savaux25_PHYCOM} that channel estimation can also be performed in the frequency domain using the special cases of \ac{AFDM} parameters described in \cite{Savaux_DFTAFDM}. 

In fact, suppose that $2N\lambda_1 \in \mathbb{Z}$ and $\frac{1}{2\lambda_1} \in \mathbb{Z}$, and that the pilot $\mathbf{x}$ in (\ref{eq:pilotsap}) is now expressed as $\mathbf{x} = [0,..,0,x_m^{\mathrm{(pil)}},0,..,0]^\mathsf{T}$. 
Then, the DFT of the \ac{AFDM} transmit signal $\mathbf{s}$ in (\ref{eq:AFDM_tx_vec}), given by
\begin{equation}
\mathbf{z} = \mathbf{F}_N \cdot\mathbf{s} = \mathbf{F}_N \cdot\mathbf{A}^\mathsf{H} \cdot \mathbf{x} \in \mathbb{C}^{N \times 1},
\label{eq:AFDM_tx_vec_freq}
\end{equation}
has the following regular structure, for $k=\{0,1,..,N-1\}$,
\begin{equation}
z_k = \begin{cases}
\zeta_k\mathbbm{1}_{(m-k)\text{mod} \text{ } 2N\lambda_1 = 0}, & \text{if $\frac{1}{2\lambda_1}$ is even} \\ 
\zeta_k\mathbbm{1}_{\frac{(m-k)}{N\lambda_1} \in \mathbb{Z} \backslash 2\mathbb{Z}}, & \text{if $\frac{1}{2\lambda_1}$ is odd}
\end{cases}~, 
\end{equation}
where $\mathbbm{1}_{\Omega}$ is the indicator function of the elements belonging to the set $\Omega$, and $\mathbb{Z} \backslash 2\mathbb{Z}$ is the subset of $\mathbb{Z}$ corresponding to the odd integers. 

\begin{figure}[t]
\centering{\includegraphics[width=0.68\columnwidth]{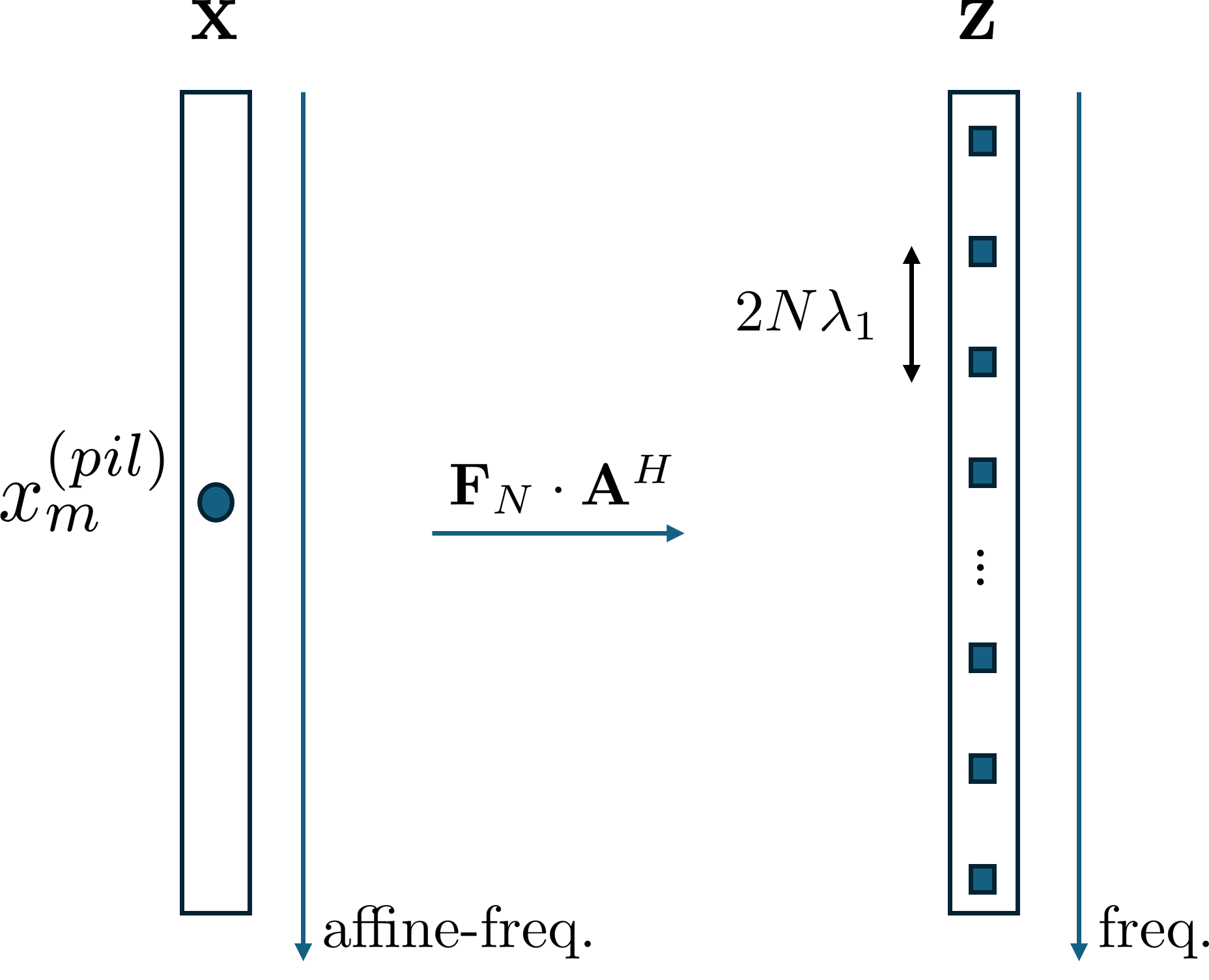}}
\caption{Regular structure of $\mathbf{z} = \mathbf{F}_N \cdot\mathbf{A}^\mathsf{H} \cdot \mathbf{x}$ where $\mathbf{x} = [0,..,0,x_m^{\mathrm{(pil)}},0,..,0]^\mathsf{T}$, in the special cases $2N\lambda_1 \in \mathbb{Z}$ and $\frac{1}{2\lambda_1} \in \mathbb{Z}$.}
\label{fig:pilot_spec_cases}
\vspace{-3.5ex}
\end{figure}

Moreover, $\zeta_k \in \mathbb{C}$ is a constant modulus value (\emph{i.e.} $|\zeta_k| = \zeta \in \mathbb{R}$ for any $k$) whose expression is detailed in \cite{savaux25_PHYCOM}, with the regular structure of $\mathbf{z}$ illustrated in Fig. \ref{fig:pilot_spec_cases}. 
It must be noticed that such equispaced pilots in the frequency domain are used in most of the communications standards based on OFDM, in particular, 4G/5G, and WiFi. 

Based on this pilot structure using special cases of \ac{AFDM} parameters, the \ac{AFDM} demodulation in \eqref{eq:AFDM_rx_freq} can be now performed using a DFT matrix $\mathbf{F}_N$ instead of $\mathbf{A}$, which yield
\begin{align}
\mathbf{y}_\mathsf{f}
&= \mathbf{F}_N \cdot \mathbf{r} \nonumber \\
&= \underbrace{\mathbf{F}_N \Big( \sum_{p=1}^{P} h_p \!\cdot\! \mathbf{\Phi}_p \!\cdot\! \mathbf{V}^{f_p} \!\cdot\! \mathbf{G}(\ell_p) \Big)\mathbf{F}_N^\mathsf{H}}_{\triangleq \,\mathbf{\Xi}^\mathrm{OFDM}\, \in\, \mathbb{C}^{N \times N}} \cdot\mathbf{F}_N \mathbf{A}^\mathsf{H} \cdot \mathbf{x} + \mathbf{F}_N\mathbf{w} \nonumber \\
&= \mathbf{\Xi}^\mathrm{OFDM} \mathbf{z} + \mathbf{F}_N\mathbf{w} \in \mathbb{C}^{N \times 1},        \label{eq:AFDM_rx_freq_DFT}
\end{align}
where $\mathbf{\Phi}_p = \mathbf{I}_N$ because $2N\lambda_1 \in \mathbb{Z}$.

The subscript $\mathsf{f}$ indicates the received AFDM pilot signal in the frequency domain.
In this special case, the usual frequency domain channel estimation used in OFDM \cite{ozdemir2007channel,6814271,savaux17IET} can be performed based on (\ref{eq:AFDM_rx_freq_DFT}). 
Data can also be multiplexed across pilot subcarriers, making AFDM fully backward-compatible with OFDM. 
Otherwise, in doubly dispersive channels, data can be multiplexed in the affine-frequency domain as previously described in (\ref{eq:pilotsap}) and Fig. \ref{fig:pilot}. 
Thus, AFDM is flexible: the modulation/demodulation and channel estimation domains and methods can be adapted to the channel severity using a single pilot scheme.

Under the generalized \ac{FDFD} channel model presented in Section~II, the effective channel matrix incorporates inter-sample coupling via the pulse-shaping kernel $\mathbf{G}(\ell_p)$, which affects both pilot footprint and the exploitable sparsity structure. The \ac{FDFD} model thus provides the structural channel description required to design accurate channel estimators in practical pulse-shaped systems. Further development of dedicated \ac{FDFD}-aware channel estimation algorithms, including low-complexity schemes that exploit delay-Doppler sparsity in this setting, is addressed in related works such as \cite{li26sbl}.

\subsection{Signal Detection and Receiver Architectures}
\label{subsec:receiver_design}

\begin{figure}[b]
\vspace{-2.5ex}
\centering{\includegraphics[width=0.575\linewidth]{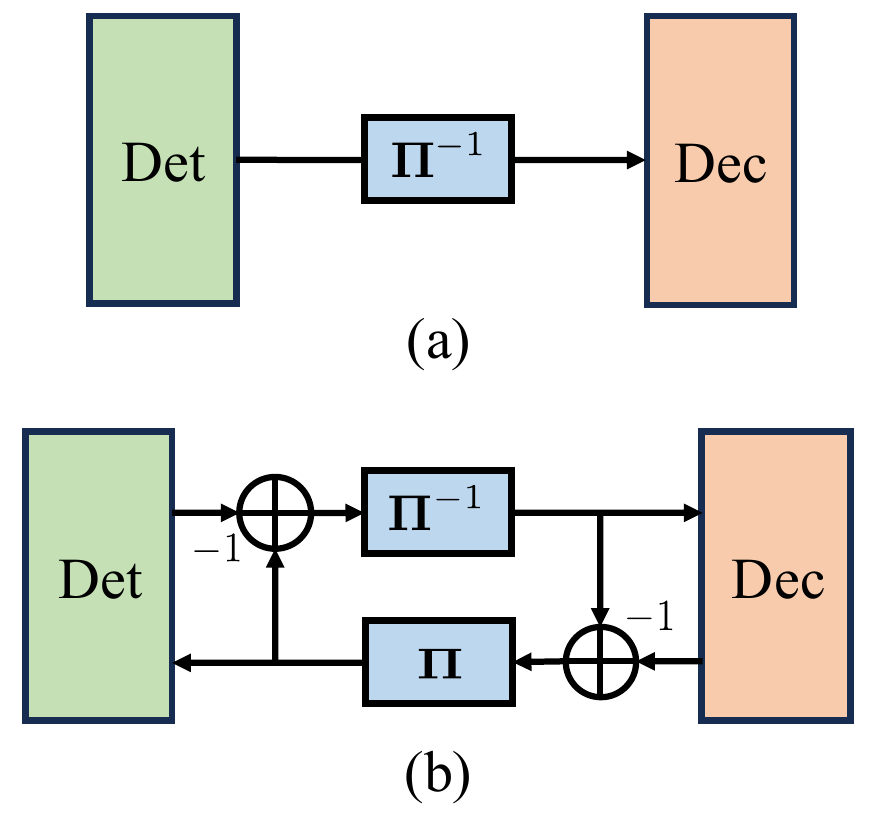}}
\vspace{-0.5ex}
\caption{Illustration of the AFDM receivers using (a) non-outer iterations and (b) turbo iterations between detectors and decoders.}
\label{fig:receiver}
\vspace{-2ex}
\end{figure}

Following the acquisition of \ac{CSI} via the estimation schemes detailed in the previous section, the receiver must recover the data vector $\mathbf{x}$ from the received affine-frequency domain signal $\mathbf{y}$. 
The structural insights provided by the generalized \ac{FDFD} model are also significant in this context, as by reformulating the complex fractional-delay channel into the structured virtual \ac{IDID} representation as defined in \eqref{eq:matrix_DDsystem_FDFD3}, the receiver can use established detection algorithms with well-defined interference cancellation targets. 
Depending on the interplay between the detector and decoder, \ac{AFDM} receivers are categorized into two architectures, as illustrated in Fig. \ref{fig:receiver}. 

The first category consists of one-shot receivers that perform detection without outer iterations for improved computational efficiency. 
In contrast, the second category comprises turbo receivers designed for coded \ac{AFDM} systems, which iteratively exchange extrinsic \acp{LLR} between the detector and the decoder to approach the channel capacity. 
We first examine \ac{AFDM} receiver architectures that operate without outer iterations, then discuss turbo receivers.

\vspace{-3ex}
\subsubsection{One-shot Receivers (No outer iterations)}

The detection algorithms of \ac{AFDM} receivers fall into two broad groups: linear detectors and sparse channel-enabled detectors. 

\emph{a) Linear detectors:} This group comprises \ac{ZF} and \ac{LMMSE} receivers \cite{11220240,11214369,11185315,10806672,yi2025non}. 
Based on the input-output relationship of \ac{AFDM} in \eqref{eq:AFDM_rx_freq}, the \ac{ZF} and \ac{MMSE} detectors can be respectively formulated as
\begin{align}
\tilde{\mathbf{x}}^\text{ZF}=\left[(\mathbf{\Xi}^\mathrm{AFDM})^\mathsf{H}\mathbf{\Xi}^\mathrm{AFDM}\right]^{-1}(\mathbf{\Xi}^\mathrm{AFDM})^\mathsf{H}\mathbf{y}, 
\end{align}
\begin{align}\label{eq:MMSE}
\tilde{\mathbf{x}}^\text{MMSE}=\left[(\mathbf{\Xi}^\mathrm{AFDM})^\mathsf{H}\mathbf{\Xi}^\mathrm{AFDM}+\sigma^2_\mathrm{w}\mathbf{I}_{N}\right]^{-1}(\mathbf{\Xi}^\mathrm{AFDM})^\mathsf{H}\mathbf{y}.
\end{align}

The complexity of calculating the \ac{MMSE} detector directly is on the order of $\mathcal{O}(N^3)$, which is excessive for large-scale \ac{AFDM} systems.
Upon exploiting matrix factorization techniques, such as LU factorization \cite{8859227}, Cholesky decomposition \cite{yi2025error}, and LDL factorization \cite{9746329}, the complexity of linear detectors can be mitigated. 
As an example, the LDL factorization-based \ac{MMSE} detector is elaborated in this section to provide further insight.

\vspace{1ex} \textbf{LDL factorization-based MMSE detector:} 
Owing to the sparsity pattern of the effective \ac{DAFT}-domain channel, the matrix $\mathbf{\Gamma}$ is Hermitian positive-definite with the lower and upper bandwidth $J$. 
Exploiting the banded structure enables the use of an $\mathrm{LDL}$ factorization and the trimming matrix {$\mathbf{T} = [\mathbf{I}_N]_{Q-(\alpha_{\max}+k_\nu):N-(\alpha_{\max}+k_\nu)-1, :}$}, where $\mathbf{\Gamma}$ is decomposed into a unit lower-triangular matrix $\mathbf{L}$ containing only $J$ sub-diagonals and a diagonal matrix $\mathbf{D}$. 
This decomposition reduces both memory usage and computational complexity to $\mathcal{O}(J^2 N)$, representing a substantial improvement over conventional \ac{MMSE} inversion. 

Once the LDL factors are obtained, the equalization step reduces to solving two triangular systems and one diagonal system, all of which can be efficiently implemented. 
The complete LDL-based low-complexity \ac{MMSE} equalization procedure is summarized in \textbf{Algorithm~\ref{algo:MMSE_detect}}.
\vspace{-3ex}

\begin{algorithm}[b]
\caption{LDL factorization-based MMSE detector}
\label{algo:MMSE_detect}
\footnotesize
\begin{algorithmic}[1]
\State Construct $\underline{\mathbf{\Xi}}^\mathrm{AFDM} = \mathbf{\Xi}^\mathrm{AFDM}\mathbf{T}^\mathsf{H}$.
\State Form the banded matrix $\mathbf{\Gamma}=\big(\underline{\mathbf{\Xi}}^\mathrm{AFDM}\big)^\mathsf{H}\underline{\mathbf{\Xi}}^\mathrm{AFDM}+\sigma^2_\mathrm{w}\mathbf{I}_{N}$.
\State Compute the $\mathrm{LDL}$ factorization $\mathbf{\Gamma} = \mathbf{L}\mathbf{D}\mathbf{L}^\mathsf{H}$, where $\mathbf{L}$ is a lower triangular matrix with $J$ sub-diagonals and $\mathbf{D}$ is diagonal.
\State Solve the lower triangular system $\mathbf{L}\mathbf{f} = \mathbf{y}$.
\State Solve the diagonal system $\mathbf{D}\mathbf{g} = \mathbf{f}$.
\State Solve the upper triangular system $\mathbf{L}^\mathsf{H}\mathbf{d} = \mathbf{g}$.
\State Compute the MMSE estimate $\tilde{{\mathbf{x}}} = (\underline{\mathbf{\Xi}}^\mathrm{AFDM})^\mathsf{H}\mathbf{d}$.
\end{algorithmic}
\vspace{-1ex}
\end{algorithm}

\emph{b) Sparse channel-enabled detectors:} 
The second group exploits the inherent sparsity of the effective DAFT-domain channel matrix in \eqref{eq:AFDM_rx_freq}. 
These detectors adopt low-complexity iterative schemes, including \ac{MRC} \cite{Bemani_AFDM}, \ac{MP} \cite{tao25,10566604,11075959,RanasingheTWC2025}, and \ac{EP} algorithms \cite{11225907}.

\begin{figure}[b]
\vspace{-1ex}
\centering{\includegraphics[width=0.6\linewidth]{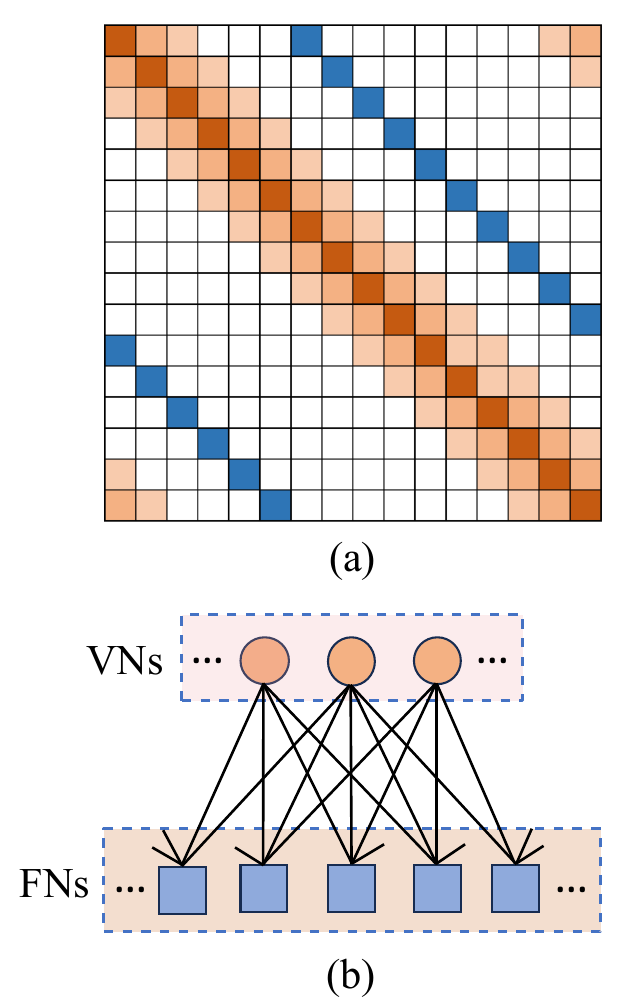}}
\caption{Illustration of (a) the effective DAFT-domain channel matrix with two paths and (b) the corresponding factor graph.}
\label{fig:channel}
\vspace{-2ex}
\end{figure}

Fig.~\ref{fig:channel} illustrates an example of the effective channel matrix and its associated factor-graph representation, where each \ac{VN} corresponds to a transmitted symbol and each \ac{FN} represents a received symbol connected to a subset of \acp{VN} determined by the channel structure. During message passing, probabilistic information is iteratively exchanged between \acp{VN} and \acp{FN} to refine the symbol estimates progressively.
Next, we provide more details on the \ac{MRC} and \ac{MP} detectors, as the main examples.

\textbf{MRC detector:} During the $t$-th iteration, each symbol $\tilde{x}^{(t)}[n]$ is sequentially updated according to
\begin{align}\label{Eq:sn}
\tilde{x}^{(t)}[n] = \frac{g^{(t)}_n}{d_n + \sigma_\mathrm{w}^2},
\end{align}
where
\begin{gather}\label{Eq:g_k_n}
g^{(t)}_n = \sum_{l \in \mathcal{L}_n} \big[\underline{\mathbf{\Xi}}^\mathrm{AFDM}\big]_{l,n}^*\, \Delta^{(t-1)}[l] + d_n\,\tilde{x}^{(t-1)}[n], \\[0.5ex]
\label{Eq:d_n}
d_n = \sum_{l \in \mathcal{L}_n} \big|\big[\underline{\mathbf{\Xi}}^\mathrm{AFDM}\big]_{l,n}\big|^2,
\end{gather}
with $\Delta^{(t-1)}$ denoting the residual vector at the $(t\!-\!1)$-th iteration and $\Delta^{(t-1)}[l]$ as its $l$-th element, and $\mathcal{L}_n$ represents the set of non-zero row indices in the $n$-th column of the truncated effective channel $\underline{\mathbf{\Xi}}^{\mathrm{AFDM}}$. 

After each symbol update, the corresponding residual entries are refined via
\begin{align}
\Delta^{(t)}[l]=\Delta^{(t-1)}[l]
- \big[\underline{\mathbf{\Xi}}^\mathrm{AFDM}\big]_{l,n}\big(\tilde{x}^{(t)}[n] - \tilde{x}^{(t-1)}[n]\big),\nonumber \\[-1ex]  \label{Eq:Delta_r}
\end{align}
for all $l\in\mathcal{L}_n$. Once all symbols have been updated, the refined estimates are used for interference cancellation in the subsequent iteration. 
This iterative procedure continues until the change between two consecutive symbol vectors becomes negligible or the maximum number of iterations $T$ is reached.
These inner iterations are not to be confused with turbo iterations between the detector and decoder, discussed later.

\textbf{MP detector:} 
From onwards, we assume that the indices $\ell_p$ and $f_p$ are integer-valued for general $p=\{1,\ldots,P\}$, either for \ac{IDID} scenarios \eqref{eq:Hp_integerdelay} or following the generalized \ac{FDFD} model of \eqref{eq:matrix_DDsystem_FDFD3}. 
First, the \ac{AFDM} input-output relation is rewritten in scalar form as
\begin{align}
y[a] = \sum_{p=1}^{P} h_p \cdot e^{j\frac{2\pi}{N}\!\left(Nc_1 \ell_p^{2}- b\ell_p + Nc_2(b^2-a^2)\right)} \cdot 
x[b] + \tilde{w}[a], \nonumber \\[-2ex]  \label{eq:AFDM-ele}
\end{align}
for $a=\{1,\ldots,N\}$. From the vectorized \ac{AFDM} model in \eqref{eq:AFDM_rx_freq}, the received vector $\mathbf{y}\in\mathbb{C}^{N}$ and noise vector $\tilde{\mathbf{w}}\in\mathbb{C}^{N}$ contain elements $y[a]$ and $\tilde{w}[a]$, respectively. 
According to \eqref{eq:AFDM-ele}, the non-zero entries of the effective matrix $\mathbf{\Xi}^{\mathrm{AFDM}}$ exhibit a structured sparsity pattern: the $a$-th row contains non-zero elements at indices $(a+\mathrm{loc}_p)_N$, while the non-zero entries of the $b$-th column appear at indices $(b-\mathrm{loc}_p)_N$. 

Then, by letting $I_a$ as the set of non-zero column indices in row $a$, and $D_b$ as the set of non-zero row indices in column $b$, the system model \eqref{eq:AFDM_rx_freq} naturally corresponds to a sparse factor graph with $N$ 
\acp{VN} representing $\mathbf{x}$ and $N$ \acp{FN} representing $\mathbf{y}$.  
Each observation node $y[a]$ connects to $P$ variable nodes $\{x[b]: b\in I_a\}$, while each variable node $x[b]$ connects to $P$ observation nodes $\{y[a]: a\in D_b\}$.

Directly from \eqref{eq:AFDM_rx_freq}, the optimal joint \ac{MAP} detector is given by
\begin{align}
\hat{\mathbf{x}}
= \arg\max_{\mathbf{x}\in\mathcal{X}^{N}} 
\Pr\!\left(\mathbf{x}\mid \mathbf{y},\mathbf{\Xi}^{\mathrm{AFDM}}\right),
\end{align}
but this approach requires exponential complexity in $N$. 

To avoid this complexity, a symbol-wise \ac{MAP} approximation is adopted as
\begin{subequations}
\begin{align}
\tilde{x}[b]
&= \arg\max_{a_j\in\mathcal{X}}
\Pr\!\left(x[b]=a_j\mid \mathbf{y},\mathbf{\Xi}^{\mathrm{AFDM}}\right) \notag \\
&= \arg\max_{a_j\in\mathcal{X}} 
\frac{1}{M}\Pr\!\left(\mathbf{y}\mid x[b]=a_j,\mathbf{\Xi}^{\mathrm{AFDM}}\right) \label{xca}\\
&\approx \arg\max_{a_j\in\mathcal{X}}
\prod_{a\in D_b}
\Pr\!\left(y[a]\mid x[b]=a_j,\mathbf{\Xi}^{\mathrm{AFDM}}\right), \label{xcb}
\end{align}
\end{subequations}
where \eqref{xca} assumes equiprobable symbols, \eqref{xcb} uses an independence approximation motivated by the sparsity of $\mathbf{\Xi}^{\mathrm{AFDM}}$, and the interference terms $\zeta^{(t)}_{b,a}$ defined in \eqref{zeta} are assumed independent for fixed $b$.

\begin{figure}[t]
\centering{\includegraphics[width=0.78\linewidth]{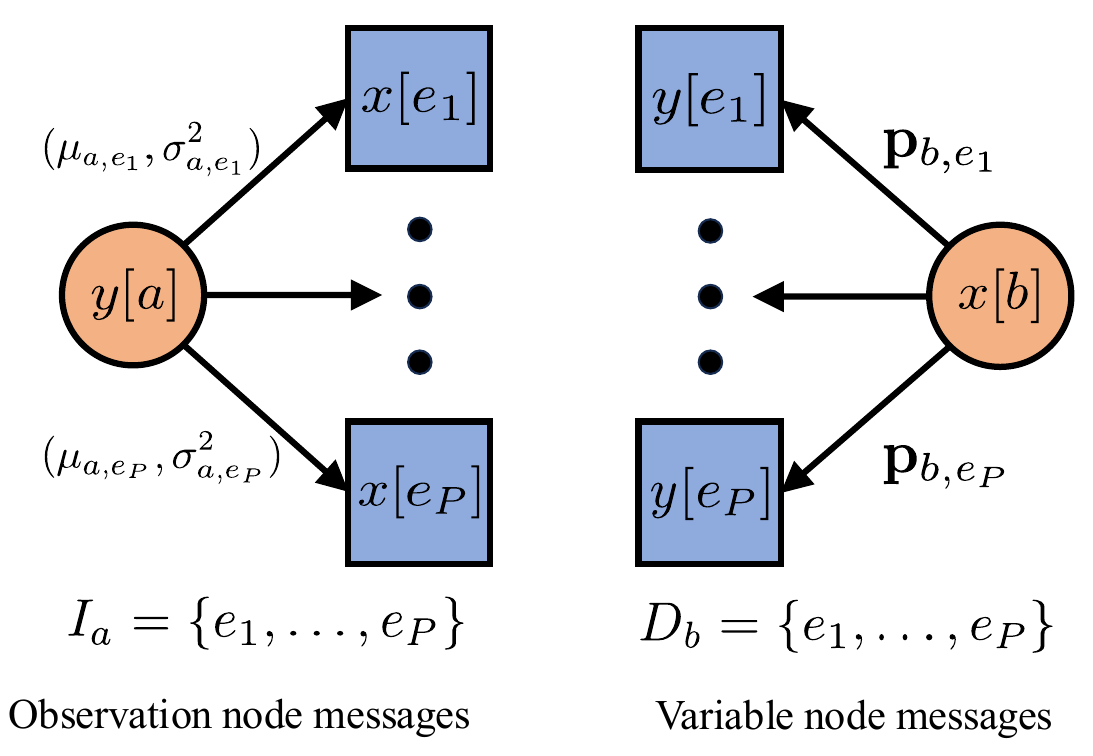}}
\caption{The factor graph structure of the MP algorithm.}
\label{fig:MP}
\vspace{-2ex}
\end{figure}

To address the intractability of the full \eqref{xcb}, an \ac{MP} detector is employed whose complexity only scales linearly with $N$.  
For each observation $y[a]$, the contribution of $x[b]$ is isolated, and the sum of remaining interference terms is approximated as a Gaussian random variable with mean and variance computed in closed form, whose factor is shown in Fig. \ref{fig:MP}. 
In the \ac{MP} framework, observation-to-variable messages consist of these Gaussian parameters, whereas variable-to-observation messages consist of \acp{PMF} over the constellation alphabet,  
$\mathbf{p}_{b,a}^{(t)}=\{p_{b,a}^{(t)}(a_j):a_j\in\mathcal{X}\}$.  

Given above, the $t$-th iteration of the MP algorithm proceeds as follows.
First, the interference term $\zeta^{(t)}_{b,a}$ is defined as
\begin{equation}\label{zeta}
\zeta^{(t)}_{a,b}
= \sum_{e\in D_b,\, e\ne b} x[e] \cdot \big[{\mathbf{\Xi}}^\mathrm{AFDM}\big]_{a,e} + \tilde{w}[a].
\end{equation}

Next, the mean $\mu_{a,b}^{(t)}$ and variance $\big(\sigma_{a,b}^{(t)}\big)^{2}$ are given by
\begin{align}\label{eq_mu}
\mu_{a,b}^{(t)}= \sum_{e\in I_a,\, e\ne b}
\sum_{j=1}^{M} p_{e,a}^{(t-1)}(a_j) \cdot a_j \cdot \big[{\mathbf{\Xi}}^\mathrm{AFDM}\big]_{a,e},
\end{align}

\begin{align}\label{eq_sigma}
\!\!\!\big(\sigma_{a,b}^{(t)}\big)^{\!2}
& \!\!=\!\!\!\!\!\! \sum_{e\in I_a,\, e\ne b} \!\!
\left(
\sum_{j=1}^{M}
p_{e,a}^{(t-1)}(a_j) \!\cdot\! |a_j|^{2} \!\cdot\!
\Big|\big[{\mathbf{\Xi}}^\mathrm{AFDM}\big]_{a,e}\Big|^{2}
\right. \\[-0.2em]
&\qquad\qquad\left.
\!\!\!\!-\!
\left|
\sum_{j=1}^{M}
p_{e,a}^{(t-1)}(a_j) \!\cdot\! a_j\!\cdot\! \big[{\mathbf{\Xi}}^\mathrm{AFDM}\big]_{a,e}
\right|^{2}
\right)
\!\!+\! \sigma_{\mathrm{w}}^{2}. \nonumber
\end{align}
From \eqref{eq:AFDM-ele}, the corresponding channel coefficient is
\begin{equation}
\big[{\mathbf{\Xi}}^\mathrm{AFDM}\big]_{a,e}
= h_p \cdot e^{j\frac{2\pi}{N}\left(Nc_1\ell_p^{2}-q\ell_p+Nc_2(q^{2}-a^{2})\right)},
\end{equation}
where $q=(a+\mathrm{loc}_p)_N$. Then, the updated \acp{PMF} are computed as
\begin{equation}\label{eq_p}
p^{(t)}_{b,a}(a_j)
= \Delta\, \tilde{p}^{(t)}_{b,a}(a_j)
+ (1-\Delta)\, p^{(t-1)}_{b,a}(a_j),
\end{equation}
where $\Delta\in(0,1]$ is a damping factor, and
\begin{align}
\tilde{p}^{(t)}_{b,a}(a_j)
&\propto \prod_{\substack{e\in D_b \\ e\ne a}}
\Pr\!\left( y[e] \mid x[b]=a_j, \mathbf{\Xi}^{\mathrm{AFDM}} \right) \\
&= \prod_{\substack{e\in D_b \\ e\ne a}}
\frac{\xi^{(t)}(e,b,j)}
{\sum_{m=1}^{M} \xi^{(t)}(e,b,m)},
\end{align}

~
\vspace{-4ex}

\noindent with
\begin{equation}
\xi^{(t)}(e,b,m)
= \exp\!\left(
-\frac{\Big|y[e]-\mu^{(t)}_{e,b} - a_m \! \cdot \! [{\mathbf{\Xi}}^\mathrm{AFDM}\big]_{a,e}\Big|^{2}}
{(\sigma^{(t)}_{e,b})^{2}}
\right), 
\end{equation}
where the channel coefficient $[{\mathbf{\Xi}}^\mathrm{AFDM}\big]_{e,b}$ is
\begin{equation}
[{\mathbf{\Xi}}^\mathrm{AFDM}\big]_{e,b}
= h_p \cdot e^{j\frac{2\pi}{N}\left(Nc_1\ell_p^{2}-b\ell_p+Nc_2(b^{2}-u^{2})\right)},
\end{equation}
with $u=(b-\mathrm{loc}_p)_N$.
Then, the symbol estimate is updated as $\tilde{x}[b]
= \arg\max_{a_j\in\mathcal{X}} p_{b}^{(t)}(a_j)$. Within the above steps, convergence is measured as
\begin{equation}
\eta^{(t)}
= \frac{1}{N}
\sum_{b=1}^{N}
\mathbb{I}\!\left(
\max_{a_j\in\mathcal{X}} p_{b}^{(t)}(a_j)
\ge 1-\gamma
\right),  
\end{equation}
for a small threshold $\gamma>0$, where
\begin{equation}
p_{b}^{(t)}(a_j)
= \prod_{e\in D_b}
\frac{\xi^{(t)}(e,b,j)}
{\sum_{m=1}^{M} \xi^{(t)}(e,b,m)}.   
\end{equation}

{
\begin{algorithm}[t]
\caption{MP-based detector}
\label{algo:MP_detect}
\footnotesize
\begin{algorithmic}[1]
\Require
Receive signal $\bf{y}$ and channel matrix $\mathbf{\Xi}^\mathrm{AFDM}$.
\State \textbf{Preparation}: ${\bf{p}}^{(0)}_{b,a}=1/M$, $b=0,\ldots,N-1$, $a\in D(b)$.
\For{$t=1$ to $T_{\text{max}}$}
\State Observation nodes $y[a]$ compute means $\mu_{a,b}^{(t)}$ and variances $\big(\sigma_{a,b}^{(t)}\big)^{\!2}$ via \eqref{eq_mu} and \eqref{eq_sigma} based on ${\bf{p}}^{(t-1)}_{b,a}$, and pass them to $\{x[b]:b\in I_a\}$.
\State Variable nodes $x[b]$ update ${\bf{p}}^{(t-1)}_{b,a}$ via \eqref{eq_p}, and pass them to $\{y[a]: a\in D_b\}$.
\State Calculate convergence indicator $\eta^{(t)}$.
\State Update the decision on the transmitted symbols $x[c]$ for $c=1,\ldots,N$, if needed.
\State \textbf{if} Stopping criteria are satisfied
\State \textbf{break}
\EndFor
\State \textbf{return} $\hat{\bf{x}}$.
\end{algorithmic}
\end{algorithm}
}

Finally, the \ac{MP} detector, as summarized in \textbf{Algorithm~\ref{algo:MP_detect}}, terminates when one of the following conditions is met at the end of each iteration: \textit{a)} $\eta^{(t)}=1$; ~ \textit{b)} $\eta^{(t)} < \eta^{(t^*)} - \epsilon$; ~\textit{c)} $t = T_{\max}$;
where $t^*$ is the iteration with the largest $\eta^{(t^*)}$ for $1\le t^*<t$;

\vspace{-2ex}
\subsubsection{Turbo Receiver}

Alternative to the one-shot receivers where there are no outer iterations between the detector and the decoder, \ac{AFDM} turbo receivers have been investigated in \cite{11150613,10566604,11214369}. 
As shown in Fig.~\ref{fig:receiver}(b), a turbo receiver operates by iteratively exchanging soft reliability information between the detector and the channel decoder, enabling both modules to refine their estimates progressively. 

Unlike conventional receivers that perform detection and decoding in isolation, as depicted in Fig.~\ref{fig:receiver}(a), a turbo receiver exploits the decoder's soft outputs to help the equalizer suppress residual interference.

Specifically, the extrinsic \ac{LLR} output by the \ac{SBC} after the detector can be formulated as \cite{tuchler2002minimum}
\begin{align}
L_e&(c_{n,j})\\
&=\ln
\frac{
\displaystyle \sum_{\forall s_i : s_{i,j} = 0} 
p(\hat{x}_n \mid \mathbf{c}_n = \mathbf{s}_i)
\prod_{\forall j' : j' \neq j} P(c_{n,j'} = s_{i,j'})
}{
\displaystyle \sum_{\forall s_i : s_{i,j} = 1} 
p(\hat{x}_n \mid \mathbf{c}_n = \mathbf{s}_i)
\prod_{\forall j' : j' \neq j} P(c_{n,j'} = s_{i,j'})
}, \nonumber
\end{align}
where $\hat{x}_n$ denotes the $n$th estimated element, $\mathbf{c}_n$ is the transmit bit sequence corresponding to ${x}_n$, $\mathbf{s}_i$ represents the $i$th bit pattern associated with the constellation point $a_i\in\mathcal{X}$, and $s_{i,j}$ is the $j$th bit of $\mathbf{s}_i$. 

This iterative feedback loop allows the system to approach optimal joint detection performance with substantially lower computational complexity.
A turbo receiver's primary advantage is exploiting coding and diversity gains simultaneously. 
After each iteration, the improved extrinsic information helps mitigate \ac{ICI}, reduce error propagation, and reshape the effective channel into a more favorable form for subsequent decoding. 
This makes turbo receivers effective in highly dispersive environments, where the detector alone is insufficient, at the cost of the additional complexity and latency of the outer iterations.

In summary, while the distinct matrix structure of \ac{AFDM} necessitates specialized internal detection logic, such as LDL-based banded equalization or sparse factor-graph processing, the high-level architectural principles remain consistent with legacy \ac{OFDM} receivers. 
Therefore, moving from \ac{OFDM} to \ac{AFDM} receivers primarily involves replacing the core detection block, while the surrounding infrastructure remains unchanged. 
This compatibility ensures that the extensive ecosystem of receiver optimization techniques developed for \ac{5G} can be directly extended to \ac{AFDM}, providing a practically viable evolutionary path for \ac{6G}.

\subsection{Impact of Fractional-Delay Model on Receiver Design}
\label{subsec:fdfd_impact}

The generalized \ac{FDFD} model of Section~\ref{sec:system_model}-\ref{subsec:matrix_form_IO} carries concrete consequences for receiver design, which we quantify here along two axes of detection performance and detector complexity.

We consider \ac{AFDM} with \ac{QPSK} over a true \ac{FDFD} channel ($N=64$, $P=3$, fractional delays $\ell_p$ and Doppler $f_p$), and compare two genie-aided \ac{MMSE} receivers given perfect knowledge of the path parameters $(h_p,\ell_p,f_p)$ that equalize the {same} received vector, differing {only} in the channel model used to construct the equalizer.
Namely, the \emph{proposed} model uses the exact \ac{FDFD} effective channel $\mathbf{A}\mathbf{H}\mathbf{A}^{\mathsf{H}}$ with $\mathbf{H}$ from \eqref{eq:Hp_FDFD_generalized}--\eqref{eq:fracCPP_interp_matrix}, whereas the \emph{conventional} model rounds the delays to the integer grid ($\ell_p \!\to\! \lfloor\ell_p\rceil$), i.e.,\ the standard \ac{IDFD} assumption that neglects inter-sample coupling.

\begin{figure}[t]
\centering
\includegraphics[width=0.925\columnwidth]{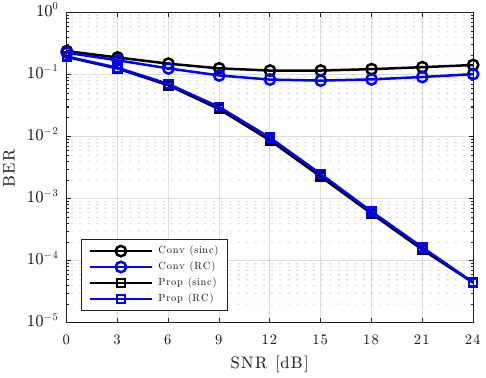}
\caption{Uncoded \ac{AFDM}-\ac{QPSK} \ac{BER} over a true \ac{FDFD} channel ($N\!=\!64$, $P\!=\!3$), comparing genie-\ac{MMSE} receivers built on the \emph{proposed} \ac{FDFD} model (fractional-delay) vs.\ the \ac{IDFD} (integer-delay) model.}
\label{fig:ber_fdfd_idfd}
\vspace{-3.5ex}
\end{figure}

Fig.~\ref{fig:ber_fdfd_idfd} shows that the conventional integer-delay receiver exhibits an irreducible \ac{BER} floor at $10^{-1}$ \emph{even with perfect parameter knowledge}, because the integer-tap model structurally cannot represent fractional-delay inter-sample coupling. The receiver built on the proposed \ac{FDFD} model removes this floor and recovers the expected waterfall. This isolates the practical necessity of the \ac{FDFD} model: ignoring fractional delay induces a \emph{modeling} error rather than an estimation error, and it dominates performance at moderate-to-high \acs{SNR} regimes. \vspace{-2ex}

In addition, the same fractional-delay coupling governs the \emph{sparsity} of the effective channel matrix, and hence the cost of the sparse detectors discussed in the previous sections. Quantifying the channel-matrix density visualized in Fig.~\ref{fig:pulse_comparison} as the average number of taps per row capturing $99\%$ of the row energy (over random \ac{FDFD} realizations, $N=64$, $P=3$), we obtain $2.6$ (rectangular), $4.6$ (raised cosine, $\alpha=0.5$), and $12.3$ (sinc) taps per row. As the per-iteration cost of message-passing and banded \ac{MMSE} detectors scales with the number of nonzero taps per row, the transmit pulse directly trades spectral containment against detector complexity: the ideal band-limited (sinc) pulse maximizes inter-sample coupling and detector load, the rectangular pulse minimizes both (at the cost of spectral leakage), and a practical raised-cosine pulse offers a favorable compromise.

\vspace{-2ex}
\subsection{Extension to MIMO-AFDM Systems}

The \ac{SISO} channel estimation and detection methods of Sections~\ref{sec:analysis_AFDMOFDM}-\ref{subsec:chanest} and~\ref{sec:analysis_AFDMOFDM}-\ref{subsec:receiver_design} extend directly to \ac{MIMO}-\ac{AFDM} via the per-antenna model of \eqref{eq:MIMO_AFDM_freq}: distributing pilots across transmit antennas as in \eqref{eq:pilotsap} allows the $N_\mathrm{r}\cdot N_\mathrm{t}$ channels to be estimated using the same \ac{SISO} techniques applied per receive antenna, following the principle established in \cite{yin24} and closely related to the multiple pilot-aided estimation of \cite{yin22}.
When the Doppler shift is small enough to be neglected, as in conventional \ac{OFDM} systems, digital beamforming (precoding in \ac{DL} or combining in \ac{UL}) can further be performed per-subcarrier in the frequency domain \cite{savaux24eusipco}, yielding \vspace{-1ex}
\begin{equation}
\mathbf{y}_{\mathsf{f},k} = \mathbf{\Xi}^\mathrm{OFDM}_k \mathbf{z}_k + \mathbf{w}_{\mathsf{f},k}  \in \mathbb{C}^{N_\mathrm{r} \times 1},\vspace{-2ex}
\label{eq:yMIMO}
\end{equation}

where $\mathbf{z}_k \in \mathbb{C}^{N_\mathrm{t} \times 1}$ is the per-subcarrier transmitted symbol vector across antennas and $\mathbf{\Xi}^\mathrm{OFDM}_k \in \mathbb{C}^{N_\mathrm{r} \times N_\mathrm{t}}$ denotes the corresponding diagonal-Doppler channel matrix, onto which standard \ac{MIMO}-\ac{OFDM} beamforming techniques (e.g., zero-forcing) apply directly.
Additionally, in the special cases of \ac{AFDM} parameters \cite{Savaux_DFTAFDM} where $2N\lambda_1 \in \mathbb{Z}$ and $\frac{1}{2\lambda_1} \in \mathbb{Z}$, channel estimation and beamforming can be performed in the frequency domain, as described in Section~\ref{sec:analysis_AFDMOFDM}-\ref{subsec:chanest}.\ref{subsec:speccaseestim}.

\subsection{Multiple Access and Coexistence with OFDM} 

The flexibility and backward compatibility of AFDM make it well suited to efficient, resilient multiple access in complex electromagnetic environments. Specifically, one can configure $c_1$ for different users based on their moving speeds to achieve robust transmissions. This supports multiuser communications in, e.g., 6G NTNs, where base stations (e.g., high-altitude platform stations and low-earth-orbit satellites) and user terminals (e.g., flying drones, connected vehicles, high-speed trains) move at varying speeds. The waveform coexistence between AFDM and OFDM was studied in \cite{Arslan-OJCOM-2025}, revealing that each waveform admits a sparse, comb-like representation in the other's native domain. By exploiting the full diversity property of AFDM, a rate-splitting multiple access (RSMA) framework was reported in \cite{Arslan-RSMA-2025}, in which the coexistence between AFDM and OFDM was investigated. A downlink AFDM-RSMA scheme was proposed in \cite{Yin-ICCC-2025} with the key idea that private messages of different users are mapped onto orthogonal chirps, while common messages span all the chirps. To support massive connectivity over high mobility channels, the integration of AFDM and sparse code multiple access (SCMA), called AFDM-SCMA, was studied in \cite{10566604}. As a code-domain non-orthogonal multiple access scheme \cite{10566604}, OFDM-based SCMA can degrade under carrier frequency offsets \cite{10342857} or Doppler shifts. AFDM-SCMA was shown to achieve favorable error-rate performance by combining the strengths of both techniques.

A complementary multiple access paradigm is \ac{DAFT}-spreading-based \ac{AFDMA} \cite{tao26afdma}, which spreads each user's data across the full \ac{DAFT} domain rather than allocating distinct subcarrier or block segments. This approach exploits the full diversity and spreading gain of the \ac{DAFT} basis to improve robustness in high-mobility environments, and can be contrasted with the per-block and per-subcarrier multiplexing schemes described below in terms of spectral efficiency, receiver complexity, and diversity exploitation.

\begin{figure}[b]
\vspace{-2ex}
\centering{\includegraphics[width=0.8\columnwidth]{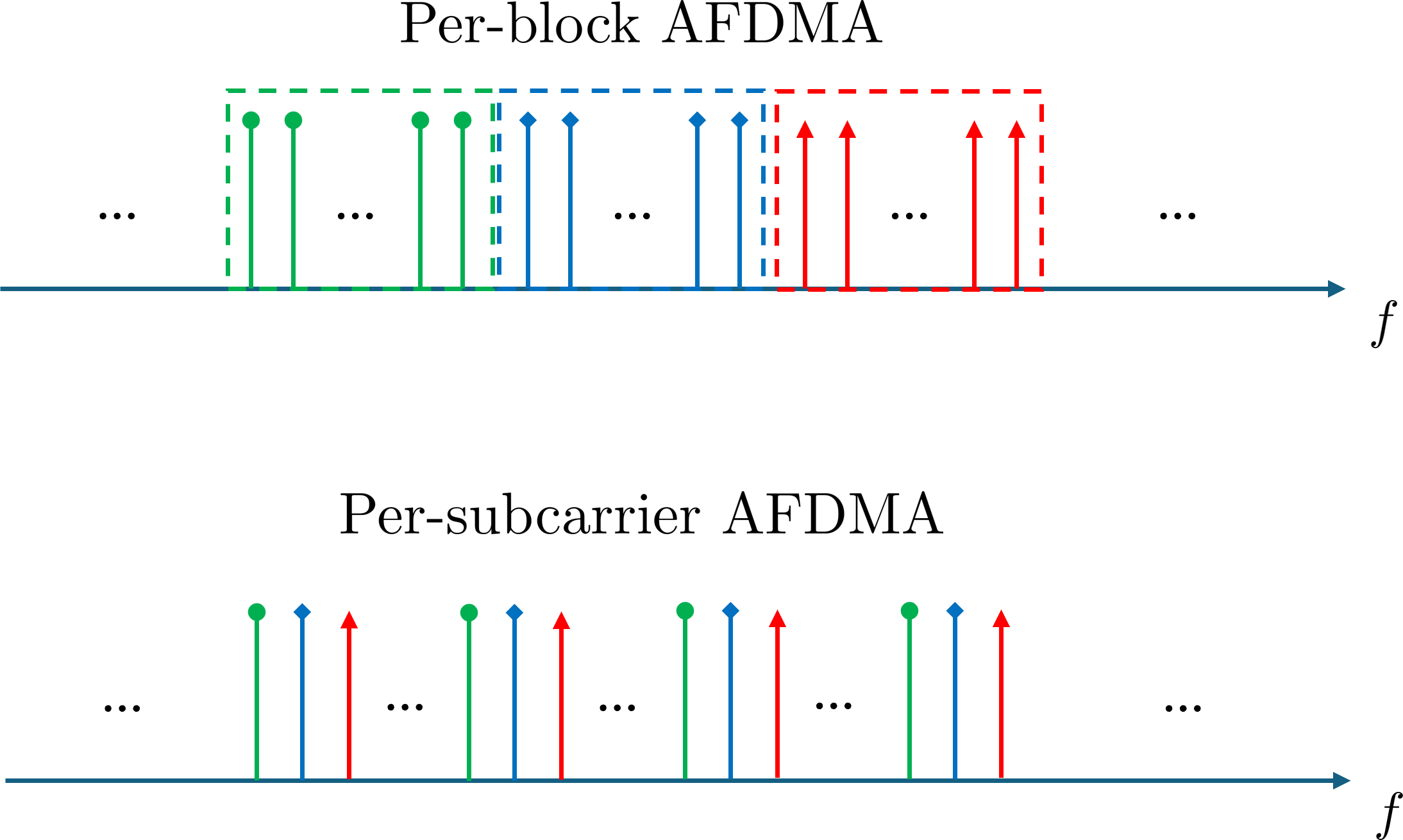}}
\vspace{-1ex}
\caption{Frequency mapping strategies for \ac{AFDMA}: per-block (top) and per-subcarrier (bottom) approaches.}
\label{fig:pbpsc_afdma}
\vspace{-2.5ex}
\end{figure}

\vspace{-2ex}
\subsubsection{General AFDMA and Coexistence with OFDMA}

Two distinct paradigms may facilitate \ac{AFDMA}, as illustrated in Fig. \ref{fig:pbpsc_afdma}: \textit{a)} per-block multiplexing and \textit{b)} per-subcarrier multiplexing. 
These schemes enable multiple \acp{UE} to share the available bandwidth while maintaining compatibility with legacy \ac{OFDMA} systems.

\paragraph{Per-Block AFDMA}

In the per-block approach, $K_\mathrm{UE}$ users share a total bandwidth of $N$ subcarriers, and each user $k_\mathrm{UE} \in \{1, \dots, K_\mathrm{UE}\}$ is allocated $N_{k_\mathrm{UE}}$ subcarriers such that $N = \sum_{k_\mathrm{UE}=1}^{K_\mathrm{UE}} N_{k_\mathrm{UE}}$. 
The composite transmit signal is expressed as
\begin{align}
\mathbf{s} = \mathbf{F}_N^\mathsf{H} \cdot \Big[ &(\mathbf{F}_{N_1} \mathbf{s}_1)^\mathsf{T}, \dots, (\mathbf{F}_{N_{K_\mathrm{UE}}} \mathbf{s}_{K_\mathrm{UE}})^\mathsf{T} \Big]^\mathsf{T}, 
\label{eq:pb_afdma}
\end{align}
where $\mathbf{s}_{k_\mathrm{UE}} = \mathbf{A}_{k_\mathrm{UE}}^\mathsf{H} \mathbf{x}_{k_\mathrm{UE}} \in \mathbb{C}^{N_{k_\mathrm{UE}} \times 1}$ represents the chirped sub-signal of the $k_\mathrm{UE}$-th user. 

At the receiver, the composite signal is decomposed in the frequency domain via an $N$-point \ac{DFT} $\mathbf{F}_N$, after which the individual user signals are recovered using UE-specific \ac{AFDM} demodulation matrices.
Given the above, a primary advantage of per-block \ac{AFDMA} is its native support for coexistence between \ac{AFDM} and \ac{OFDM}. 
Since the sub-signals are orthogonal in the frequency domain, an \ac{AFDM} sub-block can be replaced by an \ac{OFDM} sub-block ($\mathbf{x}_{k_\mathrm{UE}}$) without inducing inter-carrier interference, allowing for transparent spectral sharing between the two waveforms.

\vspace{-0.5ex}
\paragraph{Per-Subcarrier AFDMA}

As an alternative, per-subcarrier \ac{AFDMA} uses specific \ac{AFDM} parameter configurations where $2N\lambda_1 \in \mathbb{Z}$ and $1/(2\lambda_1) \in \mathbb{Z}$, as discussed in Section~\ref{sec:analysis_AFDMOFDM}-\ref{subsec:chanest}.\ref{subsec:speccaseestim}. 
Under these conditions, a data element mapped to an affine-frequency index $m$ is spread across specific frequency-domain indices $k$ satisfying $(k)_{2N\lambda_1} = (m)_{2N\lambda_1}$. 
By allocating subcarriers with index $(m)_{2N\lambda_1} = r$ to a given user, up to $K_\mathrm{UE} = 2N\lambda_1$ users can be multiplexed. 
The transmit signal is given by
\begin{align}
\mathbf{s} = \mathbf{A}^\mathsf{H} \cdot \sum_{k_\mathrm{UE}=0}^{2N\lambda_1-1} \mathbf{x}_{k_\mathrm{UE}}, 
\label{eq:psc_afdma}
\end{align}
\vspace{-1.25ex} 

\noindent where $\mathbf{x}_{k_\mathrm{UE}}$ contains non-zero symbols only at indices $k_\mathrm{UE} + n(2N\lambda_1)$. 

Unlike the per-block approach, per-subcarrier \ac{AFDMA} ensures subcarrier orthogonality in both the affine-frequency and frequency domains. 
This allows for fine-grained multiplexing of \ac{AFDM} and \ac{OFDM} symbols within the same block, as \vspace{-1ex}
\begin{align}
\mathbf{s} = \mathbf{A}^\mathsf{H} \cdot \sum_{k_\mathrm{UE} \neq i}^{2N\lambda_1-1} \mathbf{x}_{k_\mathrm{UE}} + \mathbf{F}^\mathsf{H} \mathbf{x}_i,
\label{eq:psc_afdma2}
\end{align}
where user $i$ utilizes \ac{OFDM}. 

The resulting hybrid signal is easily separable at the receiver, facilitating flexible multi-user access under coexisting \ac{OFDM} and \ac{AFDM} signaling.

\vspace{-2ex}
\subsubsection{MIMO-AFDMA} 

\begin{figure}[b]
\vspace{-2ex}
\centering{\includegraphics[width=0.9\columnwidth]{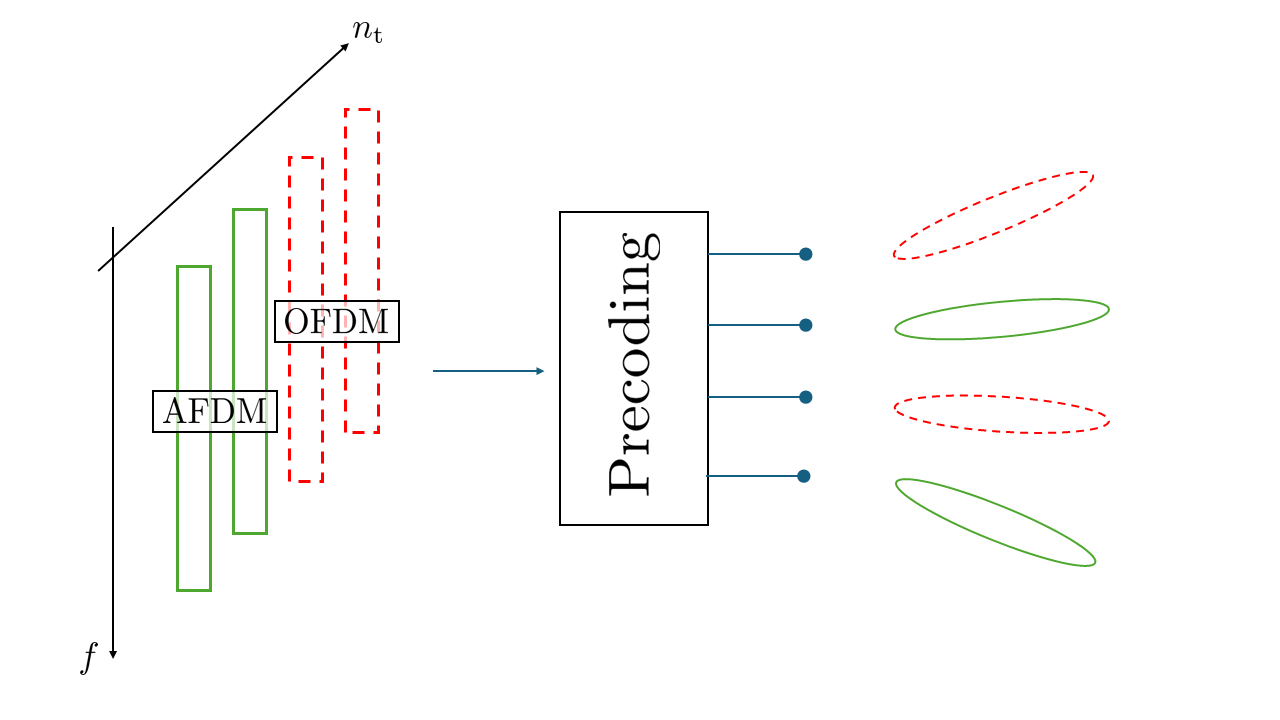}}
\caption{Spatial coexistence in MIMO-AFDM systems utilizing digital beamforming ($N_\mathrm{t}=4$).}
\label{fig:mimoafdm}
\vspace{-2ex}
\end{figure}

Building on the extension of \ac{AFDM} to \ac{MIMO} systems and the multi-user frameworks of legacy \ac{MIMO}-\ac{OFDM}, we consider its adaptation to multi-user \ac{MIMO}, referred to as \ac{MIMO}-\ac{AFDMA}.
\ac{MIMO}-\ac{AFDMA} further extends multi-user capabilities by enabling simultaneous communication between a base station and multiple \acp{UE} through spatial and domain-specific multiplexing, as illustrated in Fig. \ref{fig:mimoafdm}. 
We present a general overview of \ac{MIMO}-\ac{AFDMA} operation in both the \ac{DL} and \ac{UL}.

In the \ac{DL}, with $N_{\mathrm{r}} \geq K_\mathrm{UE}$, the first $(Q+1)N_{\mathrm{r}}$ indices of each transmit vector $\mathbf{x}_{n_\mathrm{t}}$ are reserved for pilot symbols to facilitate channel acquisition. 
The remaining indices are used for data, formatted as
\begin{equation}
\mathbf{x}_{n_\mathrm{t}} = [\underbrace{0, \dots, 0}_{(Q+1)N_{\mathrm{r}}}, \underbrace{0, \dots, 0}_{Q}, \mathbf{x}_\mathsf{d}^{(0)}, \dots, \underbrace{0, \dots, 0}_{Q}, \mathbf{x}_\mathsf{d}^{(K_\mathrm{UE}-1)}]^\mathsf{T}, 
\label{eq:xdatamimo}
\end{equation}
where $\mathbf{x}_\mathsf{d}^{(k_\mathrm{UE})}$ represents the data payload for the $k_\mathrm{UE}$-th user. 

Each data block is isolated by guard bands of $Q$ zero samples to prevent inter-user interference at the receiver. 
Since all transmit antennas broadcast the data blocks, users can exploit spatial diversity to improve detection reliability.

In the \ac{UL}, the $N$ available subcarriers are divided into subsets of contiguous subcarriers allocated to individual \acp{UE} based on throughput requirements. 
Each \ac{UE} transmits its specific payload $\mathbf{x}_{k_\mathrm{UE}}$, consisting of an embedded pilot and the data block $\mathbf{x}_\mathsf{d}^{(k_\mathrm{UE})}$. 
This domain partitioning ensures that multiple users can transmit asynchronously in high-mobility environments while minimizing mutual interference, thanks to \ac{AFDM}'s intrinsic Doppler robustness. 

AFDM and OFDM signals can also be spatially multiplexed without interference, as illustrated in Fig. \ref{fig:mimoafdm} where $n_\mathrm{t}=4$ and two UEs use OFDM. Mathematically, it corresponds to the multiplexing of AFDM and OFDM data within $\mathbf{z}_k$ in (\ref{eq:yMIMO}). This again demonstrates AFDM/OFDM coexistence within one system and AFDM's backward compatibility with OFDM, particularly at low Doppler. 

\subsection{Phase Noise and Carrier Frequency Offset}

\Acf{PHN} and \acf{CFO} are two primary hardware impairments arising from the non-idealities of practical \acp{LO} \cite{1044611}. 
In modern broadband systems, especially those operating at higher carrier frequencies, \acp{LO} are susceptible to thermal and flicker noise-induced instabilities. 
These imperfections manifest as random phase drifts and frequency deviations that distort the baseband waveform during up- and down-conversion. While these effects are highly detrimental to \ac{OFDM}, which relies on precise time-frequency alignment, the chirped subcarriers of \ac{AFDM} offer an inherent structural resilience to these distortions \cite{sui2026mimo}.

\vspace{-3.5ex}
\subsubsection{Characterization of PHN and CFO}

To characterize the statistical behavior of \ac{PHN}, a discrete-time Wiener process model is typically employed \cite{4156406}. 
In this framework, the Gaussian \ac{PHN} increments $\Delta[n]$ at a given antenna evolve as
\begin{align}
\varphi[n]=\varphi[n-1]+\Delta[n],\ n=1,\ldots,N-1,
\end{align}
where $\Delta[n]$ is a zero-mean Gaussian process with variance $\sigma^2_\Delta=4\pi^2 f_c^2 \xi T_s$, and $\xi$ represents the oscillator quality and $T_s$ denotes the sampling period \cite{10841966}. 

Given the above, the resulting distortion is modeled by the diagonal \ac{PHN} matrix as
\begin{align}
\mathbf{\Theta}=\diag\big\{e^{j\varphi[0]},\ldots,e^{j\varphi[N-1]}\big\}.
\end{align}

The spatial correlation of \ac{PHN} is determined by the \ac{LO} architecture. 
In co-located \ac{MIMO} arrays, a \ac{CLO} is typically utilized, resulting in identical \ac{PHN} across all antenna elements. 
Conversely, distributed or \ac{RF}-chain-per-antenna architectures employ \acp{SLO}, leading to spatially independent phase fluctuations. 
This distinction is critical, as \ac{CLO} induces coherent distortion, whereas \acp{SLO} can significantly degrade beamforming gains and the accuracy of spatial parameter estimation \cite{bjornson2015massive}.

\Ac{CFO} arises from frequency mismatches between the transmit and receive \acp{LO}, which are inevitable due to hardware tolerances and temperature variations. 
This mismatch induces a linearly increasing phase rotation across the time-domain samples. In multi-carrier systems, this destroys subcarrier orthogonality, leading to severe \ac{ICI}. 
For chirp-based \ac{AFDM}, \ac{CFO} manifests as an additional Doppler-like phase rotation, characterized by the diagonal matrix \cite{10342857,bemani21_iswcs}
\begin{align}
\mathbf{P}=\diag\left\{1,e^{j2\pi\theta_{\text{CFO}}/N},\ldots,e^{j2\pi\theta_{\text{CFO}}(N-1)/N}\right\},
\end{align}
where $\theta_{\text{CFO}}$ represents the normalized \ac{CFO} factor. 

In all, incorporating the impairments, the effective channel matrix from \eqref{eq:matrix_DDsystem_FDFD3} is generalized as
\begin{align}\label{eq_PHN_CFO}
\mathbf{H} = \mathbf{\Theta}\mathbf{P} \bigg( \sum_{p=1}^{P} h_p \cdot \mathbf{\Phi}_p \cdot \mathbf{V}^{f_p} \cdot \mathbf{\Pi}^{\ell_p} \bigg).
\end{align}

$~$
\vspace{-5ex}
\subsubsection{Waveform Resilience and Performance}

The impact of \ac{PHN} and \ac{CFO} differs fundamentally between the two waveforms due to their distinct basis functions. 
In \ac{OFDM}, these impairments disrupt the frequency-domain orthogonality, manifesting as a \ac{CPE} and rapidly varying \ac{ICI} that scales with the magnitude of the \ac{CFO}. 
Consequently, \ac{OFDM} requires highly accurate tracking and compensation to prevent \ac{BER} degradation.

\begin{figure}[t]
\centering{\includegraphics[width=0.97\linewidth]{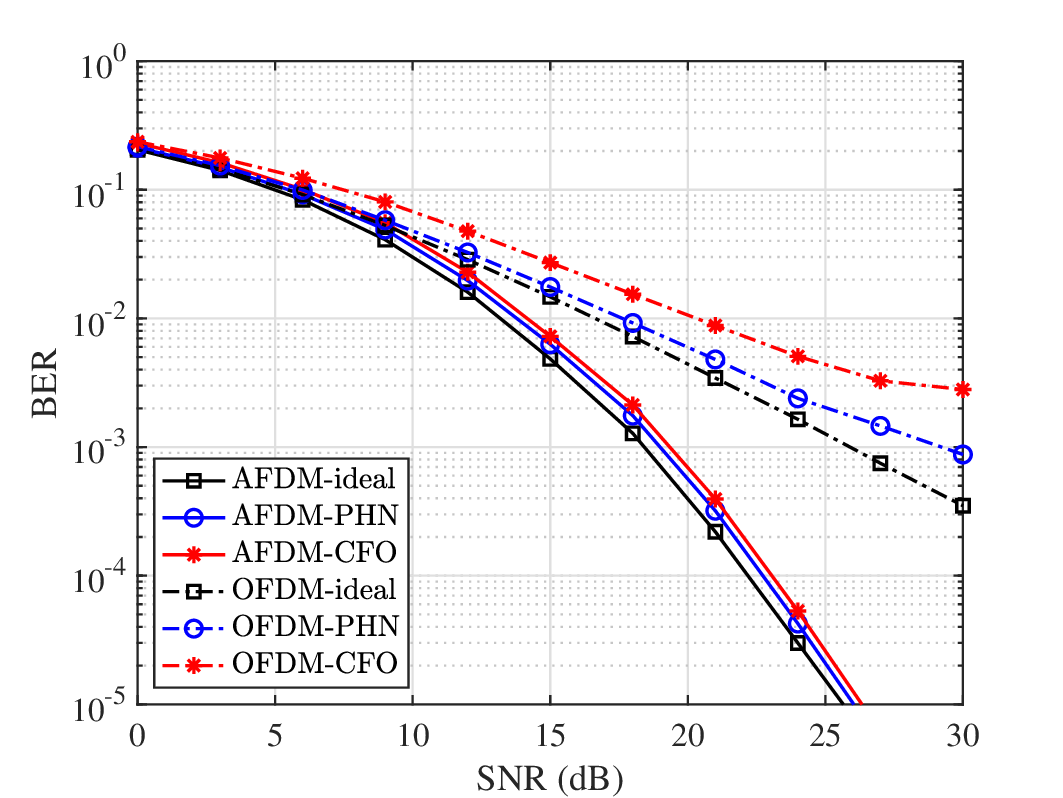}}
\caption{\ac{BER} performance of \ac{OFDM}/\ac{AFDM} under \ac{PHN} and \ac{CFO} impairments ($N\!=\!128$, \ac{QPSK}, $\theta_\text{CFO}\!=\!\frac{1}{10}$, $P\!=\!3$), with \ac{LMMSE} detection.}
\label{fig:CFO}
\vspace{-3ex}
\end{figure}

In contrast, \ac{AFDM} utilizes \ac{DAFT}-domain chirps that sweep across the time-frequency plane. 
This structure provides frequency diversity, as the symbol energy is spread across multiple frequency components. 
The phase rotation induced by \ac{CFO} partially aligns with the intrinsic linear chirp slope, rendering \ac{AFDM} significantly more tolerant to moderate \ac{CFO} levels compared to the static sinusoidal subcarriers of the conventional \ac{OFDM}.

Numerical evaluations of \ac{BER} performance, as shown in Fig. \ref{fig:CFO}, confirm these characteristics. 
Using $N=128$ subcarriers, \ac{QPSK} modulation, and $\theta_\text{CFO}=0.1$, it is observed that \ac{AFDM} performance remains nearly identical to the ideal (impairment-free) case. 
Conversely, at a \ac{BER} of $3\times10^{-3}$, \ac{OFDM} exhibits \ac{SNR} losses of approximately $2$ dB due to \ac{PHN} and $8$ dB due to \ac{CFO}. 
Furthermore, in \ac{PHN} scenarios at a \ac{BER} of $10^{-3}$, \ac{AFDM} achieves a gain of $10.5$ dB over \ac{OFDM}, consistent with improved robustness to oscillator non-idealities.

\vspace{-1ex}
\subsection{Degrees of Freedom via the Chirp Parameter Domain} 

The parameterizable nature of the \ac{DAFT}, in its two chirp parameters $\lambda_1$ and $\lambda_2$, allows \ac{AFDM} to adapt its signal structure to specific system requirements beyond conventional data transmission. 
In addition to tuning the parameters for robustness and diversity, this flexibility can also be exploited to implement functionalities such as \ac{IM}, \ac{PLS}, and \ac{PAPR} reduction.

It is essential to note that these functionalities can be implemented without altering the fundamental \ac{AFDM} framework, preserving its structural compatibility with the \ac{OFDM} legacy; \ac{OFDM} and \ac{OTFS} lack an analogous continuously tunable chirp-rate parameter that enables this without dedicated structural modification.

\vspace{-2.5ex}
\subsubsection{Index Modulation} 

\Ac{IM} schemes originally proposed for \ac{OFDM} \cite{bacsar2013orthogonal,9507331} have also been effectively adapted to the \ac{AFDM} framework \cite{tao24,tao25}. Recent works have further extended the \ac{AFDM}-\ac{IM} framework with new index dimension designs and enhanced spectral efficiency \cite{zhang26dualIM,qian25cim}.
This technique involves partitioning $N$ subcarriers into $K_{\mathrm{IM}}$ subsets, each containing $N_{\mathrm{IM}} = \frac{N}{K_{\mathrm{IM}}}$ subcarriers, and in each subset, $m_{\mathrm{IM}} < N_{\mathrm{IM}}$ subcarriers are selected as active, while the remaining $N_{\mathrm{IM}}-m_{\mathrm{IM}}$ indices are left inactive. 
Consequently, binary information is conveyed not only through the $M$-ary complex symbols but also through the selection of active indices, as also extensively studied in other \ac{IM} literature such as \ac{SM} \cite{mesleh2008spatial,HSRou_TWC2022,10129061} and STSK \cite{10250854} for \ac{MIMO} systems. More recently, by using additional bits to select the active transmit antennas, the GSM-aided AFDM scheme has been studied in \cite{11185315}.

The \ac{IM} spectral efficiency (in bps) is given by
\begin{equation}
R_{\mathrm{IM}} = K_{\mathrm{IM}} \cdot \log_2\left(\left\lfloor \binom{N_{\mathrm{IM}}}{m_{\mathrm{IM}}} \right\rfloor\right) + K_{\mathrm{IM}} m_{\mathrm{IM}} \cdot \log_2(M),
\end{equation}

\noindent offering improved \ac{BER} performance or spectral efficiency compared to standard \ac{AFDM} \cite{10342712}.

From a structural perspective, \ac{AFDM}-\ac{IM} represents an evolution of \ac{OFDM}-\ac{IM} where the modulation is performed in the \ac{DAFT} domain rather than the frequency domain. 
This backward compatibility allows the adaptation of various \ac{OFDM}-\ac{IM} enhancements, such as subcarrier power modulation \cite{Temiz_2025} and multi-mode signaling \cite{anoop25,liu25}. 

Beyond standard subcarrier selection, \ac{AFDM} introduces unique dimensions for \ac{IM} through its parameterizable chirp structure. 
For instance, the chirp parameter $\lambda_2$ can be dynamically selected from a finite set for each subcarrier subset, effectively increasing the aggregate bit rate without requiring additional bandwidth \cite{liu25ieeetwc}. 

Alternatively, information can be conveyed via the permutation of the chirp sequences themselves, a technique known as \textit{chirp-permutation} \ac{IM} (\ac{CP}-\ac{IM}) \cite{Rou_Arxiv25_CPAFDM,10943004}. 
This approach has been shown to provide significant performance gains while maintaining a complete set of active subcarriers, thereby avoiding the spectral efficiency loss typically associated with traditional \ac{IM} schemes. 

\vspace{-3ex}
\subsubsection{Physical Layer Security} 

\Ac{PLS} uses the physical dimensions of the communication system, such as time, frequency, and chirp parameters, to complement higher-layer encryption. 
\ac{AFDM} offers a unique paradigm for \ac{PLS} centered on the tunability of $(\lambda_1, \lambda_2)$. 
Research indicates that \ac{AFDM} can achieve strong physical-layer security by utilizing permutations over chirp sequences \cite{Rou_Arxiv25_CPAFDM,Rou_WCL25}, creating a combinatorial complexity that remains resilient even against quantum-accelerated eavesdropping attempts.
In addition, analytical results also demonstrate that \ac{AFDM} provides higher \ac{PLS} compared to \ac{OTFS} and \ac{OFDM}, as measured by the brute-force complexity required for unauthorized demodulation \cite{savaux26ieeewcl}.
Furthermore, parameter-hopping schemes for $(\lambda_1, \lambda_2)$ can be implemented to dynamically vary the waveform characteristics \cite{Wang_ICC25}.
Combined with its structural robustness against jamming and eavesdropping due to its wideband chirp nature \cite{savaux256Gnetafdm,savaux26ieeewcl}, \ac{AFDM} offers enhanced physical-layer security.

\begin{table*}[b]
\centering
\caption{Consolidated qualitative comparison of \ac{OFDM}, \ac{AFDM}, and \ac{OTFS} across key implementation and performance-related aspects.}
\label{tab:waveform_comparison}
\renewcommand{\arraystretch}{0.85}
\setlength{\tabcolsep}{4pt}
\fontsize{8}{8.4}\selectfont
\arrayrulecolor{gray!40}

\begin{tabular}{p{2.6cm} p{2.7cm} p{4.5cm} p{6.4cm}}
\toprule
\rowcolor{Gray!20} \textbf{Aspect} & \textbf{OFDM} & \textbf{AFDM} & \textbf{OTFS} \\
\midrule
Core transform & $N$-point IFFT/FFT & $N$-point IFFT/FFT $+$ two 1D chirp rotations & ISFFT $+$ pulse-shaped HT (or IDZT $+$ 2D DD processing) \\
\cmidrule{1-4}
Modulator complexity & $5N\log_2 N$ [FLOPS] & $5N\log_2 N + 12N$ [FLOPS] & $5N\log_2 N + 5N\log_2 L + 2N$ [FLOPS] \\
\cmidrule{1-4}
FFT-size flexibility & Arbitrary $N$ & Arbitrary $N$ & Constrained to $N=KL$ \\
\cmidrule{1-4}
Pulse shaping & 1D time-domain & 1D time-domain & 1D time-domain prototype pulse (pulse-shaped HT), or 2D DD-domain convolution (Zak-OTFS) \\
\cmidrule{1-4}
Prefix & Cyclic prefix (CP) & Chirp-periodic prefix (CPP), same length as CP & Per-frame ZP/CP (rectangular pulses); double time--frequency CP under practical pulse shaping \cite{10531762} \\
\cmidrule{1-4}
Resource grid & 1D time-frequency & 1D time-frequency & 2D delay-Doppler grid \\
\cmidrule{1-4}
Channel estimation \& pilot overhead & 1D pilots, low overhead & 1D chirp-aware pilots, comparable overhead & 2D DD pilots with 2D guard regions, higher overhead \\
\cmidrule{1-4}
Doppler robustness \& diversity & Low; loses subcarrier orthogonality & High; attains full delay-Doppler diversity & High; asymptotic diversity order one when uncoded (ML detection); full delay-Doppler diversity via phase-rotation precoding, with near-full-diversity slopes at practical SNR \cite{Surabhi_OTFSdiversity} \\
\cmidrule{1-4}
PHN \& CFO resilience & Baseline & Improved; PHN gain and reduced CFO-induced SNR loss vs.\ OFDM, demonstrated in Fig.~\ref{fig:CFO} & Improved; robust to CFO/Doppler-induced ICI; more PHN-resilient than OFDM \cite{surabhi19otfspn,abushattal23otfsrf} \\
\cmidrule{1-4}
PAPR & High; Baseline & High; Statistically identical to OFDM & Lower than OFDM/AFDM \\
\cmidrule{1-4}
Baseband PHY reuse \& standardization & Native baseline & High; implemented as a wrapper around OFDM baseband blocks & Medium; OFDM FFT core reusable (fully so under rectangular pulses), but the 2D delay-Doppler grid, pilots, and estimation/equalization require reformulation of PHY processing blocks (see Table~\ref{tab:reusability_matrix}) \\
\bottomrule
\end{tabular}
\arrayrulecolor{black}
\end{table*}

\vspace{-3ex}
\subsubsection{PAPR Reduction} 

As \ac{AFDM} is constructed from a summation of subcarriers, it exhibits a \ac{PAPR} profile statistically similar to that of \ac{OFDM}, whereas \ac{OTFS} typically achieves lower values \cite{Rou_DDWaveforms,Rou_Arxiv25_CPAFDM}. 
However, this similarity is a significant advantage for system reusability. 
Legacy \ac{OFDM} \ac{PAPR} reduction techniques such as clipping, $\mu$-law companding \cite{reddy25}, and data precoding \cite{ali25} are directly applicable to \ac{AFDM}, allowing for the reuse of established hardware infrastructure, including PA linearization and digital front-end processing.

Beyond legacy methods, specific \ac{AFDM} \ac{PAPR} reduction strategies have recently emerged. These include pre-chirp selection, where $\lambda_2$ is optimized within a finite set to minimize the peak power for each block \cite{yuan25}, and the use of block-diagonal \acp{DFT} pre-chirp matrices \cite{lu25}.

To complement the per-aspect discussions throughout this section, Table~\ref{tab:waveform_comparison} consolidates the key characteristics of \ac{OFDM}, \ac{AFDM}, and \ac{OTFS} into a single at-a-glance comparison spanning various aspects of the \ac{OFDM}-based standardization roadmap. The comparison is qualitative and intended to summarize the structural and implementation-level relationships established in this article. Detailed performance comparisons among delay-Doppler waveforms are available in dedicated studies \cite{Rou_DDWaveforms,hong2022delay,deng25unifying}.

\vspace{-1ex}
\subsection{Reusability of AFDM: A Layer-by-Layer Summary}
\label{subsec:reusability}

To consolidate the preceding technical analysis, this subsection qualitatively evaluates the reusability of \ac{AFDM} with the established \ac{OFDM} legacy.
This assessment matters for \ac{6G} standardization, where minimizing infrastructure replacement costs is as important as performance gains.
A detailed breakdown of this compatibility is presented in Table \ref{tab:reusability_matrix}.
In the \textbf{RF and Analog} domain, the waveform's statistical similarity to \ac{OFDM} allows for the {direct reuse} of PAs and transceivers without redesign.
In the \textbf{PHY Layer}, while the core \ac{FFT} hardware accelerators are preserved, the transition mainly requires {software updates} for channel estimation logic and phase rotation (chirping). 
Finally, \textbf{MAC} compatibility is achieved through minimal {extensions} to existing structure and control signaling.

\section{Open Research Challenges and Conclusion}
\label{sec:open_challenges}

While the preceding analysis establishes the compatibility and backward reusability of \ac{AFDM} relative to the \ac{OFDM} legacy, several practical aspects remain open, which we briefly summarize below.

\begin{table*}[t]
\vspace{-4ex}
\centering
\caption{Reusability Analysis: AFDM against the OFDM Legacy Infrastructure}
\label{tab:reusability_matrix}
\renewcommand{\arraystretch}{0.6}
\setlength{\tabcolsep}{5pt}
\fontsize{8}{8.6}\selectfont
\arrayrulecolor{gray!40}
\begin{tabular}{p{1.8cm} p{2.8cm} l l p{8cm}}
\toprule
\rowcolor{Gray!20} \textbf{Domain} & \textbf{Module / Function} & \textbf{AFDM} & \textbf{OTFS} & \textbf{Architectural Impact \& Rationale} \\
\midrule

\multirow{2}{*}{\textbf{RF \& Analog}}
& RF Chain \newline \textit{(PA, Antennas, Cabling)} & \cellcolor{ForestGreen!35}\textbf{Direct Reuse} & \cellcolor{ForestGreen!35}\textbf{Direct Reuse} &  AFDM retains the statistical PAPR distribution and spectral mask of OFDM. No new power back-off, PA linearization, or antenna/cable redesign is required. OTFS likewise reuses the RF front-end, and typically exhibits a lower PAPR. \\
\cmidrule{2-5}
& {Mixed Signal} \newline \textit{(ADC / DAC)} & \cellcolor{ForestGreen!35}\textbf{Direct Reuse} & \cellcolor{ForestGreen!35}\textbf{Direct Reuse} & Sampling rates, quantization noise floors, and dynamic range requirements remain identical to OFDM for both AFDM and OTFS. \\
\midrule

\multirow{5}{*}{\textbf{PHY Layer}}
& FFT / IFFT Engines & \cellcolor{ForestGreen!35}\textbf{Direct Reuse} & \cellcolor{ForestGreen!35}\textbf{Direct Reuse} & The core butterfly accelerators are preserved. AFDM is implemented as a wrapper around the legacy OFDM block, and OTFS can also reuse the core FFT but of multiple stages. \\
\cmidrule{2-5}
& Full Pulse-Shaped \newline Waveform Generation & \cellcolor{Green!15}\textbf{Soft Update} & \cellcolor{Red!10}\textbf{Logic Update} & AFDM requires insertion of two element-wise phase rotation vectors digitally (pre/post-chirp) at $\mathcal{O}(N)$ complexity, whereas OTFS requires multi-stage 2D transform and delay-Doppler pulse-shaping processing. \\
\cmidrule{2-5}
& Channel Estimation / \newline Equalization & \cellcolor{Red!10}\textbf{Logic Update} & \cellcolor{Red!10}\textbf{Logic Update} &  Requires new algorithms for the additional Doppler-domain recovery and equalization (in DAFT domain). Standard estimators (LS/MMSE) must be upgraded, or alternate novel estimators can be used. OTFS similarly requires 2D delay-Doppler estimation and equalization. \\
\cmidrule{2-5}
& Pilot Structure & \cellcolor{Goldenrod!15}\textbf{Adaptable} & \cellcolor{Red!10}\textbf{Logic Update} &  Legacy one-dimensional time-domain pilot placement and padding are reusable for AFDM, but optimal performance in high mobility may require different placement within and chirp-aware wrapping. OTFS requires dedicated 2D delay-Doppler pilots with 2D guard regions. \\
\midrule

\multirow{3}{*}{\textbf{MAC \& Control}}
& Resource Grid & \cellcolor{Green!15}\textbf{Compatible} & \cellcolor{Red!10}\textbf{Logic Update} &  Resources map to 1D streams. The fundamental time-frequency resource element (RE) grid structure is preserved for AFDM, whereas OTFS operates on a 2D delay-Doppler grid constrained to $N=KL$. \\
\cmidrule{2-5}
& Control Signaling (DCI) & \cellcolor{Goldenrod!15}\textbf{Extension} & \cellcolor{Goldenrod!15}\textbf{Extension} & Standard downlink control information (DCI) formats can apply, but new chirp parameters (chirp indices) may require additional signaling bits for AFDM, and OTFS requires grid configuration information. \\
\bottomrule
\end{tabular}
\arrayrulecolor{black}
\end{table*}

\textbf{Synchronization:} The chirped subcarrier structure alters \ac{AFDM}'s sensitivity to residual timing and frequency offsets relative to \ac{OFDM}.
While Section~\ref{sec:analysis_AFDMOFDM} characterizes steady-state \ac{PHN}/\ac{CFO} impairment, low-overhead acquisition and tracking loops that are robust to joint delay-Doppler dispersion remain immature, and whether legacy \ac{OFDM} synchronization circuitry can be reused with only chirp-aware post-processing or dedicated estimators are required remains an open question.

\textbf{Channel estimation overhead and FDFD-aware frameworks:} This article assumes perfect channel state information to isolate the effective channel's structural properties from estimation artifacts. Under the generalized \ac{FDFD} model, however, the effective channel is no longer strictly sparse, and pulse-shaping-induced inter-sample coupling increases the support an estimator must resolve. Therefore, quantifying the resulting pilot overhead and designing dedicated \ac{FDFD}-aware estimators are substantial directions, as reflected in growing related interest \cite{li26sbl}.

\textbf{Standardization pathway:} Table~\ref{tab:reusability_matrix} suggests that \ac{AFDM} can be integrated with limited modification to the \ac{OFDM} physical layer, but control-plane implications remain unspecified. For example, exact signaling of chirp parameters, coexistence of \ac{AFDM}/\ac{OFDM} numerologies on a shared resource grid, and alignment with ongoing 3GPP \ac{6G} waveform study items \cite{R1-2508043,RP-251881} are concrete open items.


\vspace{-3ex}
\subsubsection*{Conclusion}
Overall, we highlighted that \ac{AFDM} is a well-positioned waveform candidate for \ac{6G+} wireless systems, bridging extreme-mobility performance requirements and the practical demand for hardware reusability.
Built upon the established \ac{OFDM} legacy, \ac{AFDM} achieves high-fidelity communications in doubly dispersive channels through a modular, evolutionary approach rather than a radical system redesign.
The presented analysis confirms extensive reusability in the \ac{RF} front-end and the receiver's \ac{FFT}/\ac{IFFT} core, while channel estimation and equalization require logic-level updates, as summarized in Table~\ref{tab:reusability_matrix}. The tunable chirp-parameter domain further offers degrees of freedom for secondary functionalities, including \ac{IM}, \ac{PLS}, and sensing, while maintaining compatibility with legacy resources.
\ac{AFDM} thus offers a practically viable waveform candidate for \ac{6G} standardization and beyond, preserving prior engineering investments while meeting the demands of future wireless systems.

\section{Use of AI Tools}

Generative AI tools, including ChatGPT, Gemini, Claude, and Copilot, were utilized in this work for editorial refinement, code optimization, and literature scoping. 
These tools served only as auxiliary instruments to enhance clarity and presentation, and they did not generate new scientific concepts or results. 
The authors have carefully verified all AI-assisted content and assume full responsibility for the accuracy, integrity, and originality of the manuscript.

\vfil

%
%
%

\begin{IEEEbiography}[{\includegraphics[width=1in,height=1.25in,clip,keepaspectratio]{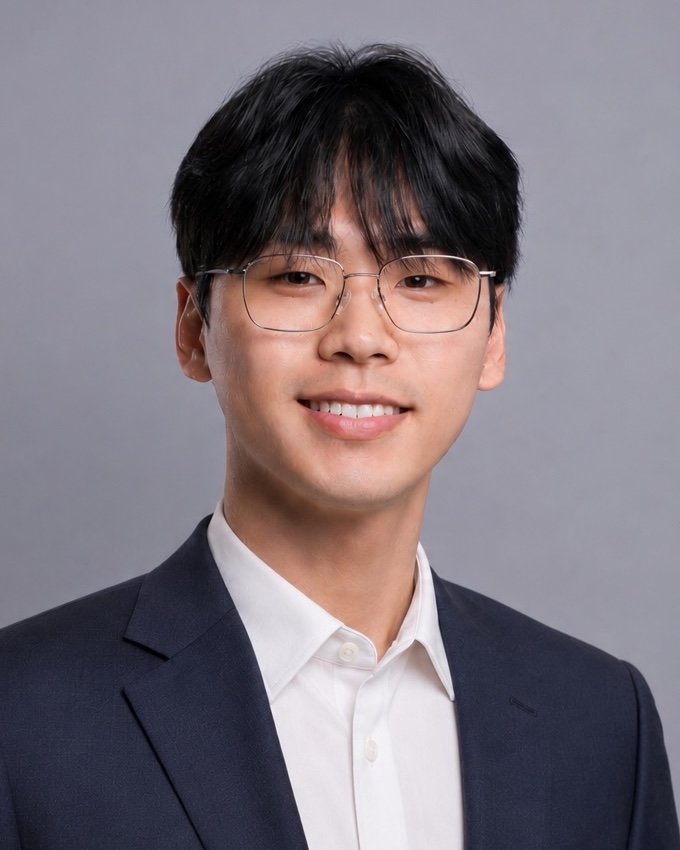}}]{Hyeon Seok Rou}\hspace*{0.33em}(Member, IEEE)
received the B.Sc. degree in electrical and computer engineering from Constructor University Bremen (formerly Jacobs University Bremen), Germany, in 2021, and the Ph.D. degree in electrical engineering from the same institution in 2024.
Since 2025, he has been a Postdoctoral Research Fellow and Lecturer with the School of Computer Science and Engineering, Constructor University Bremen, Germany.
In November 2026, he will join the School of Electrical Engineering, Korea Advanced Institute of Science and Technology (KAIST), Daejeon, South Korea, as a Research Assistant Professor.
His research interests include next-generation (B5G and 6G) wireless systems, affine frequency division multiplexing (AFDM), integrated sensing and communications (ISAC), high-mobility communications, multi-dimensional sparse modulation, over-the-air computing (AirComp), vehicle-to-everything (V2X) communications, and quantum-accelerated optimization.
In 2026, he was selected as a Brain Pool Fellow by the National Research Foundation of Korea (NRF). He has served as a Chapter Representative of the Korean Scientists and Engineers Association in the Federal Republic of Germany (VeKNI e.V.) since 2021, and was the German delegate to the Young Professionals Forum (YPF) of the 3rd World Congress of Korean Scientists and Engineers, South Korea, in 2025. In 2023, he was a Visiting Researcher with the Intelligent Communications Systems Laboratory (ICL), KAIST, and in 2022 he received the Korea Institute of Science and Technology Europe Research Scholarship Award. He also contributes to the IEEE Standards Association P3383 Working Group on Performance Metrics for Integrated Sensing and Communication.
\end{IEEEbiography}

\begin{IEEEbiography}[{\includegraphics[width=1in,height=1.25in,clip,keepaspectratio]{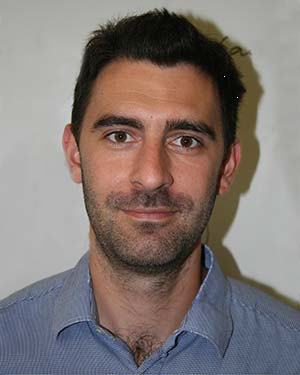}}]{Vincent Savaux}\hspace*{0.33em}(Senior Member, IEEE)
received the engineering degree from the High School of Engineering, ECAM Rennes, the M.Sc. degree from the University of Rennes 1, France, in 2010, and the Ph.D. degree in telecommunications from {\'E}cole Sup{\'e}rieure d'{\'E}lectricit{\'e}-Sup{\'e}lec in 2013.
From 2014 to 2015, he was a postdoctoral researcher with CentraleSup{\'e}lec, Campus de Rennes, France.
Since 2016, he has been a research engineer with the Institute of Research and Technology b-com, Rennes, France.
His research activities mainly focus on signal processing for wireless systems, including IoT technologies, MIMO systems, physical layer security, and new waveforms for next generations of communications systems.
He has published more than 70 papers in high-level international journals and conferences, and serves as a TPC member for several conferences.
He received the Exemplary Reviewer 2014 Appreciation from the IEEE Wireless Communications Letters and the IEEE Communications Letters, and the Best Paper Award from IEEE CSCN 2023.
\end{IEEEbiography}

\begin{IEEEbiography}[{\includegraphics[width=1in,height=1.25in,clip,keepaspectratio]{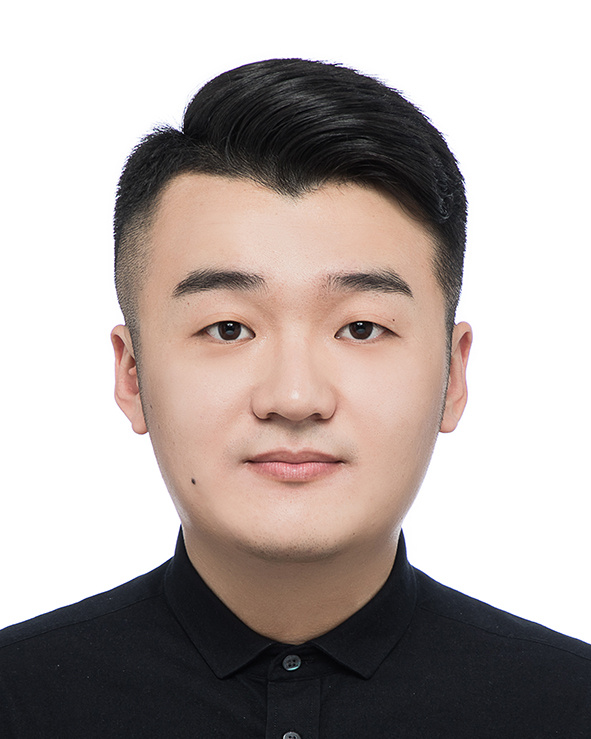}}]{Zeping Sui}\hspace*{0.33em}(Member, IEEE)
received the Ph.D. degree in signal and information processing from the University of Chinese Academy of Sciences, Beijing, China, in 2023.
From March 2022 to February 2023, he was a Research Assistant with the Institute for Infocomm Research (I2R), Agency for Science, Technology and Research (A*STAR), Singapore.
From August 2023 to August 2024, he was a Research Fellow with the Centre for Wireless Innovation (CWI), Queen's University Belfast, Belfast, U.K.
He is currently a Senior Research Officer with the School of Computer Science and Electronic Engineering (CSEE), University of Essex, Colchester, U.K.
He currently serves as an Editor for the IEEE Transactions on Communications, an Associate Editor for the IEEE Wireless Communications Letters, and a Senior Reviewer for the IEEE Open Journal of the Communications Society.
He received the Exemplary Reviewer Award from the IEEE Wireless Communications Letters in 2025.
He has served as an Organizer/Chair of several workshops and special sessions at flagship IEEE conferences, including IEEE GLOBECOM, IEEE PIMRC, IEEE WCNC, IEEE/CIC ICCC, IEEE Ucom, and IEEE SPAWC, and as a Tutorial Speaker for IEEE WCNC 2026.
His research interests span signal processing for wireless communications, waveform design, hardware-constrained communication systems, statistical signal processing, and beyond massive MIMO.
\end{IEEEbiography}

\begin{IEEEbiography}[{\includegraphics[width=1in,height=1.25in,clip,keepaspectratio]{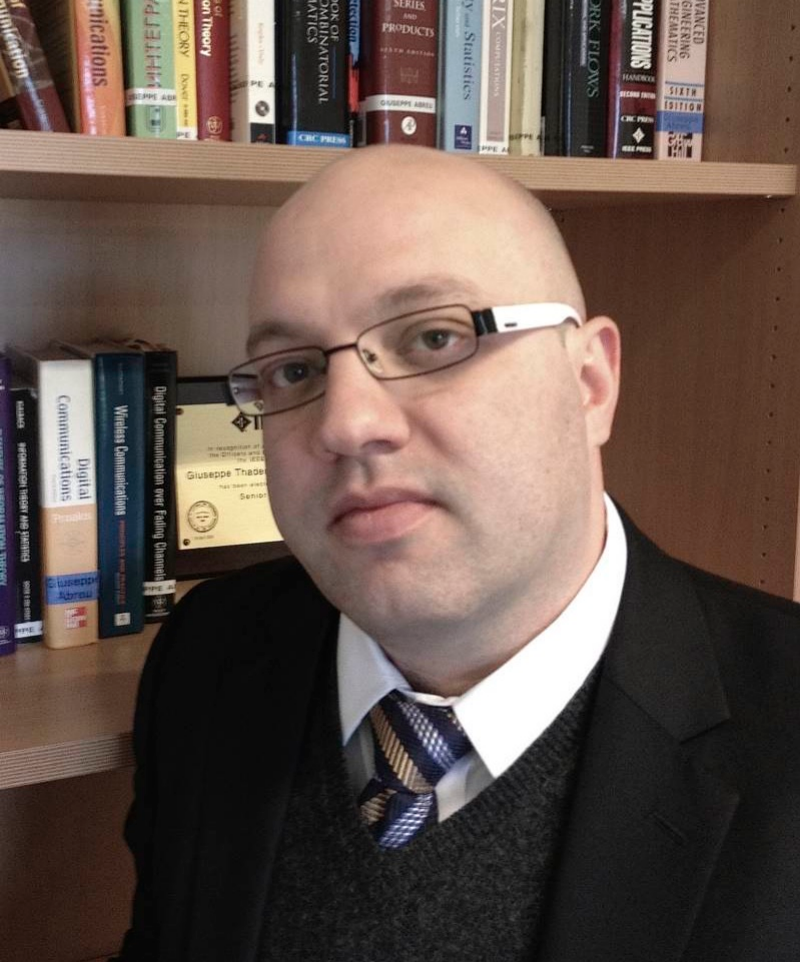}}]{Giuseppe Thadeu Freitas de Abreu}\hspace*{0.33em}(Senior Member, IEEE)
received the B.Eng. degree in electrical engineering and the specialization Lato Sensu
degree in telecommunications engineering from the
Universidade Federal da Bahia (UFBa), Salvador,
Bahia, Brazil, in 1996 and 1997, respectively, and
the M.Eng. and D.Eng. degrees in physics, electrical,
and computer engineering from Yokohama National
University, Japan, in March 2001 and March 2004,
respectively.
He was a postdoctoral fellow and later
an adjunct professor (docent) in statistical signal
processing and communications theory at the Department of Electrical and
Information Engineering, University of Oulu, Finland from 2004 to 2006
and from 2006 to 2011, respectively.
Since 2011, he has been a professor
of electrical engineering at Constructor University Bremen (formerly Jacobs University, Bremen, Germany). From April 2015 to August 2018, he simultaneously held a full tenured professorship also
at the Department of Computer and Electrical Engineering, Ritsumeikan
University, Japan.
He has been a visiting scholar at Rice University (2008), Stanford University (2009), Keio University (2010), the University of Bologna (2012 and 2014) and at the KTH Royal Institute of Technology (2025). He has also been a regular visiting professor at the University of Electro-Communications, Japan, since 2013.
He received the Uenohara Award at Tokyo University in 2000 for his master’s thesis work and was the co-recipient of awards from various international IEEE conferences, including the Asilomar in 2009, 2012 and 2014, WPNC in 2012 and 2015, ISWCS in 2016, WPMC in 2023 and ICNC in 2025.
He was also a co-recipient of the Best Journal Paper Award by the Japanese Chapter of the IEEE Signal Processing Society in 2023 and was awarded various prestigious Fellowships from the Heiwa Nakajima Foundation, the Japanese Society for the Promotion of Science (JSPS), and twice from the National Institute for Communications Technologies (NICT), respectively in 2010, 2013, 2015 and 2018.
He has been a Visiting Scholar at Stanford University, Rice University, York University, ETH Z{\"u}rich, over the years, and was a Scholar in Residence at the KTH Royal Institute of Technology, Sweden, from Sep. 2025 to January 2026.
He served as an editor for various IEEE journals including the IEEE Transactions on Wireless Communications, the IEEE Transactions on Communications, the IEEE Communications Letters, the IEEE Signal Processing Letters, and the IEEE Open Journal of the Communications Society.
He was also the executive editor for the IEEE Transactions on Wireless Communications from 2017 to 2021, and since 2026 is the Editor-at-Large for Wireless Communications of the IEEE Open Journal of the Communications Society.
His research interests span a wide range of topics on wireless communications and signal processing, including communications theory, estimation theory, optimization theory, statistical modeling, metasurfaces and electromagnetic structures, waveform design, integrated sensing and communications, over-the-air computing, wireless localization, cognitive radio, wireless security, MIMO systems, ultrawideband and millimeter wave communications, full-duplex and cognitive radio, compressive sensing, energy harvesting networks, random networks, connected vehicles networks, and many more.
\end{IEEEbiography}

\begin{IEEEbiography}[{\includegraphics[width=1in,height=1.25in,clip,keepaspectratio]{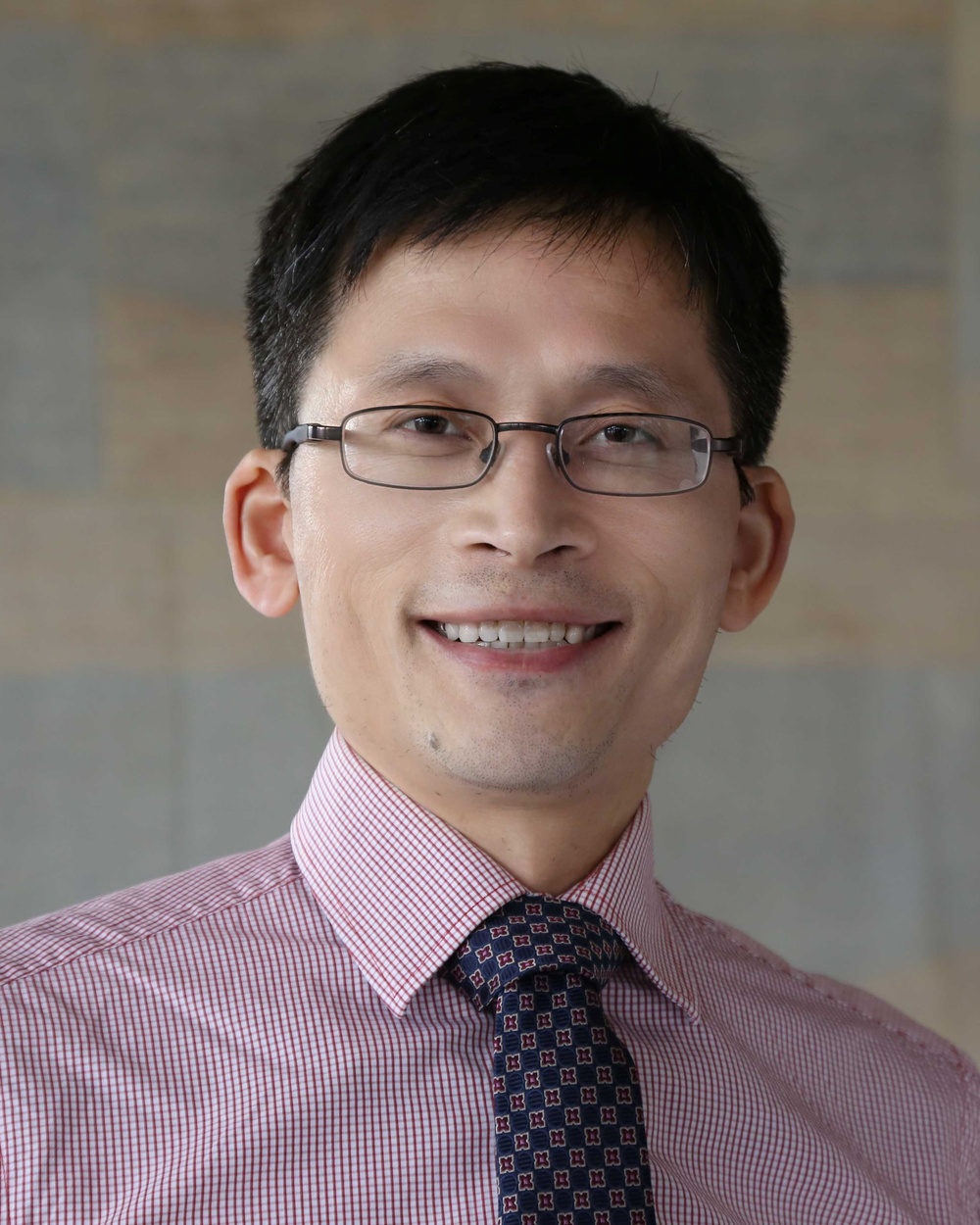}}]{Zilong Liu}\hspace*{0.33em}(Senior Member, IEEE)
is an Associate Professor, the 6G Laboratory Manager, and the 2024 Outstanding Mid-Career Researcher in the Faculty of Science and Health, University of Essex, Colchester, U.K.
He is/was a Distinguished Lecturer of the IEEE Vehicular Technology Society, a recipient of the prestigious EPSRC New Investigator Award (NIA) in the U.K., and a Consultant to the National Institute of Information and Communication Technology (NICT), a Japanese Government organisation, on 6G V2X assisted autonomous driving.
His research generally lies in the interplay of coding, signal processing, and communications, with a major objective of bridging theory and practice.
At Essex, he has led the successful development of real-time over-the-air USRP testbeds of OFDM-SCMA and AFDM, which were demonstrated at MWC'2026 and ICC'2026, respectively.
He has widely published in quality journals such as the IEEE Transactions on Information Theory, the IEEE Transactions on Signal Processing, JSAC, JS-TSP, and the Proceedings of the IEEE.
So far, he has made a series of original contributions to waveforms, sequences, and multiple access.
Prior to his current post, he was with Huazhong University of Science and Technology (Bachelor, 2000--2004), Tsinghua University (M.Sc., 2004--2007), Nanyang Technological University Singapore (Ph.D. and postdoc, 2008--2017), University of Melbourne (Visiting Ph.D., 2012), the Hong Kong University of Science and Technology (Visiting Ph.D., 2012), and the University of Surrey (Senior Research Fellow, 2018--2019).
He is/was the Hosting General Chair of IWSDA'2022 (\url{iwsda2022.github.io}) and SETA'2024 (\url{seta-2024.github.io}), an active contributor to flagship IEEE conferences (e.g., GLOBECOM'2026, PIMRC'2026, ISIT'2026, ICC'2026, ICCC'2026, WCNC'2026, VTC-Fall'2025, IWSDA'2025) in various roles (e.g., track co-chair, tutorial/keynote speaker, workshop organizer, TPC member), and an Associate Editor for several leading IEEE journals including TWC, TVT, WCL, and OJCOMS.
Details of his research can be found at \url{https://sites.google.com/site/zilongliu2357}.
\end{IEEEbiography}


\begin{thebibliography}{100}
\providecommand{\url}[1]{#1}
\csname url@samestyle\endcsname
\providecommand{\newblock}{\relax}
\providecommand{\bibinfo}[2]{#2}
\providecommand{\BIBentrySTDinterwordspacing}{\spaceskip=0pt\relax}
\providecommand{\BIBentryALTinterwordstretchfactor}{4}
\providecommand{\BIBentryALTinterwordspacing}{\spaceskip=\fontdimen2\font plus
\BIBentryALTinterwordstretchfactor\fontdimen3\font minus
\fontdimen4\font\relax}
\providecommand{\BIBforeignlanguage}[2]{{%
\expandafter\ifx\csname l@#1\endcsname\relax
\typeout{** WARNING: IEEEtran.bst: No hyphenation pattern has been}%
\typeout{** loaded for the language `#1'. Using the pattern for}%
\typeout{** the default language instead.}%
\else
\language=\csname l@#1\endcsname
\fi
#2}}
\providecommand{\BIBdecl}{\relax}
\BIBdecl

\bibitem{9349624}
W.~Jiang \emph{et~al.}, ``The road towards {6G}: A
comprehensive survey,'' \emph{IEEE Open J. Commun. Soc.}, vol.~2, pp. 334--366, 2021.

\bibitem{10054381}
C. X. Wang \emph{et~al.}, ``On the road to {6G}: Visions, requirements, key
technologies, and testbeds,'' \emph{IEEE Commun. Surveys Tuts.}, vol.~25, no.~2, pp. 905--974, 2023.

\bibitem{R1-2508043}
{3GPP TSG RAN}, ``{R1-2508043: Feature Lead summary \#3 on {6G} waveform (Source:
Nokia), RAN1 Meeting \#122bis},'' Oct. 2025.

\bibitem{22.870}
{3GPP, Group Services and System Aspects}, ``{TR 22.870: Study on {6G} Use Cases
and Service Requirements},'' Jun. 2025, {Rel. 20, v0.3.1}.

\bibitem{RP-251881}
{3GPP TSG RAN}, ``{RP-251881: New {SID}: Study on {6G} Radio},'' Jun. 2025, {RAN
Meeting \#108}.

\bibitem{dahlman20205g}
E.~Dahlman, S.~Parkvall, and J.~Skold, \emph{{5G} {NR}: The next generation
wireless access technology}.\hskip 1em plus 0.5em minus 0.4em\relax Academic
Press, 2020.

\bibitem{3gpp.36.211}
{3GPP}, ``Evolved universal terrestrial radio access ({E-UTRA}); physical
channels and modulation,'' 3rd Generation Partnership Project (3GPP),
Technical Specification (TS) TS 36.211, 2008.

\bibitem{3gpp.38.211}
{3GPP}, ``{NR}; physical channels and modulation,'' 3rd Generation Partnership
Project (3GPP), Technical Specification (TS) TS 38.211, 2018, release 15. 

\bibitem{4287203}
M.~Morelli, C.-C.~J. Kuo, and M.-O. Pun, ``Synchronization techniques for
orthogonal frequency division multiple access ({OFDMA}): A tutorial review,''
\emph{Proc. IEEE}, vol.~95, no.~7, pp. 1394--1427, Jul. 2007.

\bibitem{han2005overview}
S.~H. Han and J.~H. Lee, ``An overview of peak-to-average power ratio ({PAPR}) reduction
techniques for multicarrier transmission,'' \emph{IEEE Wireless Commun.}, vol.~12, no.~2, pp. 56--65, Apr. 2005.

\bibitem{ozdemir2007channel}
M.~Ozdemir and H.~Arslan, ``Channel estimation for wireless {OFDM} systems,''
\emph{IEEE Commun. Surveys Tuts.}, vol.~9, no.~2, pp. 18--48, 2007.

\bibitem{6814271}
Y.~Liu  \emph{et~al.}, ``Channel estimation for
{OFDM},'' \emph{IEEE Commun. Surveys Tuts.}, vol.~16, no.~4,
pp. 1891--1908, 2014.

\bibitem{Saad_6G}
W.~Saad, M.~Bennis, and M.~Chen, ``{A Vision of {6G} Wireless Systems:
Applications, Trends, Technologies, and Open Research Problems},'' \emph{IEEE
Network}, vol.~34, no.~3, pp. 134--142, May 2020.

\bibitem{Bliss_DDchannel}
D.~Bliss and S.~Govindasamy, \emph{Adaptive Wireless Communications: {MIMO}
Channels and Networks}.\hskip 1em plus 0.5em minus 0.4em\relax Cambridge
University Press, 2013.

\bibitem{Koivunen_waveform}
V.~Koivunen  \emph{et~al.},
``Multicarrier {ISAC}: Advances in waveform design, signal processing, and
learning under nonidealities,'' \emph{IEEE Signal Process. Mag.},
vol.~41, no.~5, pp. 17--30, Sep. 2024.

\bibitem{araniti2021toward}
G.~Araniti \emph{et~al.}, ``Toward {6G} non-terrestrial
networks,'' \emph{IEEE Network}, vol.~36, no.~1, pp. 113--120, Jan. 2022.

\bibitem{noor20226g}
M.~Noor-A-Rahim \emph{et~al.}, ``{6G} for vehicle-to-everything ({V2X})
communications: Enabling technologies, challenges, and opportunities,''
\emph{Proc. IEEE}, vol.~110, no.~6, pp. 712--734, Jun. 2022.

\bibitem{wang2006performance}
T.~Wang \emph{et~al.}, ``Performance degradation
of {OFDM} systems due to doppler spreading,'' \emph{IEEE Trans. Wireless
Commun.}, vol.~5, no.~6, pp. 1422--1432, Jun. 2006.

\bibitem{Liu_ISAC}
F.~Liu \emph{et~al.}, 
``{Integrated Sensing and Communications: Toward Dual-Functional Wireless
Networks for {6G} and Beyond},'' \emph{IEEE J. Sel. Areas Commun.}, vol.~40,
no.~6, pp. 1728--1767, Jun. 2022.

\bibitem{Rou_DDWaveforms}
H.~S. Rou \emph{et~al.}, ``{From Orthogonal Time-Frequency Space to Affine
Frequency-Division Multiplexing: A comparative study of next-generation
waveforms for integrated sensing and communications in doubly dispersive
channels},'' \emph{IEEE Signal Process. Mag.}, vol.~41, no.~5, pp.
71--86, Sep. 2024.

\bibitem{Bemani_AFDM}
A.~Bemani, N.~Ksairi, and M.~Kountouris, ``{Affine Frequency Division
Multiplexing for Next Generation Wireless Communications},'' \emph{IEEE
Trans. Wireless Commun.}, vol.~22, no.~11, pp. 8214--8229, Nov. 2023.

\bibitem{AFDM_6G_Rou}
H.~S. Rou \emph{et~al.}, ``Affine frequency division multiplexing
({AFDM}) for {6G}: Properties, features, and challenges,'' \emph{IEEE Commun. Stand. Mag.}, vol.~10, no.~2, pp. 216--225, Jun. 2026.

\bibitem{11003079}
Q.~Li \emph{et~al.}, ``Affine frequency
division multiplexing for {6G} networks: Fundamentals, opportunities, and
challenges,'' \emph{IEEE Network}, vol.~40, no.~1, pp. 88--97, Jan. 2026.

\bibitem{tao25}
Y.~Tao  \emph{et~al.},  ``{Affine Frequency
Division Multiplexing With Index Modulation: Full Diversity Condition,
Performance Analysis, and Low-Complexity Detection},'' \emph{IEEE J. Sel.
Areas Commun.}, vol.~43, no.~4, pp. 1041--1055, Apr. 2025.

\bibitem{liu25ieeetwc}
G.~Liu \emph{et~al.},  ``Pre-chirp-domain index modulation for full-diversity affine
frequency division multiplexing toward {6G},'' \emph{{IEEE Trans. Wireless
Commun.}}, vol.~24, no.~9, pp. 7331--7345, Sep. 2025.

\bibitem{10943004}
H.~S. Rou \emph{et~al.}, 
``{AFDM} chirp-permutation-index modulation with quantum-accelerated codebook
design,'' in \emph{Proc. Asilomar Conf. Signals Syst. Comput.}, Oct. 2024, pp. 817--821.

\bibitem{zhang26dualIM}
Y.~Zhang \emph{et~al.}, ``{Dual Index Modulation
for Affine Frequency Division Multiplexing Communications},'' \emph{IEEE
Wireless Commun. Lett.}, vol.~15, pp. 1469--1473, 2026.

\bibitem{qian25cim}
M.~Qian \emph{et~al.}, ``{Generalized Code Index
Modulation Aided {AFDM} for Spread Spectrum Systems},'' \emph{IEEE Wireless
Commun. Lett.}, vol.~14, no.~10, pp. 3229--3233, Oct. 2025.

\bibitem{savaux26ieeewcl}
V.~Savaux \emph{et~al.}, ``{On the
Robustness of {AFDM} and {OTFS} Against Passive Eavesdroppers},'' \emph{IEEE
Wireless Commun. Lett.}, vol.~15, pp. 1365--1369, 2026.

\bibitem{Wang_ICC25}
P.~Wang \emph{et~al.},  ``{A Secure Affine Frequency
Division Multiplexing for Wireless Communication Systems},'' in \emph{Proc.
IEEE ICC}, 2025, pp. 2701--2706.

\bibitem{Rou_WCL25}
H.~S. Rou and G.~T.~F. de~Abreu, ``Chirp-permuted {AFDM} for quantum-resilient
physical-layer secure communications,'' \emph{IEEE Wireless Commun. Lett.},
vol.~14, no.~8, pp. 2376--2380, Aug. 2025.

\bibitem{11322797}
M.~Ahmad  \emph{et~al.}, ``Radar-centric {AFDM} waveform with
chirp-domain index modulation for {ISAC},'' \emph{IEEE Open J. Commun. Soc.}, vol.~7, pp. 844--857, 2026.

\bibitem{sui2025multi}
Z.~Sui  \emph{et~al.}, ``Multi-functional chirp signalling for
next-generation multi-carrier wireless networks: Communications, sensing and
{ISAC} perspectives,'' \emph{arXiv preprint arXiv:2508.06022}, Aug. 2025.

\bibitem{ranasinghe2025affine}
K.~R.~R. Ranasinghe \emph{et~al.},  ``Affine filter bank modulation
({AFBM}): A novel {6G} {ISAC} waveform with low {PAPR} and {OOBE},'' \emph{IEEE Trans. Wireless Commun.}, vol.~25, pp. 12\,754--12\,769, 2026.

\bibitem{yi2025non}
Q.~Yi \emph{et~al.}, ``Non-orthogonal affine frequency
division multiplexing for spectrally efficient high-mobility
communications,'' \emph{IEEE Trans. Wireless Commun.}, vol.~25, pp. 15\,758--15\,774, 2026.

\bibitem{senger2025affine}
H.~L. Senger \emph{et~al.},  ``Affine filter bank
modulation: A new waveform for high mobility communications,'' \emph{arXiv
preprint arXiv:2505.03589}, May 2025.

\bibitem{Wei_OTFS}
Z.~Wei \emph{et~al.}, 
``{Orthogonal Time-Frequency Space Modulation: A Promising Next-Generation
Waveform},'' \emph{IEEE Wireless Commun.}, vol.~28, no.~4, pp.
136--144, Aug. 2021.

\bibitem{10183832}
Z.~Sui \emph{et~al.}, 
``Performance analysis and approximate message passing detection of
orthogonal time sequency multiplexing modulation,'' \emph{IEEE Trans.
Wireless Commun.}, vol.~23, no.~3, pp. 1913--1928, Mar. 2024.

\bibitem{9829188}
H.~Lin and J.~Yuan, ``Orthogonal delay-doppler division multiplexing
modulation,'' \emph{IEEE Trans. Wireless Commun.}, vol.~21, no.~12, pp.
11\,024--11\,037, Dec. 2022.

\bibitem{lampel2022otfs}
F.~Lampel, A.~Alvarado, and F.~M. Willems, ``On {OTFS} using the discrete zak
transform,'' in \emph{Proc. IEEE ICC Workshops}, May 2022, pp. 729--734.

\bibitem{hong2022delay}
Y.~Hong, T.~Thaj, and E.~Viterbo, \emph{Delay-Doppler Communications:
Principles and Applications}.\hskip 1em plus 0.5em minus 0.4em\relax Academic
Press, 2022.

\bibitem{deng25unifying}
Q.~Deng, Y.~Ge, and Z.~Ding, ``{A Unifying View of {OTFS} and Its Many Variants},'' \emph{IEEE Commun. Surveys Tuts.}, vol.~27, no.~6, pp. 3561--3586, 2025.

\bibitem{li2021performance}
S.~Li \emph{et~al.}, ``Performance
analysis of coded {OTFS} systems over high-mobility channels,'' \emph{IEEE
Trans. Wireless Commun.}, vol.~20, no.~9, pp. 6033--6048, Sep. 2021.

\bibitem{Surabhi_OTFSdiversity}
G.~D. Surabhi, R.~M. Augustine, and A.~Chockalingam, ``On the diversity of
uncoded {OTFS} modulation in doubly-dispersive channels,'' \emph{IEEE Trans.
Wireless Commun.}, vol.~18, no.~6, pp. 3049--3063, Jun. 2019.

\bibitem{bemani21_iswcs}
A.~Bemani \emph{et~al.}, ``{Affine Frequency
Division Multiplexing for Next-Generation Wireless Networks},'' in \emph{IEEE ISWCS}, 2021, pp. 1--6.

\bibitem{bemani2024integrated}
A.~Bemani, N.~Ksairi, and M.~Kountouris, ``Integrated sensing and
communications with affine frequency division multiplexing ({AFDM}),'' \emph{IEEE
Wireless Commun. Lett.}, vol.~13, no.~5, pp. 1255--1259, May 2024.

\bibitem{luo2025target}
Y.~Luo \emph{et~al.}, ``Target sensing with off-grid sparse
bayesian learning for {AFDM}-{ISAC} system,'' in \emph{Proc. IEEE ICC Workshops}, 2025, pp. 881--886.

\bibitem{ni2025integrated}
Y.~Ni \emph{et~al.},  ``An integrated sensing and
communications system based on affine frequency division multiplexing,''
\emph{IEEE Trans. Wireless Commun.}, vol.~24, no.~5, pp. 3763--3779, May 2025.

\bibitem{luo2025novel}
Y.~Luo \emph{et~al.},  ``A novel
angle-delay-doppler estimation scheme for {AFDM-ISAC} system in mixed
near-field and far-field scenarios,'' \emph{IEEE Internet Things J.}, vol.~12, no.~13, pp. 22\,669--22\,682, Jul. 2025.

\bibitem{rou2025normalized}
H.~S. Rou and G.~T.~F. de~Abreu, ``Normalized ambiguity function
characteristics of {OFDM, OTFS, AFDM, and CP-AFDM for ISAC},'' in \emph{Proc. IEEE ICC}, 2026.

\bibitem{ni2025ambiguity}
Y.~Ni \emph{et~al.}, ``Ambiguity function analysis of
{AFDM} under pulse-shaped random {ISAC} signaling,'' \emph{IEEE Trans. Wireless Commun.}, vol.~25, pp. 13\,619--13\,635, 2026.

\bibitem{zhu2024afdm}
J.~Zhu  \emph{et~al.},  ``{AFDM}-based bistatic
integrated sensing and communication in static scatterer environments,''
\emph{IEEE Wireless Commun. Lett.}, vol.~13, no.~8, pp. 2245--2249, Aug. 2024.

\bibitem{zhang2025afdm}
F.~Zhang \emph{et~al.}, ``{AFDM}-enabled integrated sensing and communication:
Theoretical framework and pilot design,'' \emph{IEEE J. Sel. Areas Commun.}, vol.~44, pp. 310--324, 2026.

\bibitem{ramadan2026performance}
K.~Ramadan, A.~A. Alharbi, and E.~S. Hassan, ``Performance evaluation of {AFDM}
for integrated sensing and communications in doubly dispersive channels,''
\emph{Def. Technol.}, vol.~61, pp. 110--125, Jul. 2026.

\bibitem{Ouyang_OCDM}
X.~Ouyang and J.~Zhao, ``{Orthogonal Chirp Division Multiplexing},'' \emph{IEEE
Trans. Commun.}, vol.~64, no.~9, pp. 3946--3957, Sep. 2016.

\bibitem{ranasinghe2025doubly}
K.~R.~R. Ranasinghe  \emph{et~al.},  ``Doubly-dispersive continuous {MIMO} systems: Channel
modeling and beamforming design,'' \emph{IEEE Trans. Wireless Commun.}, vol.~25, pp. 15\,441--15\,458, 2026.

\bibitem{11157883}
K.~R.~R. Ranasinghe \emph{et~al.},  ``Doubly-dispersive {MIMO} channels with stacked intelligent
metasurfaces: Modeling, parametrization, and receiver design,'' \emph{IEEE Trans. Wireless Commun.}, vol.~25, pp. 3801--3817, 2026.

\bibitem{ranasinghe2025flexible}
K.~R.~R. Ranasinghe \emph{et~al.},  ``Flexible intelligent metasurfaces in high-mobility
{MIMO} integrated sensing and communications,'' \emph{IEEE Trans. Wireless Commun.}, vol.~25, pp. 13\,319--13\,335, 2026.

\bibitem{chi26mamp}
Y.~Chi \emph{et~al.}, ``{Achievable Rate and
Coding Principle for {MIMO} Multicarrier Systems With Cross-Domain {MAMP}
Receiver Over Doubly Selective Channels},'' \emph{IEEE Trans. Wireless Commun.},
vol.~25, pp. 10354--10370, 2026.

\bibitem{Baeuml1996SLM}
R.~W. B{\"a}uml, R.~F.~H. Fischer, and J.~B. Huber, ``Reducing the
peak-to-average power ratio of multicarrier modulation by selected mapping,''
\emph{Electron. Lett.}, vol.~32, no.~22, pp. 2056--2057, Oct. 1996.

\bibitem{3gpp36211}
\emph{3GPP TS 36.211 V17.2.0: Evolved Universal Terrestrial Radio Access
({E-UTRA}); Physical Channels and Modulation}, 3rd Generation Partnership
Project (3GPP) Std., 2022, release 17.

\bibitem{IEEE80211}
\emph{IEEE Std 802.11-2012: Wireless LAN Medium Access Control ({MAC}) and
Physical Layer ({PHY}) Specifications}, IEEE Std., 2012.

\bibitem{ETSI300744}
\emph{ETSI EN 300 744 V1.6.1: Digital Video Broadcasting ({DVB}); Framing
Structure, Channel Coding and Modulation for Digital Terrestrial Television},
European Telecommunications Standards Institute (ETSI) Std., 2009, clause
4.6.2: Pilot Modulation Using {PRBS}.

\bibitem{10901415}
H.~Yin \emph{et~al.}, ``Evaluation and design criterion
for pulse-shaped {AFDM},'' in \emph{Proc. IEEE GLOBECOM}, Dec. 2024, pp. 4944--4949.

\bibitem{10531762}
M.~Mirabella, P.~Di~Viesti, and G.~M. Vitetta, ``On the use of a
two-dimensional cyclic prefix in {OTFS} modulation and its implications,''
\emph{IEEE Open J. Commun. Soc.}, vol.~5, pp.
3340--3367, May 2024.

\bibitem{yin22}
H.~Yin and Y.~Tang, ``{Pilot Aided Channel Estimation for AFDM in Doubly
Dispersive Channels},'' in \emph{Proc. IEEE/CIC ICCC}, Aug. 2022, pp. 308--313.

\bibitem{zhou24}
Y.~Zhou \emph{et~al.},  ``{GI-Free
Pilot-Aided Channel Estimation for Affine Frequency Division Multiplexing
Systems},'' \emph{arXiv preprint arXiv:2404.01088}, Apr. 2024.

\bibitem{zheng25}
K.~Zheng \emph{et~al.},  ``{Channel Estimation for AFDM
With Superimposed Pilots},'' \emph{IEEE Trans. Veh. Technol.}, vol.~74,
no.~2, pp. 3389--3394, Feb. 2025.

\bibitem{savaux25_PHYCOM}
V.~Savaux, ``Pilot design for multiple domains channel estimation in special
cases of affine frequency division multiplexing,'' \emph{Phys. Commun.}, vol.~73, p. 102863, Dec. 2025.

\bibitem{Savaux_DFTAFDM}
V.~Savaux, ``{Special Cases of {DFT}-Based Modulation and Demodulation for Affine
Frequency Division Multiplexing},'' \emph{IEEE Trans. Commun.}, vol.~72,
no.~12, pp. 7627--7638, Dec. 2024.

\bibitem{3GPP_38211r16}
3GPP, ``{NR}; {Physical channels and modulation},'' {3GPP}, Tech. Spec. TS 38.211
v16.2.0, Jul. 2020.

\bibitem{savaux17IET}
V.~Savaux and Y.~Lou\"{e}t, ``{LMMSE channel estimation in {OFDM} context: a
review},'' \emph{IET Signal Process.}, vol.~11, no.~2, pp. 123--134,
Apr. 2017.

\bibitem{li26sbl}
X.~Li \emph{et~al.},
``{Low-Complexity Channel Estimation for Internet of Vehicles {AFDM}
Communications With Sparse Bayesian Learning},'' \emph{IEEE Internet Things J.},
vol.~13, no.~5, pp. 9795--9810, Mar. 2026.

\bibitem{11220240}
C.~Shen, J.~Yuan, and J.~Tong, ``Time-domain zero-padding ({TZP}) {AFDM} with
two-stage iterative {MMSE} detection,'' \emph{IEEE Trans. Wireless Commun.},
vol.~25, pp. 6255--6269, 2026.

\bibitem{11214369}
H.~Hawkins \emph{et~al.}, ``Iterative soft-{MMSE} detection
aided {AFDM} and {OTFS},'' \emph{IEEE Open J. Veh. Technol.}, vol.~6, pp.
2944--2959, Oct. 2025.

\bibitem{11185315}
Z.~Sui \emph{et~al.}, ``Generalized spatial
modulation aided affine frequency division multiplexing,'' \emph{IEEE Trans.
Wireless Commun.}, vol.~25, pp. 4658--4673, 2026.

\bibitem{10806672}
Z.~Li \emph{et~al.}, ``Chirp
parameter selection for affine frequency division multiplexing with {MMSE}
equalization,'' \emph{IEEE Trans. Commun.}, vol.~73, no.~7, pp. 5079--5093,
Jul. 2025.

\bibitem{8859227}
S.~Tiwari, S.~S. Das, and V.~Rangamgari, ``Low complexity {LMMSE} receiver for
{OTFS},'' \emph{IEEE Commun. Lett.}, vol.~23, no.~12, pp. 2205--2209, Dec. 2019.

\bibitem{yi2025error}
Q.~Yi, Z.~Sui, and Z.~Liu, ``Error rate analysis and low-complexity receiver
design for zero-padded {AFDM},'' \emph{IEEE Trans. Veh. Technol.}, early access, 2026.

\bibitem{9746329}
A.~Bemani, N.~Ksairi, and M.~Kountouris, ``Low complexity equalization for
{AFDM} in doubly dispersive channels,'' in \emph{Proc. IEEE ICASSP}, May 2022,
pp. 5273--5277.

\bibitem{10566604}
Q.~Luo \emph{et~al.}, ``{AFDM}-{SCMA}: A
promising waveform for massive connectivity over high mobility channels,''
\emph{IEEE Trans. Wireless Commun.}, vol.~23, no.~10, pp. 14\,421--14\,436,
Oct. 2024.

\bibitem{11075959}
X.~Li \emph{et~al.}, ``Affine
frequency division multiplexing over wideband doubly-dispersive channels with
time-scaling effects,'' \emph{IEEE Trans. Wireless Commun.}, vol.~25, pp. 476--492, 2026.

\bibitem{RanasingheTWC2025}
K.~R.~R. Ranasinghe \emph{et~al.},
``Joint channel, data, and radar parameter estimation for {AFDM} systems in
doubly-dispersive channels,'' \emph{IEEE Trans. Wireless Commun.}, vol.~24,
no.~2, pp. 1602--1619, Feb. 2025.

\bibitem{11225907}
Y.~Xu \emph{et~al.}, ``{AFDM}-aided grant-free
random access for {LEO} {SIoT}: Performance analysis and near-optimal joint
detection,'' \emph{IEEE Trans. Commun.}, vol.~74, pp. 840--853, 2026.

\bibitem{11150613}
Q.~Luo \emph{et~al.}, ``Joint sparse
graph for enhanced {MIMO-AFDM} receiver design,'' \emph{IEEE Trans. Wireless
Commun.}, vol.~25, pp. 3272--3286, 2026.

\bibitem{tuchler2002minimum}
M.~Tuchler, A.~C. Singer, and R.~Koetter, ``Minimum mean squared error
equalization using \emph{a priori} information,'' \emph{IEEE Trans. Signal
Process.}, vol.~50, no.~3, pp. 673--683, Mar. 2002.

\bibitem{yin24}
H.~Yin \emph{et~al.}, ``{Diagonally Reconstructed Channel
Estimation for MIMO-AFDM With Inter-Doppler Interference in Doubly Selective
Channels},'' \emph{IEEE Trans. Wireless Commun.}, vol.~23, no.~10, pp.
14\,066--14\,079, Oct. 2024.

\bibitem{savaux24eusipco}
V.~Savaux and X.~Chen, ``{Spatial Precoding in Frequency Domain for Multi-User
MIMO Affine Frequency Division Multiplexing},'' in \emph{Proc. IEEE EUSIPCO},
Aug. 2024, pp. 2112--2116.

\bibitem{Arslan-OJCOM-2025}
R.~Y. Bir, A.~A. Boudjelal, and H.~Arslan, ``On the orthogonal coexistence of
{AFDM} and {OFDM} for joint sensing and communication,'' \emph{IEEE Open J.
Commun. Soc.}, vol.~6, pp. 10\,010--10\,022, Nov. 2025.

\bibitem{Arslan-RSMA-2025}
K.~Abela \emph{et~al.}, ``A {SIC}-free dual-domain {RSMA} strategy via {AFDM}-{OFDM} coexistence for {6G} networks,'' \emph{IEEE Trans. Commun.}, vol.~74, pp. 11\,315--11\,325, 2026.

\bibitem{Yin-ICCC-2025}
Y.~Yin \emph{et~al.}, ``Downlink {AFDM-RSMA} scheme
based on orthogonal chirps and sum-rate maximization,'' in \emph{Proc. IEEE/CIC ICCC}, Aug. 2025, pp. 1--6.

\bibitem{10342857}
H.~Liu \emph{et~al.}, ``{BER} analysis of
{SCMA-OFDM} systems in the presence of carrier frequency offset,'' \emph{IEEE
Commun. Lett.}, vol.~28, no.~1, pp. 213--217, Jan. 2024.

\bibitem{tao26afdma}
Y.~Tao \emph{et~al.}, ``{Affine Frequency
Division Multiple Access Based on {DAFT} Spreading for Next-Generation Wireless
Networks},'' \emph{IEEE Trans. Wireless Commun.}, vol.~25, pp. 4626--4641, 2026.

\bibitem{1044611}
L.~Piazzo and P.~Mandarini, ``Analysis of phase noise effects in {OFDM}
modems,'' \emph{IEEE Trans. Commun.}, vol.~50, no.~10, pp. 1696--1705, Oct.
2002.

\bibitem{sui2026mimo}
Z.~Sui \emph{et~al.},
``{MIMO-AFDM} outperforms {MIMO-OFDM} in the face of hardware impairments,''
\emph{IEEE Trans. Commun.}, vol.~74, pp. 10\,432--10\,447, 2026.

\bibitem{4156406}
D.~D. Lin and T.~J. Lim, ``The variational inference approach to joint data
detection and phase noise estimation in {OFDM},'' \emph{IEEE Trans. Signal
Process.}, vol.~55, no.~5, pp. 1862--1874, May 2007.

\bibitem{10841966}
Z.~Sui \emph{et~al.}, ``Performance analysis and
optimization of {STAR-RIS}-aided cell-free massive {MIMO} systems relying on
imperfect hardware,'' \emph{IEEE Trans. Wireless Commun.}, vol.~24, no.~4,
pp. 2925--2939, Apr. 2025.

\bibitem{bjornson2015massive}
E.~Bj{\"o}rnson, M.~Matthaiou, and M.~Debbah, ``Massive {MIMO} with non-ideal
arbitrary arrays: Hardware scaling laws and circuit-aware design,''
\emph{IEEE Trans. Wireless Commun.}, vol.~14, no.~8, pp. 4353--4368, Aug.
2015.

\bibitem{bacsar2013orthogonal}
E.~Ba{\c{s}}ar \emph{et~al.},
``Orthogonal frequency division multiplexing with index modulation,''
\emph{IEEE Trans. Signal Process.}, vol.~61, no.~22, pp. 5536--5549, Nov. 2013.

\bibitem{9507331}
Z.~Sui  \emph{et~al.},``Approximate message
passing algorithms for low complexity {OFDM-IM} detection,'' \emph{IEEE
Trans. Veh. Technol.}, vol.~70, no.~9, pp. 9607--9612, Sep. 2021.

\bibitem{tao24}
Y.~Tao \emph{et~al.}, ``{Affine Frequency Division Multiplexing
With Index Modulation},'' in \emph{Proc. IEEE WCNC}, Apr. 2024, pp. 1--6.

\bibitem{mesleh2008spatial}
R.~Y. Mesleh \emph{et~al.}, ``Spatial
modulation,'' \emph{IEEE Trans. Veh. Technol.}, vol.~57, no.~4, pp.
2228--2241, Jul. 2008.

\bibitem{HSRou_TWC2022}
H.~S. Rou \emph{et~al.}, ``Scalable quadrature spatial modulation,'' \emph{IEEE Trans. Wireless Commun.}, vol.~21, no.~11, pp. 9293--9311, Nov. 2022.

\bibitem{10129061}
Z.~Sui \emph{et~al.}, ``Low complexity
detection of spatial modulation aided {OTFS} in doubly-selective channels,''
\emph{IEEE Trans. Veh. Technol.}, vol.~72, no.~10, pp. 13\,746--13\,751,
Oct. 2023.

\bibitem{10250854}
Z.~Sui \emph{et~al.}, ``Space-time shift keying
aided {OTFS} modulation for orthogonal multiple access,'' \emph{IEEE Trans.
Commun.}, vol.~71, no.~12, pp. 7393--7408, Dec. 2023.

\bibitem{10342712}
J.~Zhu \emph{et~al.},``Design and performance
analysis of index modulation empowered {AFDM} system,'' \emph{IEEE Wireless
Commun. Lett.}, vol.~13, no.~3, pp. 686--690, Mar. 2024.

\bibitem{Temiz_2025}
M.~Temiz and C.~Masouros, ``{Affine Frequency Division Multiplexing with
Subcarrier Power-Level Index Modulation for Integrated Sensing and
Communications},'' in \emph{Proc. IEEE SPAWC}, Jul. 2025, pp. 1--5.

\bibitem{anoop25}
A.~Anoop \emph{et~al.}, ``{Dual-mode Index
Modulation based on Affine Frequency Division Multiplexing},'' \emph{Phys. Commun.}, vol.~70, p. 102628, Jun. 2025.

\bibitem{liu25}
G.~Liu \emph{et~al.}, ``{Multiple-Mode Affine
Frequency Division Multiplexing with Index Modulation},'' \emph{IEEE Wireless Commun. Lett.}, vol.~15, pp. 141--145, 2026.

\bibitem{Rou_Arxiv25_CPAFDM}
H.~S. Rou and G.~T.~F. de~Abreu, ``Chirp-permuted {AFDM}: A versatile waveform
design for {ISAC} in {6G},'' \emph{arXiv preprint arXiv:2507.20825}, Jul. 2025.

\bibitem{savaux256Gnetafdm}
V.~Savaux \emph{et~al.}, ``{On the Noise
Robustness of Affine Frequency Division Multiplexing: Analysis and
Applications},'' in \emph{Proc. 6GNet}, Dec. 2025, pp. 65--72.

\bibitem{reddy25}
V.~M. Reddy and H.~Bitra, ``{PAPR in AFDM: Upper Bound and Reduction With
Normalized $\mu$-Law Companding},'' \emph{IEEE Access}, vol.~13, pp.
86\,553--86\,561, 2025.

\bibitem{ali25}
A.~Ali, A.~Arous, and H.~Arslan, ``{Spreading the Wave: Low-Complexity {PAPR}
Reduction for {AFDM} and {OCDM} in {6G} Networks},'' \emph{IEEE Trans. Green Commun. Netw.}, vol.~10, pp. 1565--1577, 2026.

\bibitem{yuan25}
H.~Yuan \emph{et~al.}, ``{PAPR
Reduction With Pre-Chirp Selection for Affine Frequency Division
Multiplexing},'' \emph{IEEE Wireless Commun. Lett.}, vol.~14, no.~3, pp.
736--740, Mar. 2025.

\bibitem{lu25}
Z.~Lu, M.~El-Hajjar, and L.-L. Yang, ``{Augmented Affine Frequency Division
Multiplexing for Low {PAPR} Signaling and Diversity Gain Protection},''
\emph{IEEE Access}, vol.~14, pp. 69\,426--69\,442, 2026.

\bibitem{surabhi19otfspn}
G.~D. Surabhi, M.~Kollengode Ramachandran, and A.~Chockalingam, ``{OTFS}
modulation with phase noise in {mmWave} communications,'' in \emph{Proc.
IEEE VTC2019-Spring}, Kuala Lumpur, Malaysia, Apr.--May 2019, pp. 1--5.

\bibitem{abushattal23otfsrf}
A.~Abushattal \emph{et~al.}, ``A comprehensive
experimental emulation for {OTFS} waveform {RF}-impairments,'' \emph{Sensors},
vol.~23, no.~1, art.~38, Jan. 2023.

\end{thebibliography}
\end{document}